\documentclass[aps,prc,superscriptaddress,showpacs,nofootinbib]{revtex4}
\usepackage{epsfig,dcolumn,bm}

\usepackage{graphicx}
\usepackage{amsmath}
\usepackage{rotating}

\def\tstrut{\vrule height2.5ex depth0pt width0pt} % used in tables
\newcommand{\lc}{{\cal L}}

\newcommand{\oh}{\frac{1}{2}}
\newcommand{\trh}{\frac{3}{2}}
\newcommand{\fvh}{\frac{5}{2}}

\newcommand{\be}{\begin{eqnarray}}
\newcommand{\ee}{\end{eqnarray}}
\newcommand{\ba}{\begin{array}}
\newcommand{\ea}{\end{array}}
\newcommand{\bt}{\begin{tabular}}
\newcommand{\et}{\end{tabular}}
\newcommand{\btab}{\begin{table}}
\newcommand{\etab}{\end{table}}
\newcommand{\bc}{\begin{center}}
\newcommand{\ec}{\end{center}}
\newcommand{\nn}{\nonumber}

\newcommand{\Eq}[1]{Eq.~(\ref{#1})}

\newcommand{\del}{\Delta}
\newcommand{\sigs}{{\Sigma^*}}
\newcommand{\sig}{{\Sigma}}

\newcommand{\NULL}{}
\newcommand{\MeV}{{\rm MeV}}

\newcommand{\ignore}[1]{}

\begin{document}

\title{Odd parity light baryon resonances}

\author{D. Gamermann} \email{daga1@upvnet.upv.es}
\affiliation{C\'atedra Energesis de Tecnolog\'ia Interdisciplinar,
Universidad Cat\'olica de Valencia San Vicente M\'artir, \\ Guillem de
Castro 94, E-46003, Valencia, Spain} \affiliation{Instituto
Universitario de Matem\'atica Pura y Aplicada, Universidad
Polit\'ecnica de Valencia,\\ Camino de Vera 14, 46022 Valencia, Spain}

\author{C.~Garc\'ia-Recio}
\email{g_recio@ugr.es}
\affiliation{Departamento de F{\'\i}sica At{\'o}mica, Molecular y Nuclear,
  Universidad de Granada, E-18071 Granada, Spain}
\affiliation{ Instituto Carlos I de F{\'\i}sica Te\'orica y Computacional,
  Universidad de Granada, E-18071 Granada, Spain}

\author{J.~Nieves}
\email{jmnieves@ific.uv.es}
\affiliation{Instituto de F\'isica corpuscular (IFIC), Centro Mixto
Universidad de Valencia-CSIC,\\ Institutos de Investigaci\'on de
Paterna, Aptdo. 22085, 46071, Valencia, Spain}

\author{L.L.~Salcedo}
\email{salcedo@ugr.es}
\affiliation{Departamento de F{\'\i}sica At{\'o}mica, Molecular y Nuclear,
  Universidad de Granada, E-18071 Granada, Spain}
\affiliation{ Instituto Carlos I de F{\'\i}sica Te\'orica y Computacional,
  Universidad de Granada, E-18071 Granada, Spain}

\begin{abstract}
We use a consistent SU(6) extension of the meson-baryon chiral
Lagrangian within a coupled channel unitary approach in order to
calculate the $T$-matrix for meson-baryon scattering in $s$-wave. The
building blocks of the scheme are the $\pi$ and $N$ octets, the $\rho$
nonet and the $\Delta$ decuplet.  We identify poles in this unitary
$T$-matrix and interpret them as resonances. We study here the non
exotic sectors with strangeness $S=0,-1,-2,-3$ and spin $J=\oh, \trh$
and $\fvh$. Many of the poles generated can be associated with known
$N$, $\del$, $\sig$, $\Lambda$ and $\Xi$ resonances with negative
parity. We show that most of the low-lying three and four star odd
parity baryon resonances with spin $\oh$ and $\trh$ can be related 
 to multiplets of the spin-flavor symmetry group SU(6). This
study allows us to predict the spin-parity of the $\Xi(1620)$,
$\Xi(1690)$, $\Xi(1950), \Xi(2250), \Omega(2250)$ and $\Omega(2380)$
resonances, which have not been determined experimentally yet.
\end{abstract}


\date{\today}

\maketitle

%%%%%%%%%%%%%%%%
%%%%%%%%%%%%%%%%
%%%%%%%%%%%%%%%%%
%%%%%%%%%%%%%%%%%

\section{Introduction}

To understand the structure of existing baryons has been a challenging task
for many years and different approaches have been used. (See e.g.
\cite{Chao:1980em,Capstick:1986bm,Glozman:1997ag,Valcarce:2005rr,%
  Carlson:2000zr,Schat:2001xr, Goity:2003ab, Matagne:2004pm, Bijker:2000gq,
  Jido:1996zw, Oh:2007cr}.) The interpretation of baryons as three quark
states works for ground state baryons but fails in the description of many
resonances. Some examples of resonances better described by other approaches
are the $\Lambda(1405)$ which can be interpreted as a meson-baryon system
\cite{lamb1,lamb2,lamb3,GarciaRecio:2003ks} or the Roper resonance
\cite{roper}.

Since the pioneering works of Refs.~\cite{KSW95a, KSW95b}, the
Goldstone boson--baryon scattering, using for the dynamics constrains
from chiral symmetry, has been studied in several papers. From the
study of the scattering of pseudoscalars with the $J^P=\oh^+$ baryons
in the strangenessless sector, two poles that can be associated with
the $N(1535)$ and the $\del(1620)$ resonances were found in
\cite{inoue}. The $S~{\rm (strangeness)}=0$, $I~{\rm (isospin)} =\oh$
sub-sector has been also studied in \cite{juan3}, where in addition to
the $N(1535)$ pole, the $N(1650)$ resonance was also dynamically
generated. A follow up of this work \cite{lamb3} analyzed the $S=-1$,
$I=0$ sector where poles associated with the $\Lambda(1405)$ and
$\Lambda(1670)$ states were found and some of the findings of previous
works \cite{lamb1,lamb2} were also confirmed.

In \cite{Kolomeitsev:2003kt, sarkar} the interaction of the baryon
decuplet with the pseudoscalar mesons was first studied and signatures
of various $J^P=\trh^-$ baryon resonances were obtained. Some of these
resonances are the $\Xi(1820)$, $\Lambda(1520)$ and the
$\sig(1670)$. More recently the interaction of vector mesons with
baryons is being also studied within the formalism of the hidden gauge
interaction for vector mesons~\cite{Bando:1984ej,Bando:1987br}. In
\cite{eulogio} the $\rho\del$ interaction is studied and the authors
find an explanation of why there are $J^P=\oh^-,\trh^-,\fvh^-$ deltas
nearly degenerate around $1900\,{\rm MeV}$. Later, this work has been extended
in order to study all possible sectors in the interaction of the
baryon decuplet with the vector meson octet \cite{bao} and many poles
are obtained that can be associated with known experimental
states. Also the interaction of vector mesons with the baryon octet
has been studied \cite{Lutz:2001mi,angels} and again signatures of 
various states were found.

In principle there is no reason to expect that the interaction of
pseudoscalar mesons with baryons and the interaction of vector mesons
with baryons should be decoupled for channels which share $S$, $I$,
and $J^P$ (spin-parity) quantum numbers. In this work we want to
explore the consequences of coupling these sectors that have been
treated independently in previous works. In order to do that we use an
SU(6) framework which combines spin and flavor symmetries.  Thus, we
will study the $s-$wave meson-baryon interaction, where the hadrons
belong to the 35 ($\pi$-octet + $\rho$-nonet) and the 56 ($N$-octet +
$\Delta$-decuplet) SU(6) multiplets. We will use an enlarged
Weinberg-Tomozawa (WT) meson--baryon Lagrangian to accommodate vector
mesons and decuplet baryons, which guarantees that chiral symmetry is
recovered when interactions involving pseudoscalar Nambu-Goldstone
bosons are being examined\footnote{A similar study for the case of the
scattering of two mesons of the 35-plet was carried out in
Ref.~\cite{GarciaRecio:2010ki}.}.  Chiral symmetry constraints the
pseudoscalar octet--baryon decuplet interactions, and the interactions
derived here coincide with those employed in
Refs.~\cite{Kolomeitsev:2003kt} and \cite{sarkar}.  However, the
interaction of vector mesons with baryons is not constrained by chiral
symmetry, and the model presented here differs from previous
ones~\cite{bao,angels}, based in the hidden gauge formalism.

\begin{figure}[tbh]
%\vspace{-3cm}
\centerline{\hspace{-2cm}\includegraphics[height=11.cm]{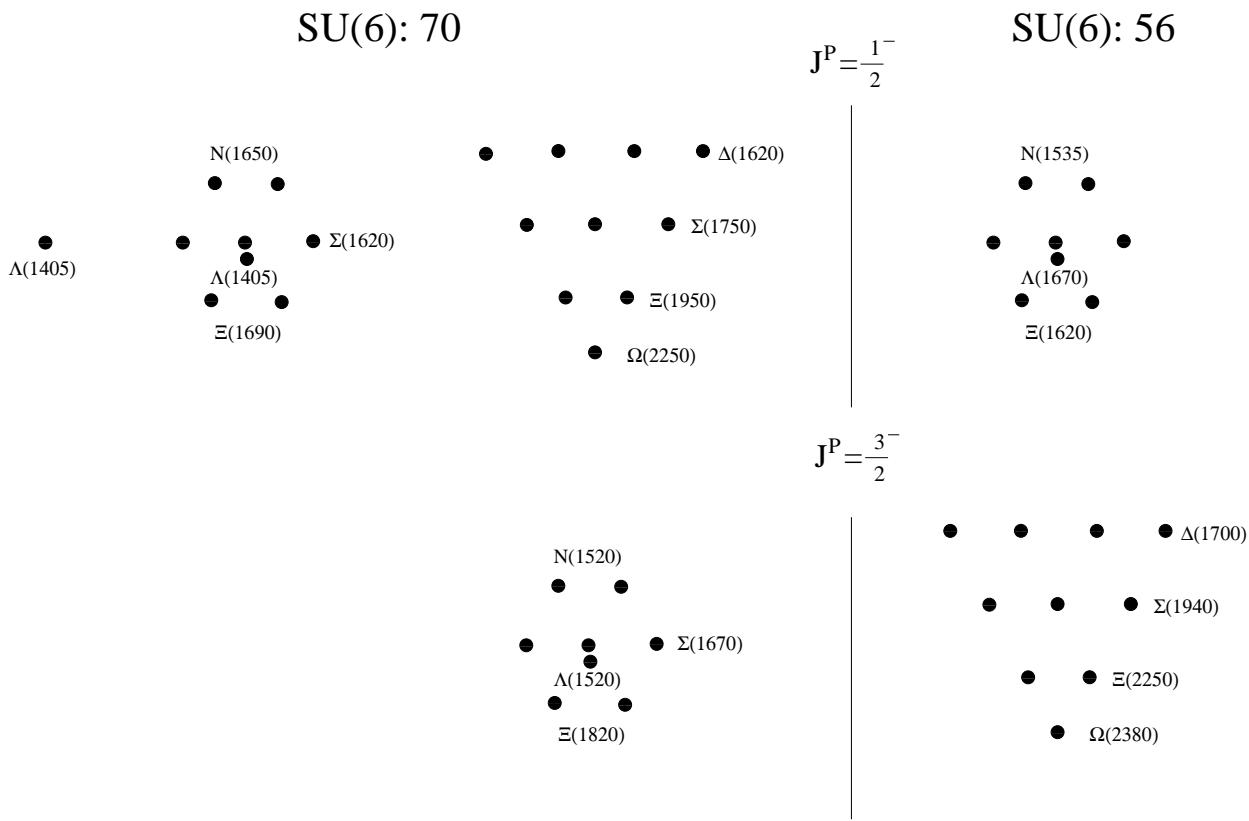}}
%\vspace{-15cm}
\caption{Qualitative SU(6) classification derived from this work for
  the experimental low-lying $\oh^-$ and $\trh^-$ baryon
  resonances. This classification is made attending to the relation of
  each resonance with the different SU(6)/SU(3) multiplets that appear
  in the spin and/or flavor symmetric scenario described in
  Sect.~\ref{sec:res}. Most of the odd parity three and four star
  resonances of the PDG are included. Few two star ($\Sigma(1620)$,
  $\Xi(2250)$, and $\Omega(2380)$) and one star ($\Xi(1620)$)
  resonances complete the SU(6) multiplets. The only exception is the missing
  $\Sigma$ state in the $\oh^-$ octet of the 56.  The classification
  predicts the experimentally not yet known spin-parity of five
  resonances: $\Xi(1620)$, $\Xi(1690)$, $\Xi(1950)$, $\Xi(2250)$,
  $\Omega(2250)$, and $\Omega(2380)$. Most of the resonances have
  large SU(6) and SU(3) mixing.}
\label{fig:summ}
\end{figure}

As a result of the present work, we show that most of the low-lying three and
four star odd parity baryon resonances with spin $\oh$ and $\trh$ can be
related to multiplets of the spin-flavor symmetry group SU(6). This can be
seen in Fig.~\ref{fig:summ}, which summarizes the better theoretically founded
set of dynamically generated resonances obtained in this work. The spin-parity
of the $\Xi(1620)$, $\Xi(1690)$, $\Xi(1950)$, $\Xi(2250)$, $\Omega(2250)$ and
$\Omega(2380)$ resonances, not experimentally determined yet,\footnote{The
  BABAR collaboration finds some evidence that the $\Xi(1950)$ has spin-parity
  $1/2^-$ in \cite{Aubert:2008ty}. Our model corroborates this assignment.}
can be read off the figure. The classification is qualitative. Actually, each
resonance displays mixing of SU(6) and SU(3) configurations (see below).

In the next section we briefly explain the SU(6) theoretical model and
the mathematical framework needed in order to calculate the $T$-matrix
and identify its poles. Also in that section, we devote a few words to
clarify the differences between our scheme and that used in
Refs.~\cite{bao,angels}. In Sec. \ref{sec:res} we present and discuss
our results and in the last section we summarize our
conclusions. There are also two appendices. The first one includes
tables with the elementary amplitudes of the model, while in the
second one, we discuss the predictions of the present model for $\pi N
$ phase shifts and inelasticities in the $S_{11}$ sector.

%%%%%%%%%%%%%%%%%%%%%%%%%%

\section{Framework}

\subsection{The SU(6) structure of the interaction}

We follow here the spin-flavor symmetric model introduced in
\cite{su6model,GarciaRecio:2006wb,juangs} for meson-baryon resonances. The
model is an SU(6) extension of the Weinberg-Tomozawa Lagrangian for
meson-baryon interactions which assumes that the quark interactions are
approximately spin and SU(3) independent.  As shown in \cite{Caldi:1975tx}
(see also \cite{GarciaRecio:2010ki}), spin-flavor and chiral symmetries are
consistent, as they can be naturally incorporated into a larger symmetry
group, corresponding to the Feynman--Gell-Mann--Zweig algebra
\cite{Feynman:1964fk}. Moreover, in the presence of heavy quarks the analogous
scheme automatically embodies heavy quark spin symmetry, another well
established approximate symmetry of QCD. The model has been extended to the
charm sector in \cite{GarciaRecio:2008dp,Gamermann:2010zz} and to the
study of meson-meson
light resonances in \cite{GarciaRecio:2010ki}.

In this SU(6) scheme, the baryons are represented by a 56-plet and the mesons
by a 35-plet plus a singlet.  The Lagrangian is a contact interaction obtained
by coupling the mesonic current ($35\otimes 35$) to the baryonic one
($56\otimes 56^*$). Such coupling takes place through an implicit
$35$-like (i.e. adjoint representation) exchange in the $t$-channel:
\be
\lc_{WT}^\textrm{SU(6)}\propto 
[[ M^\dagger\otimes M]_{35_a}\otimes [B^\dagger \otimes B]_{35}]_1 .
\label{eq:wtlag}
\ee

The 56-plet of baryons in SU(6) contains the spin $\oh^+$ and $\trh^+$ ground
state baryons while the 35-plet of mesons contains the pseudoscalar and
vector mesons. To visualize this, we show the SU(3)$\otimes$SU(2) content of
each one of these multiplets:
\be
56&\rightarrow&8_2\oplus 10_4 \\
35&\rightarrow&8_1\oplus 8_3\oplus 1_3
.
\ee
In these equations the left-hand side indicates the SU(6) content of a
multiplet and the right-hand side displays the SU(3)$\otimes$SU(2)
pattern into which it breaks. As it is standard, the regular case
number indicates the SU(3) multiplet while the subindex indicates the
number of spin states (the SU(2) content). So $8_2$ for example is a
spin $\oh$ (two spin states) SU(3) octet, while $8_1$ represents a
pseudoscalar (a single spin state) SU(3) octet.

From the point of view of SU(6), the meson-baryon interaction is represented
by the product:
\be
56\otimes 35&=& 56\oplus 70\oplus 700\oplus 1134
.
\ee
Therefore the single $35$-like coupling in the $t$-channel (cf. \Eq{eq:wtlag})
corresponds to four $s$-channel couplings. These are proportional to the
following eigenvalues \cite{su6model,GarciaRecio:2006wb}:
\be
\lambda_{56}=-12,
\quad
\lambda_{70}=-18,
\quad
\lambda_{700}=6,
\quad
\lambda_{1134}=-2
.
\label{eq:lambdas}
\ee

Under SU(3)$\otimes$SU(2) these four SU(6) multiplets break as follows:
\be
56&\rightarrow&8_2\oplus 10_4 \label{eq:56}\\
70&\rightarrow&1_2\oplus 8_2\oplus 10_2 \oplus 8_4 \label{eq:70}\\
700&\rightarrow&8_2\oplus 10_2\oplus 10^*_2\oplus 27_2\oplus 8_4\oplus 10_4\oplus 27_4\oplus 35_4\oplus 10_6\oplus 35_6 \label{eq:700}\\
1134&\rightarrow&1_2\oplus 3\times8_2\oplus 2\times10_2\oplus 10^*_2\oplus 2\times27_2\oplus 35_2\oplus 1_4
\oplus 3\times8_4\oplus 2\times10_4\oplus 10^*_4 \nn\\
&\oplus& 2\times27_4\oplus 35_4\oplus 8_6\oplus 10_6\oplus
27_6 \label{eq:1134}
.
\ee
The SU(3) multiplets interacting for each possible value of $J$ are displayed
in Table \ref{tab:SU3}.

\begin{table}[h]
\caption{SU(3) reduction of interacting multiplets. }
\label{tab:SU3}
\begin{center}
\begin{tabular}{r|rcl}
\hline
& 
$8_2\otimes 8_1$ &=& $1_2\oplus 8_2\oplus 8_2\oplus 10_2\oplus 10^*_2\oplus 27_2 $
\\
$J=\frac{1}{2}$
 & $8_2\otimes 8_3$ &=& $1_2\oplus 8_2\oplus 8_2\oplus 10_2\oplus 10^*_2\oplus 27_2 $
\\ &
$ 10_4\otimes8_3$  &=& $ 8_2\oplus 10_2\oplus 27_2\oplus 35_2$
\\ &
$ 8_2\otimes1_3$ &=& $8_2 $
\\ &
$10_4\otimes1_3$ &=& $10_2$
\\
\hline
&
$10_4\otimes 8_1$ &=& $8_4\oplus 10_4\oplus 27_4\oplus 35_4$
\\ 
$ J=\frac{3}{2}$  &
$10_4\otimes8_3$ &=& $8_4\oplus 10_4\oplus 27_4\oplus 35_4$
\\ &
$10_4\otimes1_3$ &=& $10_4$
\\ &
$8_2\otimes8_3$ &=& $1_4\oplus 8_4 \oplus 8_4 \oplus 10_4 \oplus 10^*_4
\oplus 27_4$
\\ &
$8_2\otimes1_3$ &=& $8_4$
\\
\hline
 &
$10_4\otimes8_3$ &=& $8_6\oplus10_6\oplus27_6\oplus35_6$
\\
$ J=\frac{5}{2}$ &
$10_4\otimes 1_3$ &=& $10_6$
\\
\hline
\end{tabular}
\end{center}
\end{table}

With our conventions for the potential a negative sign implies attraction. So,
attending to the eigenvalues in Eq.~(\ref{eq:lambdas}), there are two strongly
attractive multiplets ($56$ and $70$), a weakly attracting one ($1134$) and a
repulsive multiplet ($700$). The attractive sectors are candidates for
dynamically generated resonances. This can be analyzed in terms of SU(3)
multiplets and further in terms of sectors with well defined strangeness,
isospin and spin.

For $J^P=\oh^-$ the attractive SU(3) multiplets are the 35-plet, two 27-plets,
one $10^*$-plet, three 10-plets, five octets and the two singlets. In the
$J^P=\trh^-$ sector one has one 35-plet, two 27-plets, one $10^*$-plet,
three 10-plets, four octets and a singlet attractive and for $J^P=\fvh^-$
there is one 27-plet, one 10-plet and one octet attractive. These attractive
multiplets account for the attractive ones from the 56-plet, 70-plet and
1134-plet of Eqs. (\ref{eq:56}), (\ref{eq:70}) and (\ref{eq:1134}). In Table
\ref{tab:states} we show a counting of the number of states that, in
principle, one can expect to generate based on the attractive multiplets of
the model.  As usual the particle label assigned to the state refers to its
flavor quantum numbers. The states without a label in Table \ref{tab:states}
are exotic in the sense that they require SU(3) irreducible
representations (irreps)
not present in $3\otimes 3\otimes 3= 1 \oplus 8 \oplus 8 \oplus
10$. All exotic states are placed in the weakly attracting 1134 irrep,
together with all spin  $\fvh^-$ and most of the $\oh^-$ and
~$\trh^-$ non-exotic $N,\Delta,\Sigma,\Lambda, \Xi$ and $\Omega$ ones.
\btab
\bc
\caption{Expected number of states generated by the model in each
  sector. Here, $SIJ^P$ stand for strangeness, isospin and
  spin-parity, respectively. }
\label{tab:states}
\vspace{0.2cm}
\bt{ccc|ccc||ccc}
\hline
\tstrut
& &  & \multicolumn{3}{c||}{\small $56\oplus 70 \oplus 1134$} &
\multicolumn{3}{c}{\small $ 56\oplus 70$} \\ 
& &  & & $J^P$ & &\multicolumn{3}{c}{$J^P$} \\
$S$ & $I$ &  state  & ~ $\oh^-$ & ~$\trh^-$ &  ~$\fvh^-$ & ~ $\oh^-$ &
~$\trh^-$ &  \phantom{$\Big |$} \hspace{-0.1cm}$\fvh^-$  \\ 
\hline
\hline\tstrut
0 & $1/2$  & $N$ & 8 & 7 & 2 & 2 & 1 & 0\\
0 & $3/2$ & $\del$ & 6 & 6 & 2 & 1 & 1 & 0 \\
$-$1 & 1 & $\sig$ & 12 & 11 & 3 & 3 & 2 & 0\\
$-$1 & 0 & $\Lambda$ & 9 & 7 & 2 & 3 & 1 & 0\\ 
$-$2 & $1/2$ & $\Xi$ & 11 & 10 & 3 & 3 & 2 & 0\\
$-$3 & 0 & $\Omega$ & 4 & 4 & 1 & 1 & 1 & 0\\
\hline
1 & 0 &  & 1 & 1 & 0 & 0 & 0 & 0\\
1 & 1 &  & 2 & 2 & 1 & 0 & 0 & 0\\ 
1 & 2 &  & 1 & 1 & 0 & 0 & 0 & 0\\
0 & $5/2$ &  & 1 & 1 & 0 & 0 & 0 & 0 \\
$-$1 & $2$ &  & 3 & 3 & 1 & 0& 0 & 0\\
$-$2 & $3/2$ &  & 4 & 4 & 1 & 0 & 0 & 0\\
$-$3 & $1$ &  & 3 & 3 & 1 & 0 & 0 & 0\\
$-$4 & $1/2$ &  & 1 & 1 & 0 & 0 & 0 & 0\\
\hline
\et
\ec
\etab
From Table \ref{tab:states} one expects that the model generates a very rich
spectrum. This is actually the case, but not all candidate multiplets result
in poles that can be associated with physical states. Some poles appear in the
wrong Riemann sheet and, therefore, can not be associated with physical
states. In addition, though SU(6) symmetry is the driven ingredient to fix the
interaction, it is broken in the kinematics. Indeed, we implement some source
of spin-flavor symmetry breaking first by using physical masses\footnote{For
  the physical masses of the mesons we use, $m_\pi=137.5\,{\rm MeV}$,
  $m_K=496\,{\rm MeV}$, $m_\eta=548\,{\rm MeV}$, $m_\rho=775\,{\rm MeV}$,
  $m_{K^*}=894\,{\rm MeV}$, $m_\omega=783\,{\rm MeV}$ and $m_\phi=1019\,{\rm
    MeV}$ and for the physical masses of the baryons we use, $M_N=939\,{\rm
    MeV}$, $M_\Lambda=1116\,{\rm MeV}$, $M_\Sigma=1193\,{\rm MeV}$,
  $M_\Xi=1318\,{\rm MeV}$, $M_\Delta=1210\,{\rm MeV}$, $M_{\Sigma^*}=
  1385\,{\rm MeV}$, $M_{\Xi^*}=1533\,{\rm MeV}$ and $M_\Omega=1672\,{\rm
    MeV}$.} for all the mesons and baryons and second by the use of different
meson decay constants. Both types of symmetry breaking terms induce SU(6)
violations not only at the level of kinematics, but also in the interactions.
The values for the decay constants of the mesons we use are as follows (see
Table II of Ref.~\cite{GarciaRecio:2008dp}):
\be
f_{\pi}=92.4\, {\rm MeV}, \quad f_K=113\,{\rm MeV},  
\quad f_\eta=111 \, {\rm MeV}, \quad
f_\rho=f_{K^*}=153 \, {\rm MeV}, \quad f_\omega=138 \, {\rm MeV},
\quad f_\phi=163 \, {\rm MeV}
\label{eq:efes}
\ee

We assume an ideal mixing in the isoscalar vector meson sector, namely,
$\omega=\sqrt{\frac{2}{3}}\omega_1+\frac{1}{\sqrt 3}\omega_8$ and
$\phi=\sqrt{\frac{2}{3}}\omega_8-\frac{1}{\sqrt 3}\omega_1$.

\subsection{Coupled channels and unitarization}

All meson-baryon pairs coupled to the same $SIJ$ quantum numbers span the
coupled channel space. The $s$-wave tree level amplitudes between two
channels for each $SIJ$ sector are given by:
\be
V_{ij}^{SIJ}
&=&
\xi_{ij}^{SIJ}\frac{2\sqrt{s}-M_i-M_j}{4f_if_j} 
\sqrt{\frac{E_i+M_i}{2M_i}}\sqrt{\frac{E_j+M_j}{2M_j}}, 
, 
\label{eq:pot}
\ee
where $\sqrt{s}$ is the center of mass (C.M.) energy of the system, $E_i$ and
$M_i$ are, respectively, the C.M. energy and mass of the baryon in the channel
$i$, $f_i$ is the decay constant of the meson in the channel $i$, finally
$\xi_{ij}^{SIJ}$ are coefficients coming from the $\rm{SU}(6)$ group structure
of the couplings. That is,
\be \xi^{SIJ}_{ij}=
\sum_r\lambda_r \,[P_r]^{SIJ}_{ij}
,
\label{eq:xi}
\ee
where $r=56,70,700,1134$, $P_r$ denotes the projector on the SU(6) irreducible
representation $r$ and the eigenvalues $\lambda_r$ are given in
\Eq{eq:lambdas}.  Tables for the coefficients $\xi$ can be found in the
appendix \ref{app:tables}.

We use the matrix $V^{SIJ}$ as kernel to calculate the $T$-matrix:
\be
T^{SIJ}&=&(1-V^{SIJ}G^{SIJ})^{-1}V^{SIJ}\label{eq:bse},
\ee
where $G^{SIJ}$ is a diagonal matrix containing the two particle propagator
for each channel. Explicitly
\be
G^{SIJ}_{ii} = 2M_i
(\bar{J}_0(\sqrt{s};M_i,m_i) - \bar{J}_0(\mu^{SI};M_i,m_i)).\label{eq:subs}
\ee
$m_i$ is the mass of the meson in the channel $i$. The loop function
$\bar{J}_0$ can be found in the appendix of \cite{juan3} for the different
relevant Riemann sheets. The two particle propagator diverges logarithmically
and to make it finite we have adopted the prescription of 
~\cite{hofmann,hofmann2} which we now describe. The loop is renormalized by a
subtraction constant such that
\be
G_{ii}^{SIJ}=0 \quad\text{at~~} \sqrt{s}=\mu^{SI}. 
\label{eq:musi}
\ee
To fix the subtraction point $\mu^{SI}$, all sectors with a common $SI$ and
different $J$ and all corresponding channels are considered. Then $\mu^{SI}$
is taken as $\sqrt{m_{\rm{th}}^2+M_{\rm{th}}^2}$, where $m_{\rm{th}}$ and
$M_{\rm{th}}$, are respectively, the masses of the meson and baryon producing
the lowest threshold (minimal value of $m_{\rm{th}}+M_{\rm{th}}$). Therefore
the subtraction point takes a common value for all sectors $SIJ$ with equal
$SI$. Results from this renormalization scheme (RS), but involving only the
mesons and baryons of the pion and nucleon octets were
already obtained in \cite{GarciaRecio:2003ks}.

With all these ingredients we look for the poles of the $T$-matrix on the
$\sqrt{s}$ complex plane. Following extended practice, the poles on the first
Riemann sheet on the real axis and below threshold will be called bound
states. Poles on the second Riemann sheet (SRS) below the real axis and above
threshold will be called resonances. Poles on the SRS on or below the real
axis but below threshold will be called virtual states. Poles appearing in
different positions than the ones mentioned can not be associated with
physical states and are, therefore, artifacts.  The real part of the pole
position on the $\sqrt{s}$ complex plane is associated with the mass of the
state, and the imaginary part is associated with minus one half of its
width. Further information that can be extracted from the poles of the
$T$-matrix is the residue, related to the couplings of the states to their
coupled channels. Close to a pole the $T$-matrix can be written as:
\be
T^{SIJ}_{ij}(z) \approx \frac{g_ig_j}{z-z_R}, 
\ee
where $z_R$ is the pole position in the $\sqrt{s}$ plane and the $g_k$ is the
dimensionless coupling of the resonance to channel $k$. So, by calculating the
residues of the $T$-matrix at the pole, one obtains the product of the
couplings $g_ig_j$.

Some of the channels considered have mesons or baryons which are themselves
resonances.  The meson-baryon resonances with large couplings to these
channels and with mass close to these thresholds may have their width enhanced
by the decay of its components. Following \cite{roca, meuax}, we take this
effect into account by convoluting the loop function $G^{SIJ}$ of these
channels with the spectral function of the unstable particles in the
channel. We use for the width of the unstable particles the values:

\begin{equation}
\Gamma_\rho=150 \textrm{ MeV}, \quad
\Gamma_{K^*}=50 \textrm{ MeV}, \quad
\Gamma_\Delta=120 \textrm{ MeV}, \quad 
\Gamma_\sigs=35 \textrm{ MeV} 
\end{equation}

\subsection {SU(6) spin--flavor vs hidden gauge formalism for
  vector interactions}

As mentioned in the introduction, the $s-$wave interaction of vector mesons
with the octet of stable spin-parity $\oh^+$ baryons and with the resonances
of $\Delta(1232)$ decuplet has been previously studied in
Refs.~\cite{bao,angels}. Both works are based on the local hidden gauge
formalism for vector interactions and use a coupled channel unitary
approach. In this subsection, we devote a few words to clarify the main
theoretical differences between the schemes based on hidden gauge approach and
the scheme based on spin-flavor employed here. Namely

\begin{itemize}
\item[i)] \underline{Dynamics}: As we commented above and explained in
more detail in Ref.~\cite{su6model}, the interaction in our model is
of the form $\left[({ 35}\otimes { 35})_{{ 35_a}}\otimes({
56^*}\otimes { 56})_{{ 35}}\right]_{ 1}$ in the
$t-$channel. This can be regarded as the zero-range $t-$channel
exchange of a full $35$ irreducible representation of SU(6), carried by
an octet of spin 0 and a nonet of spin 1 even parity mediators.

\begin{figure}[tbh]
%\vspace{-3cm}
\centerline{\hspace{-2cm}\includegraphics[height=5.cm]{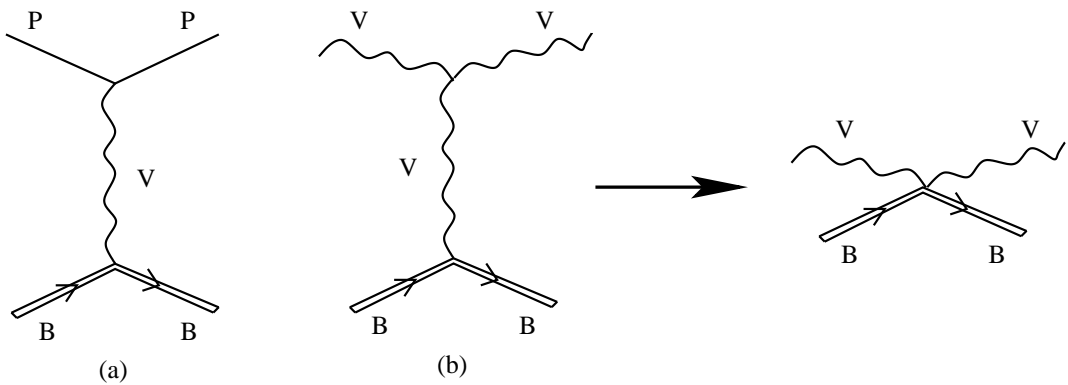}}
%\vspace{-15cm}
\caption{Diagrams contributing to the pseudoscalar-baryon (a) or
vector- baryon (b) interaction via the exchange of a vector meson
leading to the effective vector-baryon contact interaction.}
\label{fig:hidd}
\end{figure}

In Refs.~\cite{bao,angels}, the interaction mechanism is the $t-$channel
exchange of vector mesons (see diagram (b) of
Fig.~\ref{fig:hidd}). Contributions from $u-$ and $s-$channel mechanisms are
neglected as they are argued to be small at threshold. Furthermore, the
$t-$channel vector exchange is evaluated with certain approximations, which
amount to neglect both $q^2/m^2_V$ and the three-momentum of the external
vector mesons versus their masses. In these circumstances, the interaction
becomes of contact type and it depends only on the vector meson energies, and
it does not depend on three-momenta (see Eq.(9) of Ref.~\cite{bao} or Eq.~(12)
of Ref.~\cite{angels}). Actually, it is originated by the $t-$exchange of the
{\em time component} of vector mesons, which has certain resemblance with our
zero-range exchange of $0^+$ mediators.

More specifically, consider the diagram (a) of Fig.~\ref{fig:hidd} near
threshold where only $s-$wave couplings survive.  Then, in any scheme, parity
and angular momentum conservation implies that the pseudoscalar meson can only
exchange a $0^+$ mediator with the baryon. This is effectively simulated by
the time component of the exchanged vector meson in the hidden gauge
scheme. Our interaction and that derived from the hidden gauge
Lagrangians~\cite{sarkar, Kolomeitsev:2003kt} turn out to be identical in the
pseudoscalar-baryon decuplet sector. Of course, both approaches also reduce to
the SU(3) WT term in the pseudoscalar-baryon octet sector. Indeed, the
interaction of soft pions with heavy sources (octet and decuplet baryons) is
completely fixed by the WT theorem \cite{Weinberg:1966kf,Tomozawa:1966jm}
(leading order in the chiral expansion) and that should be accomplished by all
schemes.

Consider now the diagram (b) of Fig.~\ref{fig:hidd} near threshold. In this
case parity and angular momentum conservation implies that the vector meson
and the baryon can exchange $0^+$, $1^+$ and $2^+$ mediators in the
$t-$channel. The $2^+$ exchange is absent in our model and in that of
Refs.~\cite{bao,angels}.  The $0^+$ exchange is not absent but it is similar
in both schemes. The main differences between the two models arise because of
the $1^+$ exchange, which is present in our scheme, as required by SU(6)
symmetry, while it is not present in the hidden gauge formalism adopted in
Refs.~\cite{bao,angels}.  We do not see a priori any compelling reason to
favor any of the two approaches. It is interesting to point out that the
hidden gauge vector mesons are naturally of the Proca type and so with
off-shell content $(1^-,0^+)$. Vector mesons in the antisymmetric tensor
formulation have instead a $(1^-,1^+)$ off-shell content. Such formulation
would naturally allow a $1^+$ exchange in schemes based on vector meson
exchange mechanisms, however, the $0^+$ exchange mechanism is missing and
ought to be added as a contact term. The scheme analyzed in the present work
does not rely on explicit meson exchange mechanisms, instead it is based on
using the minimal low energy effective interaction consistent with both chiral
and spin-flavor symmetries. This yields $0^+$ and $1^+$ $t-$channel exchange
mechanisms simultaneously. Nevertheless, it should be noted that chiral
symmetry combined with spin-flavor favors low lying vector mesons of the 35 of
SU(6) of the antisymmetric tensor type \cite{Caldi:1975tx,GarciaRecio:2010ki}.

\item[ii)] \underline{Coupled channel space}: Though the pseudoscalar
  meson-decuplet baryon $\to$ pseudoscalar meson-decuplet baryon interactions
  used here are the same as those employed in \cite{Kolomeitsev:2003kt} or
  \cite{sarkar}, our results will not necessarily coincide with those obtained
  in these two references. This is not only due to possible differences in the
  adopted RS or in the adopted pattern of flavor symmetry, that we will
  address next, but also because the coupled channel spaces are different as a
  consequence of the overall different dynamics. Actually, in the works of
  Refs.~\cite{Kolomeitsev:2003kt,sarkar}, the space does not contain channels
  with $J^P=\trh^-$ that can be constructed out of vector-decuplet or
  vector-octet baryon pairs. This can be extended to all sectors, for
  instance, interactions of the type $\rho N \to \rho \Delta$, that will
  connect the coupled spaces used in \cite{bao} and \cite{angels} are ignored
  in these two references. This has some effects that we will address in the
  next section (see for instance the discussion of the $N(1650)$ or $N(1520)$
  resonances). A similar situation occurs in the context of even parity
  low-lying meson resonances, where the hidden gauge scheme also prevents
  mixings among vector--vector, vector--pseudoscalar and
  pseudoscalar--pseudoscalar sectors~\cite{GarciaRecio:2010ki}. Such forbidden
  mixings in coupled channel space are beyond general QCD requirements and are
  idiosyncratic of the hidden gauge model.

\item[iii)] \underline{Renormalization}: The RS used here fixes for each
  $IS$ sector the subtraction constant to some specific quantity
  determined by the masses of the hadrons (see
  Eq.~(\ref{eq:musi})). This is in contrast with the RS advocated in
  other works~\cite{lamb1, lamb3, inoue, juan3, Oset:2001cn,
  sarkar,bao,angels}, which allows for some free variations in the
  subtraction constants of each of the coupled channels that enter in
  any $JIS-$sector.

\item[iv)] \underline{Symmetry breaking}: We use $f_V \ne f_P$ for those
channels which involve vector mesons, while a universal $1/f^2$
coupling is assumed for all channels in
Refs.~\cite{bao,angels}. Therefore the interaction involving vector
meson is weakened in our model (since $f_V > f_P$). 

\end{itemize}

%%%%%%%%%%%%%%%%%%%%%%%%%%%%

\section{Dynamically Generated Poles} \label{sec:res}

In order to attach each pole to definite $\rm{SU}(6)$ and $\rm{SU}(3)$
multiplets we use the following prescription. We start from an $\rm{SU}(6)$
symmetric scheme by setting the masses of all particles belonging to the same
$\rm{SU}(6)$ multiplet to a common value. In this $\rm{SU}(6)$ limit we use
the following values for the masses, which are approximately the average value
of the mass in each multiplet, $m_{35}=0.575\,{\rm GeV}$ for the mesons and
$M_{56}=1.2\, {\rm GeV}$ for the baryons.  To gradually break $\rm{SU}(6)$
symmetry down to flavor $\rm{SU}(3)$ we write the mass of the hadrons as a
function of a parameter $x$ such that
\be
m(x)=\bar{m}+x(m_{\rm{SU(3)}}-\bar{m})\label{eq:su6bre},
\ee
where $\bar{m}=m_{35}, M_{56}$ is the mass of the hadron in the $\rm{SU}(6)$
limit and $m_{\rm{SU(3)}}$ is the mass of the particle in an $\rm{SU}(3)$
flavor symmetric scheme ($m_{8_1}$=0.3 GeV and $m_{8_3}=m_{1_3}=0.85\,{\rm
  GeV}$ for the mesons and $M_{8_2}=1\,{\rm GeV}$ and $M_{10_4}=1.4\,{\rm
  GeV}$ for the baryons). In this way, we vary $x$ between 0 and 1, 0 being
the $\rm{SU}(6)$ limit and 1 the $\rm{SU}(3)$ limit. (Note that the modified
hadron masses are also used for the subtraction points, cf
Eq.~(\ref{eq:musi}).) In the SU(6) limit, we also use a common value of
$\bar{f}=125\,{\rm MeV}$ for all pseudoscalar ($f_P$) and vector ($f_V$) meson
decay constants, and change $f_{V}$ to gradually deviate from $f_P$ when $x$
increases towards 1. Namely,
\be
f(x)=\bar{f}+x(f_{\rm{SU(3)}}-\bar{f})\label{eq:su6bre-f},
\ee
with $f_{\rm{SU(3)}}= f_P, f_V$, that in the flavor 
limit take the values $f_P = 100\,{\rm MeV}$ and $f_V = 150\,{\rm MeV}$.
Next, we break the
$\rm{SU}(3)$ symmetry down to isospin $\rm{SU}(2)$, and now write the
mass of the hadrons
as a function of a parameter $y\in [0,1]$ such that
\be
m(y)=m_{\rm{SU(3)}}+y(m_{\rm{phys}}-m_{\rm{SU(3)}})\label{eq:su3bre},
\ee
where $m_{\rm{phys}}$ is the physical mass of the particle.  We also change
$f_{P}$ and $f_V$ to gradually approach the physical values of
Eq.~(\ref{eq:efes}) when $y$ reaches 1. Some states of the
Table~\ref{tab:states} are lost (do not show up as poles in the appropriated
Riemann sheets) when we move from the fully SU(6) symmetric model to our
scheme, which incorporates a certain symmetry breaking pattern. This specially
occurs for states that belong to the 1134-plet.

The procedure just described is a prescription to assign SU(6) and SU(3)
labels to the states. Within this prescription the only ambiguity could come
from the choice of symmetric points. In any case this assignment is
qualitative since the SU(6) and SU(3) symmetries are approximated ones and
some mixing between irreps necessarily exists.

The WT term is a first-order $s$-wave potential
and therefore our results could be modified to some extent by higher
order terms and higher order even waves. The strong interacting
multiplets, belonging to the 56-plet and 70-plet of SU(6), are tightly
bound and the poles generated from these multiplets should be rather
robust against perturbations caused by higher order terms. On the
other hand, the poles coming from the weakly bound 1134-plet might be
subject to larger relative corrections or even disappear by the
consideration of such terms.  In addition, resonances well above their
decay threshold could receive important corrections from $d$-wave
interactions and therefore such predictions should be less
reliable. Resonances with a large $d$-wave component cannot be
properly described within this model.

Tables \ref{tab:jp12n}-\ref{tab:jp52o} show the position of the
resonances generated by the model in non exotic sectors. Other
properties displayed are the SU(3) and SU(6) multiplets assigned to them
and the coupling of the resonance to each channel calculated through
the residues of the poles.

For the poles which are strongly affected by the consideration of the width of
unstable particles in the channels ($K^*$, $\rho$, $\Delta$ or $\sigs$) we
show in squared brackets the new pole position when this effect is taken into
account. For each resonance an up arrow is used to indicate the position of
the pole, so channels before (i.e., at the left of) the up arrow are open for
decay. The main channels are indicated using boldface.

We have assigned to some poles a tentative identification with known
states from the PDG \cite{pdg}. This identification is made by
comparing the data from the PDG on these states with the information
we extract from the poles, namely the mass, width and, most important,
the couplings. The couplings give us valuable information on the
structure of the state and on the possible decay channels and their
relative strength. It should be stressed that there will be mixings
between states with the same $SIJ^P$ quantum numbers but belonging to
different SU(6) and/or SU(3) multiplets, since these symmetries are
broken both within our approach and in nature. Additional breaking of 
SU(6) (and SU(3)) is expected to take place not only in the kinematics but
also in the interaction amplitudes. This will occur when using more
sophisticated models going beyond the (hopefully dominant) lowest order
retained here. No such explicit symmetry breaking has been included in our
model in the interaction. Also no re-fitting of subtraction points is made to
achieve better agreement in masses and widths of the resonances or
in phase shifts and inelasticities (we will briefly address this issue
in the appendix \ref{app:s11}).

In the following we comment on the identifications made in each sector
separately.

\begin{table}
\begin{center}
\caption{Properties for $J^P=\oh^-$ nucleon resonances generated by
  the model. The value in bracket is the new pole position after
  inclusion of the width of the unstable particles in the channels. An
  up arrow indicates the position of the pole; channels at the left of
  the up arrow are open for decay. The channels with largest couplings
  are highlighted with boldface. A question mark symbol expresses our
  doubts on the assignment, while states highlighted in boldface stem
  from  the strongly attractive 70 and 56 SU(6)
  irreps. }\vspace{0.2cm}
\label{tab:jp12n}
\begin{tabular}{c|c|c|c|c}
\hline
SU(3) &  &  &   &  \\
(SU(6))&  Pole    & $|g_i|$    & possible ID & status PDG\\
irrep & position [MeV] & &  \\
\hline

\hline 27 & 2160$-$70i & $\NULL{g_{\pi N}=0.2}$, $\NULL{g_{\eta N}=0.5}$,
$\NULL{g_{K\Lambda}=0.7}$, $\NULL{g_{K\Sigma}=0.4}$, &\\ & & $\NULL{g_{\rho
    N}=0.3}$, $\NULL{g_{\omega N}=0.4}$, $\NULL{g_{\phi N}=0.9}$,
$\NULL{g_{\rho\Delta}=0.6}$, &$N(2090)$ ? & $\star$\\ (1134) & &
$\NULL{g_{K^*\Lambda}=0.7}$, $\NULL{g_{K^*\Sigma}=0.2}$
$\big\uparrow$
$\mathbf{g_{K^*\sigs}=3.5}$ & \\

\hline 8 & 2082$-$30i & $\NULL{g_{\pi N}=0.1}$, $\NULL{g_{\eta N}=0.1}$,
$\NULL{g_{K\Lambda}=0.5}$, $\NULL{g_{K\Sigma}=0.2}$, & \\ & [2070$-$109i] &
$\NULL{g_{\rho N}=0.1}$, $\NULL{g_{\omega N}=0.2}$, $\mathbf{g_{\phi N}=1.2}$,
$\NULL{g_{\rho\Delta}=0.2}$, & $N(2090)$ ? & $\star$ \\ (1134) & &
$\NULL{g_{K^*\Lambda}=0.8}$, $\mathbf{g_{K^*\Sigma}=2.6}$
$\big\uparrow$
$g_{K^*\sigs}=0.7$ & \\

\hline 8 & 1795$-$80i & $\NULL{g_{\pi N}=0.1}$,
$\NULL{g_{\eta N}=0.6}$, $\NULL{g_{K\Lambda}=0.6}$, $\mathbf{g_{K\Sigma}=1.9}$,
& \\ & [1793$-$98i] & $\NULL{g_{\rho N}=0.5}$, $\NULL{g_{\omega N}=0.4}$
$\big\uparrow$
$g_{\phi N}=1.1$, $\mathbf{g_{\rho\Delta}=3.4}$, &  &  \\ (1134) & &
$g_{K^*\Lambda}=1.5$, $g_{K^*\Sigma}=1.5$, $g_{K^*\sigs}=0.9$ & \\

\hline 8 & 1706$-$70i & $\NULL{g_{\pi N}=1.0}$, $\NULL{g_{\eta N}=2.0}$,
$\NULL{g_{K\Lambda}=1.5}$, $\NULL{g_{K\Sigma}=1.1}$
$\big\uparrow$ 
& \\ & [1693$-$105i]
& $g_{\rho N}=1.9$, $\mathbf{g_{\omega N}=3.2}$, $g_{\phi N}=1.5$,
$\mathbf{g_{\rho\Delta}=3.0}$, &   \\ (1134) & & $g_{K^*\Lambda}=2.1$,
$g_{K^*\Sigma}=1.2$, $g_{K^*\sigs}=0.8$ &  \\ 

\hline {\bf 8} & 1639$-$38i & $\NULL{g_{\pi N}=1.2}$, $\NULL{g_{\eta N}=0.8}$,
$\NULL{g_{K\Lambda}=0.6}$
$\big\uparrow$
$g_{K\Sigma}=1.7$, & \\ & & $g_{\rho N}=0.2$, $\mathbf{g_{\omega N}=2.9}$,
$g_{\phi N}=0.7$, $\mathbf{g_{\rho\Delta}=2.6}$, &  ${\bf N(1650)}$ & 
$\star\star\star\,\star$\\ {\bf (70)} & &
$g_{K^*\Lambda}=1.2$, $g_{K^*\Sigma}=0.4$, $g_{K^*\sigs}=1.1$ & \\

\hline {\bf 8} & 1556$-$47i & $\NULL{g_{\pi N}=0.6}$, $\mathbf{g_{\eta N}=2.1}$
$\big\uparrow$
$g_{K\Lambda}=1.7$, $\mathbf{g_{K\Sigma}=2.4}$, & \\ & & $g_{\rho N}=0.6$, $g_{\omega
  N}=0.9$, $g_{\phi N}=0.3$, $\mathbf{g_{\rho\Delta}=2.6}$, &  $ {\bf N(1535)}$& 
$\star\star\star\,\star$\\ {\bf (56)} & &
$g_{K^*\Lambda}=1.9$, $g_{K^*\Sigma}=0.9$, $g_{K^*\sigs}=1.4$ & \\ 

\hline
\end{tabular}
\end{center}
\end{table}

\begin{table}
\begin{center}
\caption{Same as Table \ref{tab:jp12n} for $\trh^-$ nucleon
  resonances.}
\vspace{0.1cm}
\label{tab:jp32n}
\begin{tabular}{c|c|c|c|c}
\hline
SU(3) & &  &  \\
(SU(6))&   Pole     & $|g_i|$    & possible ID  & status PDG\\
irrep & position [MeV] & &  \\
\hline

27 & 2228$-$41i & $\NULL{g_{\pi \del}=0.2}$, $\NULL{g_{\rho N}=0.1}$,
$\NULL{g_{\omega N}=0.2}$, $\NULL{g_{K\sigs}=0.2}$, & \\ & [2232$-$94i] &
$\NULL{g_{\phi N}=0.8}$, $\NULL{g_{\rho \del}=0.5}$,
$\NULL{g_{K^*\Lambda}=0.9}$, $\NULL{g_{K^*\Sigma}=0.1}$
$\big\uparrow$ & $N(2080)$ ? & $\star\star$
 \\ (1134) & & $\mathbf{g_{K^*\sigs}=3.0}$  \\

\hline $10^*$&2083$-$4i& $\NULL{g_{\pi \del}=0.1}$, $\NULL{g_{\rho N}<0.1}$,
$\NULL{g_{\omega N}=0.2}$, $\NULL{g_{K\sigs}=0.1}$, & \\ &  &
$\NULL{g_{\phi N}=0.2}$, $\NULL{g_{\rho \del}=0.3}$,
$\NULL{g_{K^*\Lambda}=0.3}$
$\big\uparrow$
$\mathbf{g_{K^*\Sigma}=1.8}$, & $N(2080)$ ? & $\star\star$\\ (1134) & & $g_{K^*\sigs}=0.1$ & \\

\hline 8 & 1895$-$72i & $\NULL{g_{\pi \del}=0.4}$, $\NULL{g_{\rho N}=1.4}$,
$\NULL{g_{\omega N}=0.9}$, $\NULL{g_{K\sigs}=1.6}$
$\big\uparrow$
& \\ &[1894-106i] & $g_{\phi N}=1.0$, $\mathbf{g_{\rho \del}=3.1}$, $g_{K^*\Lambda}=1.1$,
$g_{K^*\Sigma}=1.1$, & $N(1700)$ ? & $\star\star\star$ \\ (1134)& & $g_{K^*\sigs}=0.3$ & \\

\hline 8 & 1832$-$106i& $\NULL{g_{\pi \del}=0.7}$, $\mathbf{g_{\rho N}=2.1}$,
$\NULL{g_{\omega N}=0.8}$
$\big\uparrow$
$g_{K\sigs}=1.1$, & \\ &[1829-158i] & $g_{\phi N}=0.3$, $\mathbf{g_{\rho \del}=3.3}$,
$g_{K^*\Lambda}=0.3$, $g_{K^*\Sigma}=1.0$, & $N(1700)$ ? & $\star\star\star$ \\ (1134)& &
$g_{K^*\sigs}=1.0$ & \\

\hline {\bf 8} & 1348$-$20i & $\mathbf{g_{\pi \del}=2.4}$
$\big\uparrow$
$g_{\rho N}=1.2$, $g_{\omega N}=0.3$, $g_{K\sigs}=0.3$, & \\ & [1373-43i] &
$g_{\phi N}<0.1$, $\mathbf{g_{\rho \del}=1.8}$, $g_{K^*\Lambda}<0.1$,
$g_{K^*\Sigma}=0.7$, &  ${\bf N(1520)}$ & $\star\star\star\,\star$\\ {\bf (70)} & & $g_{K^*\sigs}=0.3$ & \\

\hline
\end{tabular}
\end{center}
\end{table}

\begin{table}
\begin{center}
\caption{Same as Table \ref{tab:jp12n} for $\fvh^-$ nucleon resonances.
The (*) denotes a virtual state.}\vspace{0.1cm}
\label{tab:jp52n}
\begin{tabular}{c|c|c|c|c}
\hline
SU(3) & &  &  \\
(SU(6))&  Pole    & $|g_i|$    & possible ID & status PDG  \\
irrep & position [MeV] & &  \\
\hline
27   & 2264     & $g_{\rho\del}=0$
$\big\uparrow$
$\mathbf{g_{K^*\sigs}=2.1}$  &  $N(2200)$ ? & $\star\star$\\
(1134) &  [2259$-$28i]  & &  \\
\hline
8    & 1981(*)  & 
$\big\uparrow$
$\mathbf{g_{\rho\del}=1.6}$, $g_{K^*\sigs}=0$ & $N(2200)$ ? & $\star\star$ \\
(1134) &  [1994-392i]  & &  \\
\hline
\end{tabular}
\end{center}
\end{table}

\subsection{Nucleons ($N$)}

Results for the nucleon-like resonances are summarized in Tables
\ref{tab:jp12n}, \ref{tab:jp32n}, and \ref{tab:jp52n}, for $J^P=\oh^-$,
$J^P=\trh^-$, and $J^P=\fvh^-$ respectively.
\begin{itemize}
\item[i)] In the PDG there are three $J^P=\oh^-$ resonances, namely, $N(1535)$,
$N(1650)$ and $N(2090)$.  The existence of the first two is firmly
established, while the latest GWU partial-wave analysis of $\pi N$
data~\cite{Arndt:2006bf} finds no evidence for the one star resonance
placed above 2 GeV. Experimentally there is no information on
branching fractions for the decays of this state and there is a huge
uncertainty in its mass. Some analysis suggest that it could be as
small as $1822\pm43\,{\rm MeV}$ \cite{n20901} or as large as
$2180\pm80\,{\rm MeV}$ \cite{n20902}. Any of the two poles in
Table~\ref{tab:jp12n} located in the region of 2 GeV might have some
relation with this $N(2090)$, if it exists. From the theoretical point
of view these two poles come from the weakly attracting 1134 irrep and
we have already expressed our concerns on the real existence of states
stemming from this SU(6) part of the interaction. Nonetheless in both
cases, the couplings of these two poles to the open light channels are
small, in special to $\pi N$, which might explain the lack of evidence
of their existence in the GWU analysis. Moreover, they are placed very
close to the $K^*\Sigma^*$ and $K^*\Sigma$ thresholds, respectively,
but, precisely these poles strongly couple to these channels, having
the largest couplings among all showed in Table~\ref{tab:jp12n}. From
this perspective, these two poles might point out
the actual existence of some physical states at these energies.  The works of
Refs.~\cite{bao} and ~\cite{angels} find also similar states.

 It seems natural to associate the 56-plet pole obtained at
$\sqrt{s}=1556-47{\rm i}\,{\rm MeV}$ with the $N(1535)$ resonance,
 since its mass and width are in close agreement with the experimental state
and because the dominant decay channel observed for this resonance,
the $N\eta$ channel, is the one to which the pole has a large
coupling. 

In the region of the $N(1650)$ resonance, our model generates two or
even perhaps three poles that could be contributing to this
resonance. It looks reasonable to identify the one at
$\sqrt{s}=1639-38{\rm i}\,{\rm MeV}$, related to the strongly
attractive 70 irrep, with the four star $N(1650)$ resonance. Actually,
its mass, width and main decay modes agree reasonably well with those
reported in the PDG~\cite{pdg}.  The other two 1134 poles, if it
happens that one or both of them exist, might induce some mixings,
most likely difficult to disentangle from the experimental point of
view.  In Ref.~\cite{angels}, the dynamics of the state identified
there as the $N(1650)$ resonance is dominated by a large $\rho N$
component, which in our case is almost negligible.  In contrast, our
state couples directly to the pseudoscalar meson--baryon octet $\pi N$
(dominant decay mode), $K \Sigma$, $K \Lambda$  and $\eta N$
channels\footnote{As in the recent work of Ref.~\cite{Bruns:2010sv}, 
the $K\Sigma$ one is dominant among these type of components.}  
that do not appear in the scheme of \cite{angels}, and it also has a
large coupling to the $\omega N$ channel, which turns out to be very
small in \cite{angels}.  Yet, we also notice a large coupling of our
pole to the closed channel $\rho\Delta$, which is absent in the
analysis of Ref.~\cite{angels}, as well. The $\pi N$ decay mode, and
the width itself, of this pole will become larger when mechanisms like
those depicted in Fig.~\ref{fig:box}, constructed out of the strong
$p-$wave $\rho\pi\pi$ and $\Delta N\pi$ couplings, will be taken into
account.

The discussion on the $N(1650)$ might serve to illustrate one of the
differences between this study and the previous ones carried out in
Refs.~\cite{bao} and ~\cite{angels}. In the first of these two
references, the vector octet--baryon decuplet interaction is
considered, and the second one deals with the vector octet--baryon
octet interaction. But, within the tree level hidden gauge scheme
adopted in these two references, both sectors are disconnected. These
two sectors in turn are also disconnected from the pseudoscalar
octet--baryon octet one\footnote{The pseudoscalar octet--baryon
decuplet sector, that does not contribute to $J^P= 1/2^-$ in $s-$wave,
is also separated from the other three sectors, and treated
independently in Ref.~\cite{sarkar} by the same group.}. For instance
in Refs.~\cite{bao,angels}, couplings of the type $\pi N \to \rho N$
or $\rho\Delta \to \rho N$ do not exist, and thus within the scheme of
these references, the $\pi N$ and $\rho \Delta$ channel do not
enter into the coupled channel dynamics that gives rise to the state
identified in ~\cite{angels} as the $N(1650)$ resonance. Though in
this case, the $\rho \Delta$ threshold is sufficiently above the mass
of the $N(1650)$ resonance to keep small the influence of this channel
in the position of the pole in the SRS, this might not be the case in
other sectors that we will discuss below. Nonetheless, as pointed out
above, and because the large width of the $\rho$ and $\Delta$
resonances and their large $p-$wave couplings to the $\pi\pi$ and $\pi
N$ pairs respectively, the $\rho \Delta$ component should enhance the
$\pi N$ decay mode of the $N(1650)$ resonance. The $\pi N$ decay mode
is reported to be large in \cite{pdg} and about 80\% of a total width
of around $165\,{\rm MeV}$. Note that in our case, the coupling of the
$N(1650)$ resonance to the $\rho \Delta$ channel is an additional
source for its $\pi N$ decay mode (we have a direct sizeable coupling
to $\pi N$, $g_{\pi N}=1.2$, see Table~\ref{tab:jp12n}). In the scheme
of Ref.~\cite{angels}, the $N^*(1650)\to \pi N$ decay should
predominantly proceed through a diagram similar to that depicted in the
left panel of Fig.~\ref{fig:box}, but replacing the intermediate
$\Delta (1232)$ by a $N(940)$. Although, it is true that the latter
baryon will be less off-shell than the former one, one should also bear
in mind that the $p-$wave coupling $\pi N\Delta$ is more than twice
larger than the corresponding $\pi N N$ one.

Likely, there will be also some mixing between the $J^P = \oh^-$
nucleon states that we have identified here as the $N(1535)$ and
the $N(1650)$ resonances, since they were originated by 
different SU(6) multiplets, and this spin--flavor symmetry is
broken in nature with additional terms to those considered in the
present approach.

The $N(1535)$ and $N(1650)$ resonances, neglecting the influence of
the decuplet baryons and the nonet of vector mesons, were previously
studied within a similar RS in Ref.~\cite{GarciaRecio:2003ks} using
the chiral SU(3) WT amplitude as a kernel to calculate the $T$-matrix
(see Eq.~(\ref{eq:bse})). The results obtained in this reference for
the $N(1535)$ compare rather well with those discussed
here\footnote{The $N(1535)$ alone has received a lot of attention
since the pioneering work of Ref.~\cite{KSW95a,KSW95b}, and several
groups~\cite{inoue, Lutz:2001yb} have also found a fair description of
its dynamics starting from the tree level SU(3) chiral
Lagrangian.}. However in \cite{GarciaRecio:2003ks} a clear signal for
the $N(1650)$ was not found. This is common for all studies that use
only the tree level SU(3) WT amplitude~\cite{KSW95a,KSW95b,inoue,
Lutz:2001yb}. Indeed, this latter resonance is dynamically generated
in pion-nucleon scattering analysis that either use a unitarized
chiral effective Lagrangian including all dimension two contact
terms~\cite{Bruns:2010sv}, going in this manner beyond the leading
order WT term, or if the WT is taken as the kernel, when it was used
within a different RS that embodies some more
counter-terms~\cite{juan3}. In this latter case, the extra
counter-terms mimic the effect of higher order terms in the kernel of
the Bethe-Salpeter equation, and might be also related to the extra
channels induced by the vector mesons and decuplet baryons included
here. The inclusion of the extra counter-terms, besides allowing for a
reasonable description of the properties of both $N^*$ (1535 and 1650)
resonances, lead also to a reasonable description of the $\pi N$
$S_{11}$ phase shift and inelasticity from threshold to about $\sqrt
s \sim 1.9$ GeV, together with cross section data for $\pi^- p \to
n\eta$ and $\pi^- p \to K^0 \Lambda$ in the respective threshold
regions~\cite{juan3}.

It is also illustrative to pay attention to the predictions for
phase-shift and inelasticities deduced from the simple model
presented here. In general, the model provides a poor description of
these observables, though it hints to the gross features of the
amplitude. This should not be surprising, since we have not fitted any
parameter and we have just retained here the SU(3) WT lowest order
contribution to fix the SU(6) interaction. Moreover, additional
breaking of SU(6) (and SU(3)) is expected to take place not only in
the kinematics but also in the interaction amplitudes. We briefly
address this issue in the appendix \ref{app:s11} for the case of $(J^P
= \oh^-, I=\oh)$ $\pi N$ scattering, though conclusions are
similar for other sectors.
\begin{figure}[tbh]
%\vspace{-3cm}
\centerline{\hspace{-2cm}\includegraphics[height=3.cm]{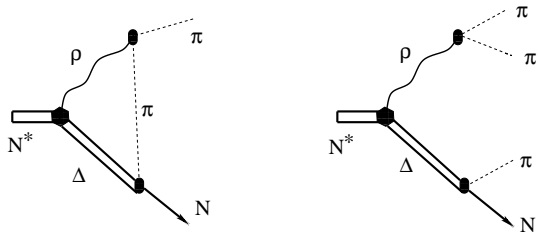}}
%\vspace{-15cm}
\caption{%\footnotesize 
Resonance ($N^*$) decay to $\pi N$ (left) or $\pi\pi\pi N$
    (right)  through its
    $s-$wave (hexagon) coupling to  $\Delta \rho$ and the
    $p-$wave coupling (ovals) of these latter hadrons to two
    pions and to a $\pi N$ pair.}
\label{fig:box}
\end{figure}

\item[ii)] For the $J^P=\trh^-$ resonances there are also three possible
observed states quoted in the PDG, the $N(1520)$, the $N(1700)$ and
the $N(2080)$. The existence of the first state is firmly established,
while latest GWU partial-wave analysis of $\pi N$
data~\cite{Arndt:2006bf} finds no evidence for the two star resonance
$N(2080)$.

For this $J^P$ quantum number we expect a worst
description of the experimental states since our model does not take
into account the $d$-wave pseudoscalar-baryon channels which can give
important contributions on the decays of the $J^P=\trh^-$ resonances.

  The lightest pole found in this sector at $\sqrt{s}=1348-20{\rm
  i}\,{\rm MeV}$,  stemming from the attractive 70 SU(6) irrep, could be
  associated with the $N(1520)$. Our model misses here the
  contribution from the $d$-wave $\pi N$ which should be important for
  this resonance (the branching fraction quoted in ~\cite{pdg} is
  around 60\% for this mode) and could bring the pole closer to the
  experimental position. The pole found here has large couplings to
  the $\pi \Delta$ and $\rho N$ channels, which account for the bulk
  of the approximately remaining 40\% of the branching fraction quoted
  in the PDG~\cite{pdg} for the $N(1520)$.  The $\pi \Delta$ and the
  $\rho N$ channels are considered independently in
  Refs.~\cite{sarkar} and \cite{angels}, respectively. Neither in the
  former nor in the latter of these works, signals for the four star
  $N(1520)$ resonance are found. We believe this is because this state
  appears as a result of the coupled channel dynamics involving both
  the $\pi \Delta$ and $\rho N$ channels, which our SU(6) model
  provides. Note that the pseudoscalar octet--baryon decuplet
  interaction is determined by chiral symmetry and therefore is the
  same here as that used in Ref.~\cite{sarkar}. The interaction of
  vector mesons with baryons is not constrained by chiral symmetry and
  the hidden--gauge scheme of Refs.~\cite{bao} and \cite{angels}
  predicts different potentials than the ones used here and deduced
  from spin-flavor SU(6) symmetry. We should finally mention that in
  Ref.~\cite{Kolomeitsev:2003kt}, this $N(1520)$ resonance was also
  dynamically generated with a large $\pi \Delta$ component.

The rest of the poles predicted by the SU(6) model are related to 
the 1134 irrep, and thus are subject to much larger
uncertainties. Perhaps, the $N(1700)$ can be associated to one or both
of the poles at
$\sqrt{s}=1895-72{\rm i}\,{\rm MeV}$ and 
$\sqrt{s}=1832-106{\rm i}\,{\rm MeV}$. The observed state has as most
important decay channel the $N\pi\pi$, which is a result of the decay
to $\del\pi$ and $N\rho$ channels to which these two poles generated by the
model have important couplings.  

The data on the heavier $N(2080)$ state are ambiguous and more
experimental information is needed in order to do a proper association
of the 1134 heavier poles displayed in Table~\ref{tab:jp32n} with the
$N(2080)$. Comments here are similar to those we made above in the
case of the spin 1/2 $N(2090)$ resonance. Nevertheless, we would like
to point out that a recent study~\cite{Xie:2010yk} finds indirect
hints of the existence of the nucleon resonance $N(2080)$ in the
recent data of the LEPS collaboration on the $\vec{\gamma} p \to
K^+\Lambda(1520)$ reaction~\cite{Muramatsu:2009zp,
Kohri:2009xe}. Actually in Ref.~\cite{Xie:2010yk}, it is shown 
that the inclusion of this resonance, with a sizable coupling to the
$\Lambda(1520)K$ pair, leads to a fairly good description of a bump
structure in the differential cross section at forward $K^+$ angles
observed in the new LEPS differential cross-section
data~\cite{Kohri:2009xe}.  The pole at $\sqrt{s}=2228-41{\rm i}\,{\rm
MeV}$, related to a SU(3) 27-plet, will naturally provide a sizable
$\Lambda(1520)K N(2080)$ decay thanks to: i) its dominant
$K^*\Sigma^*$ coupling, ii) the large $K^*\to K\pi$ width, and iii) that, as we will
discuss in the Subsect.~\ref{sec:lambs}, the $\Lambda(1520)$ resonance
has a large $\pi\Sigma^*$ component (actually, it might be a
$\pi \Sigma^*$ quasi--bound state).

\item[iii)] Two $J^P=\fvh^-$ states are compiled in the PDG. The
firmly established (four stars) $N(1675)$ resonance has a mass far too
low to be associated with any of the two poles generated by our model
here from the weakly bound 1134-plet. Moreover, the main decay modes
of this state are the $d-$wave $N\pi$ and $\Delta\pi$
channels~\cite{pdg}, which are out of the scope of our approach. 

On the other hand, from the data on the unsettled two star $N(2200)$
resonance any of the two poles in Table~\ref{tab:jp52n} could be
associated with it (note that the SU(6) transition potential $\rho \Delta
\to K^* \Sigma^*$ is zero in this sector). The latest GWU partial-wave
analysis of $\pi N$ data~\cite{Arndt:2006bf} finds no evidence for
this $N^*$ resonance either.

\end{itemize}

\begin{table}
\begin{center}
\caption{Same as Table \ref{tab:jp12n} for $\oh^-$ $\Delta$ resonances.}
\vspace{0.1cm}
\label{tab:jp12d}
\begin{tabular}{c|c|c|c|c}
\hline
SU(3) &  &  &  \\
(SU(6))& Pole      & $|g_i|$    & possible ID & status PDG \\
irrep & position [MeV] & &  \\
\hline
35 & 2244-18i & $g_{\pi N}=0.1$, $g_{K \Sigma}=0.1$, $g_{\rho N}=0.1$, 
                $g_{\rho \Delta}=0.5$, &  \\
(1134) & [2243-45i]&  $g_{\omega\Delta}=0.4$, 
                $g_{K^*\Sigma}=0.4$, $g_{\phi\Delta}=0.8$
                $\big\uparrow$
                $\mathbf{g_{K^*\sigs}=2.7}$ & $\del(2150)$ ? & $\star$\\
\hline
10 & 2187-50i & $g_{\pi N}=0.1$, $g_{K \Sigma}=0.8$, $g_{\rho N}=0.1$, 
                $g_{\rho \Delta}=0.3$, & \\
(1134) & [2178-104i] & $g_{\omega\Delta}=0.3$, 
                $\mathbf{g_{K^*\Sigma}=1.4}$
                $\big\uparrow$
                $\mathbf{g_{\phi\Delta}=2.6}$,
                $\mathbf{g_{K^*\sigs}=1.8}$  & $\del(2150)$ ? & $\star$\\
\hline
27 & 2025-88i & $g_{\pi N}=0.1$, $g_{K \Sigma}=1.7$, $g_{\rho N}=0.1$, 
                $g_{\rho \Delta}=0.7$, & \\
(1134) & [2028-101i]& $g_{\omega\Delta}=0.6$
                $\big\uparrow$ 
                $\mathbf{g_{K^*\Sigma}=2.7}$, $\mathbf{g_{\phi\Delta}=2.2}$,
                $g_{K^*\sigs}=1.5$  &  $\del(2150)$ ? & $\star$\\
\hline
10 & 1935-51i & $g_{\pi N}=0.8$, $g_{K \Sigma}=0.3$, $\mathbf{g_{\rho N}=1.3}$
                $\big\uparrow$
                $\mathbf{g_{\rho \Delta}=1.7}$, & \\
(1134) & [1929-144i] &  $\mathbf{g_{\omega\Delta}=2.8}$, 
                $g_{K^*\Sigma}=0.4$, $g_{\phi\Delta}=0.3$,
                $g_{K^*\sigs}=0.3$ & $\del(1900)$ ? & $\star \star$\\
\hline
27 & 1732-91i & $g_{\pi N}=1.0$, $g_{K \Sigma}=0.8$, $\mathbf{g_{\rho N}=2.1}$
                $\big\uparrow$
                $\mathbf{g_{\rho \Delta}=2.5}$, & \\
(1134) & [1763-144i]  & $\mathbf{g_{\omega\Delta}=2.7}$, 
                $g_{K^*\Sigma}=1.4$, $g_{\phi\Delta}=0.8$,
                $g_{K^*\sigs}=1.6$  & $\del(1900)$ ? & $\star \star$\\
\hline
{\bf 10} & 1472-77i & $g_{\pi N}=1.7$
                $\big\uparrow$
                $g_{K \Sigma}=1.3$, $\mathbf{g_{\rho N}=2.4}$, 
                $\mathbf{g_{\rho \Delta}=3.7}$, & \\
{\bf (70)} & &  $g_{\omega\Delta}=1.5$, 
                $g_{K^*\Sigma}=0.7$, $g_{\phi\Delta}=0.2$,
                $g_{K^*\sigs}=1.2$  & ${\bf \del(1620)}$ &
$\star\star\star\,\star $ \\
\hline
\end{tabular}
\end{center}
\end{table}

\begin{center}
\begin{table}
\caption{Same as Table \ref{tab:jp12n} for $\trh^-$ $\Delta$.
The (*) denotes a virtual state.}
\vspace{0.1cm}
\label{tab:jp32d}
\begin{tabular}{c|c|c|c|c}
\hline SU(3) &  & & \\
      (SU(6))& Pole & $|g_i|$ & possible ID & status PDG  \\ irrep & position
       [MeV] & &  &  \\
\hline
 10 & 2226-17i & $g_{\pi\Delta}<0.1$, $g_{\rho N}<0.1$,
$g_{\eta\Delta}=0.2$, $g_{K\sigs}=0.2$, & \\ (1134) & [2214-68i] &
$g_{\rho\Delta}<0.1$, $g_{\omega\Delta}<0.1$, $g_{K^*\Sigma}=1.0$
$\big\uparrow$
$\mathbf{g_{\phi\Delta}=1.9}$, $\mathbf{g_{K^*\sigs}=1.6}$ & \\

\hline 35 & 2172-49i & $g_{\pi\Delta}<0.1$, $g_{\rho N}<0.1$,
$g_{\eta\Delta}=0.5$, $g_{K\sigs}=1.2$, & \\ (1134) & [2170-65i] &
$g_{\rho\Delta}=0.1$, $g_{\omega\Delta}<0.1$, $g_{K^*\Sigma}<0.1$
$\big\uparrow$
$\mathbf{g_{\phi\Delta}=1.9}$, $\mathbf{g_{K^*\sigs}=2.8}$ &  \\

\hline 10 & 1915-40i & $g_{\pi\Delta}=0.8$, $g_{\rho N}=0.2$,
$g_{\eta\Delta}=1.0$, $g_{K\sigs}=0.4$ & \\ (1134) & [1912-88i] &
$\big\uparrow$
$\mathbf{g_{\rho\Delta}=1.6}$, $\mathbf{g_{\omega\Delta}=2.8}$, $g_{K^*\Sigma}=0.5$,
$g_{\phi\Delta}=0.2$, $g_{K^*\sigs}=0.1$ & $\del(1940)$ ? & $\star$ \\

\hline 27 & 1712-46i (*) & $g_{\pi\Delta}=1.1$
$\big\uparrow$
$g_{\rho N}=0.5$, $\mathbf{g_{\eta\Delta}=2.3}$, $\mathbf{g_{K\sigs}=2.5}$, &
\\ (1134) & & $g_{\rho\Delta}=0.9$, $g_{\omega\Delta}=1.3$,
$\mathbf{g_{K^*\Sigma}=2.9}$, $g_{\phi\Delta}=0.3$, $g_{K^*\sigs}=2.0$ & 
\\
\hline 
{\bf 10} & 1426-75i & $\mathbf{g_{\pi\Delta}=2.3}$
$\big\uparrow$
$\mathbf{g_{\rho N}=2.5}$,
$g_{\eta\Delta}=0.1$, $g_{K\sigs}=0.8$, & \\ {\bf (56)} & [1439-80i]& $g_{\rho\Delta}=1.3$,
$g_{\omega\Delta}=0.5$, $g_{K^*\Sigma}=1.0$, $g_{\phi\Delta}=1.6$,
$g_{K^*\sigs}=0.7$ & ${\bf \del(1700)\, ?}$ & $\star\star\star\,\star$\\

\hline
\end{tabular}
\end{table}
\end{center}

\begin{center}
\begin{table}
\caption{Same as Table \ref{tab:jp12n} for $\fvh^-$ $\Delta$.
The (*) denotes a virtual state.}
\vspace{0.1cm}
\label{tab:jp52d}
\begin{tabular}{c|c|c|c|c}
\hline
SU(3) &  &  &  \\
(SU(6))&  Pole    & $|g_i|$    & possible ID & status PDG  \\
irrep & position [MeV] & &  \\
\hline

\hline 27 & 2229 & 
 & \\ (1134) &
[2238-115i]& $g_{\rho\Delta}=0$, $g_{\omega\Delta}=0$
$\big\uparrow$
$\mathbf{g_{\phi\Delta}=0.7}$, $\mathbf{g_{K^*\sigs}=0.5}$ & $\del(2350)$? & $\star$ \\

\hline 10 & 1974-15i (*) & 
 & \\ (1134) & [1994-311i] &$\big\uparrow$
$\mathbf{g_{\rho\Delta}=3.5}$, $\mathbf{g_{\omega\Delta}=3.6}$, $g_{\phi\Delta}=0$,
$g_{K^*\sigs}=0$ & $\del(1930)$ ? &
$\star\star\star$ \\

\hline
\end{tabular}
\end{table}
\end{center}

\subsection{Deltas ($\del$)}

The results for the $\Delta$ resonances are shown in Tables
\ref{tab:jp12d}, \ref{tab:jp32d} and \ref{tab:jp52d}. In the PDG,
there are only two firmly established four star, $\del(1620)$ and
$\del (1700)$, resonances with spin-parity $\oh^-$ and $\trh^-$,
respectively.  In addition, there exist~\cite{pdg} a three star $\del$
state ($\del(1930)$) with spin-parity $\fvh^- $, and four other
resonances (two star $\del(1900)$ and  the one star
$\del(2150)$, $\del (1940)$ and $\del(2350)$) for which there exist little
evidence of their existence, as confirmed in the latest GWU
partial-wave analysis of $\pi N$ data~\cite{Arndt:2006bf}.

\begin{itemize}
\item[i)] There are three $\del$ resonances in the PDG  with
$J^P=\oh^-$.  To the $\del(1620)$ we associate the lightest pole
generated by our model at $\sqrt{s}=1472-77{\rm i}\,{\rm MeV}$ that
comes from an SU(3) decuplet of the SU(6) 70 irrep. We see a sizeable
coupling to the $N\rho$ channel and different potential sources of
decay into a $\pi N$ pair (direct coupling to $\pi N$ and large
coupling to the $\Delta\rho $ channel) in agreement with some known
features of this state.  However, the model misses the presumably
sizable contribution that the $\pi\del$ $d$-wave channels should have
in this state, and that might help to understand the difference
between our prediction for its position and the mass value reported in
\cite{pdg}. From the data on the PDG those channels should be
responsible for between 30\% to 60\% of the decay width of this
state. The works of Refs.~\cite{bao} and \cite{angels} do not find
this $\del(1620)$ state. This is not surprising because here, it
appears mostly as the result of the coupled channel dynamics of the
$\pi N$, $\rho N$ and $\rho \del$ pairs, and these channels are
treated separately in the framework set up in
~\cite{inoue,sarkar,bao,angels}. Unitarized
chiral perturbation theory studies that do not include three-body
$\pi\pi N$ states do not generate any resonance in the $S_{31}$
partial wave at low energies either~\cite{juan3,Bruns:2010sv}. The effects
of these latter states were taken into account, within certain
approximations, in \cite{inoue} and there some signatures of the
$\del(1620)$ were reported.

The rest of poles generated by our model in this sector are associated
with the SU(6) 1134-plet, and thus deciding on their real existence and
on their possible relation with physical states becomes
cumbersome. This task is even more speculative given the few existing
experimental evidences for the rest of $\del$ states reported in the
PDG. Thus, we might associate the pole at $\sqrt{s}=1935-51{\rm
i}\,{\rm MeV}$ with the $\del(1900)$ state because of the proximity in
mass and width. Moreover from the PDG data the decay of this state
into $N\pi$ is between 10\% and 30\% and the dynamically generated
state has as its most important decay channels the $N\rho$ and
$\del\rho$ channels. Yet, the pole at $\sqrt{s}=1732-91{\rm i}\,{\rm
MeV}$ could also be part of the $\del(1900)$ state or it might mix
with the lightest of the poles that we identified above with the
$\del(1620)$ resonance.

The data on the $\del(2150)$ is poor and either of the poles at
$\sqrt{s}=2244-18{\rm i}\,{\rm MeV}$, $\sqrt{s}=2187-50{\rm i}\,{\rm
MeV}$ or at $\sqrt{s}=2025-88{\rm i}\,{\rm MeV}$, if they are real, could
be associated with it.

\item[ii)] For the $J^P=\trh^-$ $\Delta$ resonances, at first sight it
 might seem natural to identify the four star $\del(1700)$ with the
 pole at $\sqrt{s}=1712-46{\rm i}\,{\rm MeV}$, although this is
 actually a virtual state in our model; it appears in the second
 Riemann sheet but below the $N\rho$ threshold. However, we rather
 think that the correct identification of the $\del(1700)$ should be
 done with the 56-plet pole at $\sqrt{s}=1426-75{\rm i}\,{\rm MeV}$,
 because of its large $\Delta \pi$ and $\rho N$ couplings. Indeed, the
 $s-$wave $\Delta \pi$ and $\rho N$ decay modes are known to be large
 around 25-50\% and 5-20\%, respectively~\cite{pdg}. As in the case of
 the $\Delta (1620)$, other decay modes in $d-$wave ($\pi N$ and also
 $\Delta \pi$ and $\rho N$) are important, which points out the
 importance of these components in the inner structure of this
 resonance, and that could also explain the existing discrepancy
 between the pole position predicted here and that reported in
 \cite{pdg}. Nevertheless, in this latter reference, the real part of
 the pole position is quoted to be well below $1700\,{\rm MeV}$ (1620~to~1680),
 while the resonance is quite broad, $|2\,{\rm Im}(\sqrt{s})| = 160 \text{~to~} 240\,{\rm MeV}$,
which makes less important the difference for the mass. This
 interpretation does not coincide with that of
 Refs.~\cite{Kolomeitsev:2003kt, sarkar}, where a second pole on top
 of the $\Delta \eta$ threshold and with large $\Sigma^* K$ and $\del
 \eta$ couplings was identified with the $\del(1700)$. This state
 would correspond to the virtual state found here. The wide pole
 placed below 1.5 GeV and with a strong coupling to the $\Delta \pi$
 channel, that we have assigned here to the $\Delta(1700)$, is
 associated in \cite{sarkar} to a missing resonance, with no known
 counterpart in the PDG, that could be searched experimentally.

The other observed state for $J^P=\trh^-$ is the one star
$\del(1940)$, which we might associate with the pole at
$\sqrt{s}=1912-88{\rm i}\,{\rm MeV}$. Although there is no data on the
partial decay widths of this state the mass and width of the state are
fairly close to the ones obtained from the pole position.

\item[iii)] For $J^P=\fvh^-$, we have only structures coming from the 1134
  irrep. Experimentally, there is a three star resonance $\del(1930)$
  more or less well established and  presumably with a small $N\pi$
  $d-$wave component. It is quite wide, with a Breit-Wigner full width of
  around $400\,{\rm MeV}$ or larger, and it has a mass of around 1950
  MeV~\cite{pdg}. Tentatively, we could identify the pole at
  $\sqrt{s}=1994-311{\rm i}\,{\rm MeV}$ with the observed $\del(1930)$.  The
  pole at $\sqrt{s}=2238-115{\rm i}\,{\rm MeV}$ would then correspond to the
  $\del(2350)$. There is no information on the branching fractions for
  this state and huge uncertainty on its mass and width.
\end{itemize}

\bc
\btab
\caption{Same as Table \ref{tab:jp12n} for $\oh^-$ $\Sigma$ resonances.}
\vspace{0.1cm}
\label{tab:jp12s}
\bt{c|c|c|c|c}
\hline
SU(3) &  &  &  \\
(SU(6))&Pole      & $|g_i|$    & possible ID & status PDG  \\
irrep & position [MeV] & &  \\
\hline
 35 & 2369-16i & $g_{\pi\Lambda}=$ 0.1, $g_{\pi\sig}=$ 0.1,
$g_{\bar{K}N}=$ 0.1, $g_{\eta\sig}=$ 0.2, & \\ & & $g_{K\Xi}=$ 0.1,
$g_{\bar{K}^*N}=$ 0.1, $g_{\rho\Lambda}=$ 0.1, $g_{\rho\sig}<$0.1, & \\ (1134)
& & $g_{\omega\sig}<$0.1, $g_{\bar{K}^*\del}=$ 0.5, $g_{\rho\sigs}=$ 0.5,
$g_{\omega\sigs}=$ 0.3, & \\ & & $g_{K^*\Xi}<$0.1, $g_{\phi\sig}<$0.1
$\big\uparrow$
$\mathbf{g_{\phi\sigs}= 2.2}$, $\mathbf{g_{K^*\Xi^*}= 1.7}$ & \\

\hline 10 & 2277-66i & $g_{\pi\Lambda}=$ 0.1, $g_{\pi\sig}<$0.1,
$g_{\bar{K}N}<$0.1, $g_{\eta\sig}=$ 0.6, & \\ & & $g_{K\Xi}=$ 0.4,
$g_{\bar{K}^*N}=$ 0.1, $g_{\rho\Lambda}<$0.1, $g_{\rho\sig}=$ 0.5, & \\ (1134)
& & $g_{\omega\sig}=$ 0.4, $g_{\bar{K}^*\del}=$ 0.3, $g_{\rho\sigs}=$ 0.6,
$g_{\omega\sigs}=$ 0.4, & \\ & & $g_{K^*\Xi}=$ 0.7, $g_{\phi\sig}=$ 0.9
$\big\uparrow$
$\mathbf{g_{\phi\sigs}= 2.1}$, $\mathbf{g_{K^*\Xi^*}= 2.7}$ & \\

\hline 27 & 2144-55i & $g_{\pi\Lambda}=$ 0.1, $g_{\pi\sig}=$ 0.1,
$g_{\bar{K}N}=$ 0.1, $g_{\eta\sig}=$ 0.8, & \\ & [2141-68i] & $g_{K\Xi}=$ 1.2,
$g_{\bar{K}^*N}=$ 0.1, $g_{\rho\Lambda}=$ 0.2, $g_{\rho\sig}=$ 0.1, &
\\ (1134) & & $g_{\omega\sig}=$ 0.1, $g_{\bar{K}^*\del}=$ 0.2
$\big\uparrow$
$g_{\rho\sigs}=$ 1.1, $g_{\omega\sigs}=$ 1.0, & \\ & & $\mathbf{g_{K^*\Xi}= 1.6}$,
$\mathbf{g_{\phi\sig}= 2.1}$, $\mathbf{g_{\phi\sigs}= 1.5}$, $g_{K^*\Xi^*}=$ 1.2 & \\

\hline 10 & 2093-48i & $g_{\pi\Lambda}=$ 0.6, $g_{\pi\sig}=$ 0.3,
$g_{\bar{K}N}=$ 0.6, $g_{\eta\sig}=$ 0.2, & \\ & [2090-65i] & $g_{K\Xi}=$ 0.4,
$g_{\bar{K}^*N}=$ 0.6, $g_{\rho\Lambda}=$ 0.8, $g_{\rho\sig}=$ 0.4, &
\\ (1134) & & $g_{\omega\sig}=$ 0.1
$\big\uparrow$
$g_{\bar{K}^*\del}=$ 0.3, $\mathbf{g_{\rho\sigs}= 1.7}$,
$\mathbf{g_{\omega\sigs}= 2.8}$, & \\ & & $g_{K^*\Xi}=$ 0.4, $g_{\phi\sig}=$
0.6, $g_{\phi\sigs}=$ 0.4, $g_{K^*\Xi^*}=$ 0.6 & \\

\hline 8 & 1972-31i& $g_{\pi\Lambda}=$ 0.3, $g_{\pi\sig}=$ 0.2,
$g_{\bar{K}N}=$ 0.2, $g_{\eta\sig}=$ 0.6, & \\ & [1969-167i] & $g_{K\Xi}=$
0.5, $g_{\bar{K}^*N}=$ 0.9, $g_{\rho\Lambda}=$ 0.4, $g_{\rho\sig}=$ 0.5
$\big\uparrow$
&  \\ (1134) & & $\mathbf{g_{\omega\sig}= 2.3}$,
$\mathbf{g_{\bar{K}^*\del}= 1.2}$, $g_{\rho\sigs}=$ 0.3, $g_{\omega\sigs}=$
0.5, & \\ & & $g_{K^*\Xi}=$ 0.8, $g_{\phi\sig}=$ 0.8, $g_{\phi\sigs}=$ 0.5,
$g_{K^*\Xi^*}=$ 0.5 & \\

\hline 8 & 1895-63i & $g_{\pi\Lambda}=$ 0.4, $g_{\pi\sig}=$ 0.7,
$g_{\bar{K}N}=$ 0.4, $g_{\eta\sig}=$ 1.1, & \\ & [1870-153i] & $g_{K\Xi}=$
0.5, $g_{\bar{K}^*N}=$ 1.1, $g_{\rho\Lambda}=$ 0.8
$\big\uparrow$
$g_{\rho\sig}=$ 0.7, & \\ (1134) & & $g_{\omega\sig}=$ 1.0,
$\mathbf{g_{\bar{K}^*\del}= 2.0}$, $\mathbf{g_{\rho\sigs}= 2.1}$,
$g_{\omega\sigs}=$ 1.5, & \\ & & $\mathbf{g_{K^*\Xi}= 2.3}$, $g_{\phi\sig}=$
0.9, $g_{\phi\sigs}=$ 0.9, $\mathbf{g_{K^*\Xi^*}= 1.9}$ & \\

\hline 8 & 1867-36i & $g_{\pi\Lambda}=$ 0.3, $g_{\pi\sig}=$ 0.6,
$g_{\bar{K}N}=$ 0.9, $g_{\eta\sig}=$ 0.5, & \\ & [1873-53i] & $g_{K\Xi}=$ 0.2,
$g_{\bar{K}^*N}=$ 0.3
$\big\uparrow$
$g_{\rho\Lambda}=$ 1.5, $g_{\rho\sig}=$ 0.5, &
\\ (1134) & & $g_{\omega\sig}=$ 1.2, $\mathbf{g_{\bar{K}^*\del}= 3.0}$,
$\mathbf{g_{\rho\sigs}= 1.8}$, $g_{\omega\sigs}=$ 0.9, & \\ & & $g_{K^*\Xi}=$ 1.1,
$g_{\phi\sig}=$ 0.5, $g_{\phi\sigs}=$ 0.9, $g_{K^*\Xi^*}=$ 1.0 & \\

\hline $10^*$ & 1754-74i & $g_{\pi\Lambda}=$ 1.1, $g_{\pi\sig}=$ 0.9,
$g_{\bar{K}N}=$ 1.1, $g_{\eta\sig}=$ 0.9
$\big\uparrow$
& \\ & & $\mathbf{g_{K\Xi}= 1.7}$, $\mathbf{g_{\bar{K}^*N}= 1.8}$,
$g_{\rho\Lambda}=$ 1.3, $g_{\rho\sig}=$ 1.0, &  \\ (1134) & &
$\mathbf{g_{\omega\sig}= 2.4}$, $g_{\bar{K}^*\del}=$ 0.9, $g_{\rho\sigs}=$
1.3, $g_{\omega\sigs}=$ 0.8, & \\ & & $\mathbf{g_{K^*\Xi}= 1.9}$, $g_{\phi\sig}=$ 0.7,
$g_{\phi\sigs}=$ 0.2, $g_{K^*\Xi^*}=$ 0.7 & \\

\hline 
{\bf 10} & 1599-61i & $g_{\pi\Lambda}=$ 1.3, $g_{\pi\sig}=$ 0.2,
$g_{\bar{K}N}=$ 1.1
$\big\uparrow$
$g_{\eta\sig}=$ 1.3, & \\ & & $g_{K\Xi}=$ 0.8, $\mathbf{g_{\bar{K}^*N}= 1.9}$,
$g_{\rho\Lambda}=$ 1.2, $g_{\rho\sig}=$ 1.2, & ${\bf \sig(1750)}$ &
${\bf \star\star\star}$ \\ {\bf (70)} & &
$g_{\omega\sig}=$ 0.5, $\mathbf{g_{\bar{K}^*\del}= 3.1}$,
$\mathbf{g_{\rho\sigs}= 2.4}$, $g_{\omega\sigs}=$ 0.6, & \\ & & $g_{K^*\Xi}=$
0.3, $g_{\phi\sig}=$ 0.5, $g_{\phi\sigs}=$ 0.9, $g_{K^*\Xi^*}=$ 1.2 & \\

\hline 
{\bf 8} & 1489-117i & $g_{\pi\Lambda}=$ 1.4, $g_{\pi\sig}=$ 1.9,
$g_{\bar{K}N}=$ 1.0
$\big\uparrow$
$g_{\eta\sig}=$ 0.9, & \\ & & $g_{K\Xi}=$ 1.4,
$g_{\bar{K}^*N}=$ 1.3, $\mathbf{g_{\rho\Lambda}= 2.3}$,
$\mathbf{g_{\rho\sig}= 2.7}$, & \\ {\bf (70)} &
& $g_{\omega\sig}=$ 0.7, $\mathbf{g_{\bar{K}^*\del}= 2.1}$, $g_{\rho\sigs}=$ 0.4,
$g_{\omega\sigs}=$ 0.4, & ${\bf \sig(1620)}$ & ${\bf \star\star}$ \\ & & $g_{K^*\Xi}=$ 1.8, $g_{\phi\sig}=$ 0.1,
$g_{\phi\sigs}=$ 0.5, $g_{K^*\Xi^*}=$ 0.1 & \\

\hline
\et
\etab
\ec

\bc
\btab
\caption{Same as Table \ref{tab:jp12n} for $\trh^-$ $\Sigma$ resonances.}
\vspace{0.1cm}
\label{tab:jp32s}
\bt{c|c|c|c|c}
\hline
SU(3) &  &  &  \\
(SU(6))& Pole      & $|g_i|$    & possible ID & status PDG  \\
irrep & position [MeV] & &  \\
\hline

\hline 10 & 2352-38i & $g_{\pi\sigs}=$ 0.1, $g_{\bar{K}\del}=$ 0.2,
$g_{\bar{K}^*N}=$ 0.2, $g_{\rho\Lambda}=$ 0.1, & \\ & & $g_{\eta\sigs}<$0.1,
$g_{\rho\sig}=$ 0.1, $g_{\omega\sig}=$ 0.1, $g_{K\Xi^*}=$ 0.1, & \\ (1134) & &
$g_{\bar{K}^*\del}=$ 0.5, $g_{\rho\sigs}=$ 0.2, $g_{\omega\sigs}=$ 0.2,
$g_{K^*\Xi}=$ 1.0, & \\ & & $g_{\phi\sig}=$ 0.6
$\big\uparrow$
$\mathbf{g_{\phi\sigs}= 2.2}$,
$\mathbf{g_{K^*\Xi^*}= 2.1}$ & \\

\hline 35 & 2295-48i & $g_{\pi\sigs}<$0.1, $g_{\bar{K}\del}<$0.1,
$g_{\bar{K}^*N}<$0.1, $g_{\rho\Lambda}<$0.1, & \\ & & $g_{\eta\sigs}=$ 0.7,
$g_{\rho\sig}<$0.1, $g_{\omega\sig}<$0.1, $g_{K\Xi^*}=$ 1.0, & \\ (1134) & &
$g_{\bar{K}^*\del}=$ 0.1, $g_{\rho\sigs}=$ 0.1, $g_{\omega\sigs}=$ 0.1,
$g_{K^*\Xi}=$ 0.1, &  \\ & & $g_{\phi\sig}=$ 0.1
$\big\uparrow$
$\mathbf{g_{\phi\sigs}= 2.4}$,
$\mathbf{g_{K^*\Xi^*}= 2.5}$ &\\

\hline 8 & 2207-2i & $g_{\pi\sigs}<$0.1, $g_{\bar{K}\del}=$ 0.1,
$g_{\bar{K}^*N}=$ 0.1, $g_{\rho\Lambda}<$0.1, & \\ & & $g_{\eta\sigs}=$ 0.1,
$g_{\rho\sig}=$ 0.1, $g_{\omega\sig}=$ 0.1, $g_{K\Xi^*}=$ 0.1, & \\ (1134) & [2210-10i]&
$g_{\bar{K}^*\del}=$ 0.3, $g_{\rho\sigs}=$ 0.1, $g_{\omega\sigs}=$ 0.1
$\big\uparrow$
$\mathbf{g_{K^*\Xi}= 1.1}$, & \\ & & $\mathbf{g_{\phi\sig}= 1.3}$,
$g_{\phi\sigs}=$ 0.1, $g_{K^*\Xi^*}=$ 0.1 & \\

\hline 27 & 2150-24i & $g_{\pi\sigs}=$ 0.1, $g_{\bar{K}\del}<$0.1,
$g_{\bar{K}^*N}=$ 0.6, $g_{\rho\Lambda}=$ 0.9, & \\ & [2132-107i] &
$g_{\eta\sigs}=$ 0.2, $g_{\rho\sig}=$ 0.3, $g_{\omega\sig}=$ 0.1,
$g_{K\Xi^*}=$ 0.1, & \\ (1134) & & $g_{\bar{K}^*\del}=$ 0.2
$\big\uparrow$
$\mathbf{g_{\rho\sigs}= 2.0}$, $\mathbf{g_{\omega\sigs}= 1.3}$,
$g_{K^*\Xi}<$0.1, & \\ & & $g_{\phi\sig}=$ 0.2, $g_{\phi\sigs}<$0.1,
$g_{K^*\Xi^*}=$ 0.1 & \\

\hline 10 & 2070-46i & $g_{\pi\sigs}=$ 0.6, $g_{\bar{K}\del}=$ 0.6,
$g_{\bar{K}^*N}=$ 0.2, $g_{\rho\Lambda}=$ 0.2, & \\ & & $g_{\eta\sigs}=$ 0.9,
$g_{\rho\sig}=$ 0.3, $g_{\omega\sig}=$ 0.4, $g_{K\Xi^*}=$ 0.6
$\big\uparrow$
& \\ (1134) & & $g_{\bar{K}^*\del}=$ 0.7, $\mathbf{g_{\rho\sigs}= 1.5}$,
$\mathbf{g_{\omega\sigs}= 2.7}$, $g_{K^*\Xi}=$ 0.4, & \\ & & $g_{\phi\sig}=$
0.8, $g_{\phi\sigs}<$0.1, $g_{K^*\Xi^*}=$ 0.1 & \\

\hline 8 & 2015-46i & $g_{\pi\sigs}=$ 0.4, $g_{\bar{K}\del}=$ 0.9,
$g_{\bar{K}^*N}=$ 0.5, $g_{\rho\Lambda}=$ 0.3, & \\ & [2001-72i] &
$g_{\eta\sigs}=$ 0.5, $g_{\rho\sig}=$ 0.6, $g_{\omega\sig}=$ 0.8
$\big\uparrow$
$g_{K\Xi^*}=$ 0.9, & \\ (1134) & & $\mathbf{g_{\bar{K}^*\del}= 3.0}$,
$g_{\rho\sigs}=$ 0.1, $g_{\omega\sigs}=$ 0.7, $g_{K^*\Xi}=$ 0.5, & \\ & &
$\mathbf{g_{\phi\sig}= 1.1}$, $g_{\phi\sigs}=$ 0.6, $g_{K^*\Xi^*}=$ 0.4 & \\

\hline 8 & 1932-50i & $g_{\pi\sigs}=$ 0.9, $g_{\bar{K}\del}=$ 0.3,
$g_{\bar{K}^*N}=$ 1.0, $g_{\rho\Lambda}=$ 1.2, & \\ & [1929-82i] &
$g_{\eta\sigs}=$ 1.2
$\big\uparrow$ $g_{\rho\sig}=$ 0.8, $g_{\omega\sig}=$ 0.9,
$\mathbf{g_{K\Xi^*}= 1.5}$, &  \\ (1134) & & $g_{\bar{K}^*\del}=$ 0.7,
$g_{\rho\sigs}=$ 1.2, $\mathbf{g_{\omega\sigs}= 1.4}$, $g_{K^*\Xi}=$ 1.0, & \\ & &
$g_{\phi\sig}=$ 0.4, $g_{\phi\sigs}=$ 0.3, $g_{K^*\Xi^*}=$ 1.1 & \\

\hline 
{\bf 10} & 1605-55i & $\mathbf{g_{\pi\sigs}= 2.3}$
$\big\uparrow$
$g_{\bar{K}\del}=$ 1.2, $\mathbf{g_{\bar{K}^*N}= 1.5}$, $g_{\rho\Lambda}=$
1.9, & \\ & & $g_{\eta\sigs}=$ 0.7, $g_{\rho\sig}=$ 0.9, $g_{\omega\sig}=$
0.3, $g_{K\Xi^*}=$ 1.1, & \\ {\bf (56)} & & $g_{\bar{K}^*\del}=$ 1.1,
$g_{\rho\sigs}=$ 1.3, $g_{\omega\sigs}=$ 0.6, $g_{K^*\Xi}=$ 0.7, &
${\bf \sig(1940)}$ ? & ${\bf \star\star\star}$  \\ & &
$g_{\phi\sig}=$ 0.6, $g_{\phi\sigs}=$ 0.5, $g_{K^*\Xi^*}=$ 1.0 & \\

\hline 
{\bf 8} & 1571-8i & $g_{\pi\sigs}=$ 1.1
$\big\uparrow$
$\mathbf{g_{\bar{K}\del}= 3.1}$,
$g_{\bar{K}^*N}=$ 1.5, $g_{\rho\Lambda}=$ 0.9, & \\ & & $g_{\eta\sigs}=$ 1.2,
$g_{\rho\sig}=$ 0.2, $g_{\omega\sig}=$ 0.2, $g_{K\Xi^*}=$ 0.4, & ${\bf
  \sig(1670)}$
& ${\bf \star\star\star\,\star}$\\ {\bf (70)} & & $\mathbf{g_{\bar{K}^*\del}= 2.7}$, $g_{\rho\sigs}=$ 0.8, $g_{\omega\sigs}=$
0.5, $g_{K^*\Xi}=$ 0.3, & \\ & & $g_{\phi\sig}=$ 1.0, $g_{\phi\sigs}=$ 0.9,
$g_{K^*\Xi^*}=$ 0.4 & \\

\hline
\et
\etab
\ec

\bc
\btab
\caption{Same as Table \ref{tab:jp12n} for $\fvh^-$ $\Sigma$ resonances.}
\vspace{0.1cm}
\label{tab:jp52s}
\bt{c|c|c|c|c}
\hline
SU(3) & Pole &  &  \\
(SU(6))& Pole     & $|g_i|$    & possible ID & status PDG \\
irrep & position [MeV] & &  \\
\hline
27 & 2394 & $g_{\bar{K}^*\del}=$ 0, $g_{\rho\sigs}=$ 0,
$g_{\omega\sigs}=$ 0
$\big\uparrow$
$\mathbf{g_{\phi\sigs}= 1.5}$, & \\ (1134) & [2390-9i] &
$\mathbf{g_{K^*\Xi^*}= 1.6}$ & $\sig(2250)$ ?
& $\star\star\star$ \\

\hline 10 & 2159-0.0i & $g_{\bar{K}^*\del}=$ 0.1
$\big\uparrow$
$\mathbf{g_{\rho\sigs}= 0.9}$,
$\mathbf{g_{\omega\sigs}= 0.7}$, $g_{\phi\sigs}=$ 0, &
 \\ (1134) & [2162-151i] &
$g_{K^*\Xi^*}=$ 0 &\\

\hline 8 & 2100 (*) & $\big\uparrow$ $\mathbf{g_{\bar{K}^*\del}= 1.8}$
$g_{\rho\sigs}=$ 0.6,
$\mathbf{g_{\omega\sigs}= 1.3}$, $g_{\phi\sigs}=$ 0,& \\ (1134) & [2128-178i] &
$g_{K^*\Xi^*}=$ 0 &\\

\hline
\et
\etab
\ec

\subsection{Sigmas ($\sig$)}

Tables \ref{tab:jp12s}, \ref{tab:jp32s} and \ref{tab:jp52s} display
the results for the $\sig$ states generated by the model. This sector
is specially complex to analyze, and the situation here is unclear.
On the experimental side, there are only two firmly established (four stars) odd
parity resonances. These are the $\sig(1670)$ and the $\sig(1775)$
states, with spin $3/2$ and $5/2$, respectively. The latter one cannot
be described by our model, since the decays of this state reveal a
fundamental role of $d-$wave interactions between the $N\bar K$,
$\Lambda \pi$, $\Sigma \pi$ and $\Sigma^* \pi$ pairs. Besides these
two resonances, there exists scarce trustworthy information in the PDG
on $s-$ and $d-$wave $\Sigma$'s resonances: three additional 3-star
resonances (one of them with yet undetermined $J^P$) and a
plethora of one and two star states and undetermined spin-parity
bumps from which is difficult to draw any robust conclusion. On the
theory side, there are many channels participating in the dynamics (16
and 15 for $J=\oh$ and $J=\trh$, respectively), and as consequence 
the model provides a rich
spectrum in this sector. Indeed,  attending to Table~\ref{tab:states}, we
might expect as many as 26 different states (five if we limit the study
to the 56 and 70 irreps) which are difficult to identify.

\begin{itemize}
\item[i)] With $J^P=\oh^-$, the only state for which there is experimental
  information on its decays is the three star $\sig(1750)$
  resonance. It has decays into $\bar{K}N$, $\eta\sig$, $\pi\Lambda$
  and $\pi\sig$, being the latter one likely suppressed. These
  features seems to agree with the couplings of the pole at
  $\sqrt{s}=1599-61{\rm i}\,{\rm MeV}$ that stems from an SU(3)
  decuplet of the attractive SU(6) 70 irrep. Note that from the decays
  compiled in \cite{pdg}, one might expect some $d-$wave $\Sigma^*
  \pi$ component in the structure to the $\sig(1750)$ resonance that
  is not considered within our scheme. This might, at least partially,
  explain the disagreement between the mass predicted by our model and
  that quoted in the PDG. Nevertheless, one should also bear in mind
  that this state is quite wide (full width $\Gamma= 60$ to 160
  MeV)~\cite{pdg}, and thus the difference in the real part of the
  position of the pole becomes less relevant. On the other hand, little
  is known on the two star $\sig(1620)$ state, besides it might have a
  width of the order of few tens of MeV and that the $\pi\sig$ decay
  mode might be sizable. We associate to this state the lowest lying
  pole ($\sqrt{s}=1489-117{\rm i}\,{\rm MeV}$) generated in our scheme
  and that also has its origin in the attractive 70 irrep. We find in agreement
  with Ref.~\cite{Oset:2001cn} that the $\Sigma(1620)$ resonance has
  couplings of normal size to all pseudoscalar--baryon octet channels,
  and, given the large phase space available, it has a sizable decay
  width into any of the channels and hence a considerably large total
  width. Identifying any of the remaining states of
  Table~\ref{tab:jp12s}, that come from the 1134 irrep, with the one
  star $\Sigma(2000)$ or some of the one and two star bumps listed in
  \cite{pdg} would be too speculative and we refrain from doing it.
It is noteworthy that the $\Sigma$~$\oh^-$ state needed to complete the
56-plet does appear in the SU(6) limit, but, when going to the physical
point (physical masses and decay constants) the pole moves to unphysical
regions of the $\sqrt{s}$ Riemann surface.

\item[ii)] For $J^P=\trh^-$, and besides the bumps, there are three
$\Sigma$ states compiled in the PDG. The decay width of one of them,
the four star $\sig(1670)$ resonance, comes mostly from $d$-wave
pseudoscalar-baryon channels, though this state also decays into a
$s-$wave $\Sigma^* \pi$ pair~\cite{pdg}. Therefore our model should
have some problems to predict correctly its mass and width. Nevertheless, its
$\Sigma^* \pi$ decay mode and the
fact that it is relatively narrow suggests that our pole placed at 
$\sqrt{s}=1571-8{\rm i}\,{\rm MeV}$
could be identified with this $\sig(1670)$ state. Indeed, this pole
appears in the evolution of an SU(3) octet of the attractive 70 SU(6)
irrep, and it clearly coincides (position and couplings) with
that assigned to the $\sig(1670)$ resonance in
\cite{Kolomeitsev:2003kt,sarkar}. 

The second pole generated by our model at $\sqrt{s}=1605-55{\rm
  i}\,{\rm MeV}$ stems from an SU(3) decuplet of the 56 irrep, and it
  is also obtained in \cite{sarkar}, but it is not mentioned
  in~\cite{Kolomeitsev:2003kt}. It is wider than the first one because
  it has a larger $\pi \Sigma^*$ coupling, and it is not associated with
  any state in \cite{sarkar}. Indeed, it is not straightforward to
  assign any state of the PDG to this pole. It seems unappropriated
  its identification with the one star $\Sigma (1580)$, because this
  state might not exist and also because in the PDG, only $d-$wave
  decay modes ($N \bar K$, $\Lambda \pi$ and $\Sigma \pi$) are quoted
  for this resonance. It could be associated with some of the bumps
  listed in the PDG, or perhaps this pole corresponds to the three
  star $\Sigma(1940)$ resonance. As we have argued before in the
  case of the $\Delta(1700)$, the $\Sigma(1940)$ state is very wide
  (full width $\Gamma$ = 150 to 300 MeV) and the actual position of
  the pole should be influenced by genuine $d-$wave channels ($N \bar
  K$, $\Lambda \pi$, $\Sigma \pi$, \dots) that have not been
  considered in the present model. The fact that this resonance has
  been granted with three stars~\cite{pdg} implies that its existence
  ranges from very likely to certain, which in our scheme would
  naturally fit with it being related to the strongly attractive 56
  irrep.  Furthermore, as possible decay modes of this $\Sigma(1940)$
  state, the $s-$wave $\Sigma^* \pi$, ${\bar K}^* N$ and ${\bar K}
  \Delta$ pairs are also given in ~\cite{pdg}, being the latter one
  relatively sizable ($\sim 16 \%$). This could be easily accommodated
  in our model if we identify the $\Sigma(1940)$ with our 56-plet
  pole. In Refs.~\cite{Kolomeitsev:2003kt,sarkar}, instead the
  $\Sigma(1940)$ resonance is identified with a pole much closer in mass
  to the $1.9\,{\rm GeV}$ region, which couples strongly to $\Xi^* K$, but very
  weakly to ${\bar K} \Delta$. This would be similar to our 1134 irrep
  $\sqrt{s}=1932-50{\rm i}\,{\rm MeV}$ pole.

\item[iii)] The obtained states with $J^P=\fvh^-$ are placed in the weakly
  attractive 1134 and are too heavy to be associated with the
  PDG $\sig(1775)$ state. Some of them or other of the 1134-plet states with
  $J^P=\oh^-$ and $J^P=\trh^-$, could fit some of the $\Sigma$
  resonances which have not had their $J^P$ quantum numbers
  identified yet, but more experimental information is needed to do a proper
  identification. Other poles obtained here might disappear when higher
  order terms are taken into account in the potential.
 
  Perhaps, the $\sig (2250)$ resonance is of special relevance for our
  discussion here. It is
  classified with three stars,  some experiments see two
  resonances, one of them with $J^P=\fvh^-$ and with a sizable $\Xi K$
  decay mode~\cite{Bellefon:1900zz}. Attending to this latter feature,
  we might assign the highest pole predicted by our model in 
 Table~\ref{tab:jp52s} to this $\sig (2250)$ state. Indeed, this is
  the only pole among the three compiled in this table that has a
  non-vanishing coupling to the $K^*\Xi^*$ channel. This decay mode
  would provide  $\Xi K$ events through a mechanism similar to that
  sketched in the left panel of Fig.~\ref{fig:box}.

\end{itemize}

\bc
\btab
\caption{Same as Table \ref{tab:jp12n} for $\oh^-$ $\Lambda$ resonances.}
\label{tab:jp12l}
\vspace{0.1cm}
\bt{c|c|c|c|c}
\hline
SU(3) &  &  &  \\
(SU(6))& Pole      & $|g_i|$    & possible ID & status PDG \\
irrep & position [MeV] & &  \\
\hline

\hline 27 & 2254-74i & $g_{\pi\sig}=$ 0.2, $g_{\bar{K}N}=$ 0.2,
$g_{\eta\Lambda}=$ 0.6, $g_{K\Xi}=$ 0.6, & \\ & & $g_{\bar{K}^*N}=$ 0.6,
$g_{\omega\Lambda}=$ 0.3, $g_{\rho\sig}=$ 0.2, $\mathbf{g_{\phi\Lambda}= 1.0}$, &
\\ (1134) & & $g_{\rho\sigs}=$ 0.8, $g_{K^*\Xi}=$ 0.3
$\big\uparrow$
$\mathbf{g_{K^*\Xi^*}= 3.6}$ &
\\

\hline 27 & 2182-27i & $g_{\pi\sig}=$ 0.2, $g_{\bar{K}N}=$ 0.1,
$g_{\eta\Lambda}=$ 0.2, $g_{K\Xi}=$ 0.4, & \\ & [2177-52i] & $g_{\bar{K}^*N}=$
0.3, $g_{\omega\Lambda}=$ 0.3, $g_{\rho\sig}=$ 0.8, $g_{\phi\Lambda}=$ 0.5, &
\\ (1134) & & $g_{\rho\sigs}=$ 0.3
$\big\uparrow$
$\mathbf{g_{K^*\Xi}= 2.8}$, $g_{K^*\Xi^*}=$ 0.8 &
\\

\hline 8 & 2104-56i & $g_{\pi\sig}<$0.1, $g_{\bar{K}N}=$ 0.1,
$g_{\eta\Lambda}=$ 1.2, $g_{K\Xi}=$ 1.0, & \\ & & $g_{\bar{K}^*N}=$ 0.4,
$g_{\omega\Lambda}=$ 0.1, $g_{\rho\sig}<$0.1
$\big\uparrow$
$\mathbf{g_{\phi\Lambda}= 2.9}$, &
\\ (1134) & & $g_{\rho\sigs}=$ 0.3, $g_{K^*\Xi}=$ 1.3, $\mathbf{g_{K^*\Xi^*}= 2.1}$ &
\\

\hline 8 & 1929-44i & $g_{\pi\sig}=$ 0.2, $g_{\bar{K}N}=$ 0.3,
$g_{\eta\Lambda}=$ 1.0, $g_{K\Xi}=$ 1.0, & \\ & [1914-57i] & $g_{\bar{K}^*N}=$
0.1, $g_{\omega\Lambda}=$ 0.2
$\big\uparrow$
$\mathbf{g_{\rho\sig}= 1.4}$, $g_{\phi\Lambda}=$ 0.3, & \\ (1134) & &
$\mathbf{g_{\rho\sigs}= 3.4}$, $g_{K^*\Xi}=$ 0.2, $\mathbf{g_{K^*\Xi^*}= 1.4}$
& \\

\hline 8 & 1870-27i & $g_{\pi\sig}=$ 0.8, $g_{\bar{K}N}=$ 0.2,
$g_{\eta\Lambda}=$ 0.5, $g_{K\Xi}=$ 0.5, & \\ & & $g_{\bar{K}^*N}=$ 0.1
$\big\uparrow$
$\mathbf{g_{\omega\Lambda}= 2.4}$, $\mathbf{g_{\rho\sig}= 1.8}$,
$g_{\phi\Lambda}=$ 0.2, & $\Lambda(1800)$? &$\star\star\star$ \\ (1134) & & $g_{\rho\sigs}=$ 1.4,
$g_{K^*\Xi}=$ 1.3, $g_{K^*\Xi^*}=$ 1.0 & \\

\hline 1 & 1826-42i & $g_{\pi\sig}=$ 0.2, $\mathbf{g_{\bar{K}N}= 1.4}$,
$g_{\eta\Lambda}=$ 0.2, $g_{K\Xi}=$ 0.4
$\big\uparrow$
& \\ & [1824-115i] & $\mathbf{g_{\bar{K}^*N}= 2.5}$,
$\mathbf{g_{\omega\Lambda}= 1.4}$, $\mathbf{g_{\rho\sig}= 1.4}$,
$g_{\phi\Lambda}=$ 0.8, & $\Lambda(1800)$?  & $\star\star\star$ \\ (1134) & & $g_{\rho\sigs}=$ 1.2,
$g_{K^*\Xi}=$ 0.6, $g_{K^*\Xi^*}=$ 0.4 & \\

\hline {\bf 8} & 1691-26i & $g_{\pi\sig}=$ 0.5, $g_{\bar{K}N}=$ 0.9,
$g_{\eta\Lambda}=$ 0.8
$\big\uparrow$
$\mathbf{g_{K\Xi}= 2.8}$, & \\ & & $g_{\bar{K}^*N}=$ 1.0, $g_{\omega\Lambda}=$
0.2, $\mathbf{g_{\rho\sig}= 2.5}$, $g_{\phi\Lambda}=$ 0.3, & 
${\bf \Lambda(1670)}$ & ${\bf
  \star\star\star\,\star}$
\\ {\bf (56)} & & $g_{\rho\sigs}=$ 1.2, $g_{K^*\Xi}=$ 1.4, $g_{K^*\Xi^*}=$ 1.2 & \\

\hline 
{\bf 8} & 1430-3i & $g_{\pi\sig}=$ 0.5
$\big\uparrow$
$\mathbf{g_{\bar{K}N}= 1.8}$, $g_{\eta\Lambda}=$ 0.9, $g_{K\Xi}=$ 0.1, & \\ &
& $\mathbf{g_{\bar{K}^*N}= 2.2}$, $g_{\omega\Lambda}=$ 0.5, $g_{\rho\sig}=$
0.3, $g_{\phi\Lambda}=$ 0.9, & ${\bf \Lambda(1405)}$ & ${\bf
  \star\star\star\,\star}$ \\ {\bf (70)} & & $g_{\rho\sigs}=$
0.4, $g_{K^*\Xi}=$ 0.2, $g_{K^*\Xi^*}=$ 0.1 & \\

\hline 
{\bf 1} & 1374-85i & $\mathbf{g_{\pi\sig}= 2.6}$
$\big\uparrow$
$g_{\bar{K}N}=$ 1.0, $g_{\eta\Lambda}=$ 0.2, $g_{K\Xi}=$ 0.5, & \\ & &
$g_{\bar{K}^*N}=$ 0.6, $g_{\omega\Lambda}=$ 0.2, $\mathbf{g_{\rho\sig}= 1.7}$,
$g_{\phi\Lambda}=$ 0.2, & ${\bf \Lambda(1405)}$ & ${\bf
  \star\star\star\,\star}$  \\ {\bf (70)} & & $g_{\rho\sigs}=$ 0.6,
$g_{K^*\Xi}=$ 0.9, $g_{K^*\Xi^*}=$ 0.3 & \\

\hline
\et
\etab
\ec

\bc
\btab
\caption{Same as Table \ref{tab:jp12n} for $\trh^-$ $\Lambda$ resonances.}
\label{tab:jp32l}
\vspace{0.1cm}
\bt{c|c|c|c|c}
\hline
SU(3) &  &  &  \\
(SU(6))& Pole      & $|g_i|$    & possible ID & status PDG \\
irrep & position [MeV] & &  \\
\hline

\hline 27 & 2338-54i & $g_{\pi\sigs}=$ 0.3, $g_{\bar{K}^*N}=$ 0.1,
$g_{\omega\Lambda}=$ 0.4, $g_{\rho\sig}=$ 0.3, & \\ & & $g_{K\Xi^*}=$ 0.1,
$\mathbf{g_{\phi\Lambda}= 1.2}$, $g_{\rho\sigs}=$ 0.7, $g_{K^*\Xi}=$ 0.4
$\big\uparrow$
& $\Lambda(2325)$ ? & $\star$ \\ (1134) & & $\mathbf{g_{K^*\Xi^*}= 3.2}$ & \\

\hline 8 & 2206-11i & $g_{\pi\sigs}=$ 0.1, $g_{\bar{K}^*N}=$ 0.1,
$g_{\omega\Lambda}=$ 0.1, $g_{\rho\sig}<$0.1, & \\ & [2198-75i] &
$g_{K\Xi^*}=$ 0.3, $\mathbf{g_{\phi\Lambda}= 1.0}$, $g_{\rho\sigs}=$ 0.3
$\big\uparrow$
$\mathbf{g_{K^*\Xi}= 2.1}$, & \\ (1134) & & $g_{K^*\Xi^*}=$ 0.4 & \\

\hline 8 & 2090-65i & $g_{\pi\sigs}=$ 0.2, $g_{\bar{K}^*N}=$ 0.8,
$g_{\omega\Lambda}=$ 0.6, $\mathbf{g_{\rho\sig}= 1.1}$, & \\ & [2076-125i] &
$\mathbf{g_{K\Xi^*}= 1.1}$
$\big\uparrow$
$\mathbf{g_{\phi\Lambda}= 1.0}$, $\mathbf{g_{\rho\sigs}= 2.8}$,
$\mathbf{g_{K^*\Xi}= 1.0}$, & \\ (1134) & & $g_{K^*\Xi^*}=$ 0.6 & \\

\hline 1 & 2023-79i & $g_{\pi\sigs}=$ 0.5, $g_{\bar{K}^*N}=$ 0.9,
$g_{\omega\Lambda}=$ 0.5, $\mathbf{g_{\rho\sig}= 2.1}$
$\big\uparrow$
& \\ & [2029-144i] & $\mathbf{g_{K\Xi^*}= 1.6}$, $g_{\phi\Lambda}=$ 0.7,
$\mathbf{g_{\rho\sigs}= 2.7}$, $g_{K^*\Xi}=$ 0.6, & \\ (1134) & &
$g_{K^*\Xi^*}=$ 1.2 & \\

\hline 8 & 1894-120i & $g_{\pi\sigs}=$ 0.7, $\mathbf{g_{\bar{K}^*N}= 2.8}$
$\big\uparrow$
$g_{\omega\Lambda}=$ 1.7, $g_{\rho\sig}=$ 1.5, & \\ & [1890-140i] &
$g_{K\Xi^*}=$ 1.1, $g_{\phi\Lambda}=$ 0.3, $\mathbf{g_{\rho\sigs}= 2.6}$,
$g_{K^*\Xi}=$ 0.5, & \\ (1134) & & $g_{K^*\Xi^*}=$ 1.0 & \\

\hline 27 & 1879-32i & $g_{\pi\sigs}=$ 1.2, $g_{\bar{K}^*N}=$ 0.3
$\big\uparrow$
$\mathbf{g_{\omega\Lambda}= 2.3}$, $g_{\rho\sig}=$ 1.1, & \\ & & $g_{K\Xi^*}=$
1.3, $g_{\phi\Lambda}=$ 0.3, $\mathbf{g_{\rho\sigs}= 1.6}$, $g_{K^*\Xi}=$ 0.4,
& $\Lambda(1690)$? & $\star\star\star\,\star$\\ (1134) & & $g_{K^*\Xi^*}=$ 1.1 & \\

\hline {\bf 8} & 1542-37i & $\mathbf{g_{\pi\sigs}= 2.3}$
$\big\uparrow$ $g_{\bar{K}^*N}=$ 0.9, $g_{\omega\Lambda}=$ 0.4,
$g_{\rho\sig}=$ 1.2, & \\ & & $g_{K\Xi^*}=$ 0.6,
$g_{\phi\Lambda}<$0.1, $\mathbf{g_{\rho\sigs}= 1.6}$, $g_{K^*\Xi}=$
0.6, & ${\bf \Lambda(1520)}$ & {\bf $\star\star\star\,\star$} \\ {\bf
(70)} & & $g_{K^*\Xi^*}=$ 0.6 & \\

\hline
\et
\etab
\ec

\bc
\btab
\caption{Same as Table \ref{tab:jp12n} for $\fvh^-$ $\Lambda$ resonances.}
\label{tab:jp52l}
\vspace{0.1cm}
\bt{c|c|c|c|c}
\hline
SU(3) &  &  &  \\
(SU(6))& Pole      & $|g_i|$    & possible ID & status PDG  \\
irrep & position [MeV] & &  \\
\hline

\hline 27 & 2404 & $g_{\rho\sigs}=$ 0
$\big\uparrow$
$\mathbf{g_{K^*\Xi^*}= 2.4}$ &  \\ (1134) & [2399-11i] & &\\

\hline 8 & 2160 & $\big\uparrow$
$\mathbf{g_{\rho\sigs}= 0.6}$, $g_{K^*\Xi^*}=$ 0 & \\ (1134) &
       [2167-276i] & & \\

\hline
\et
\etab
\ec

\subsection{Lambdas ($\Lambda$)}
\label{sec:lambs}
Tables \ref{tab:jp12l}, \ref{tab:jp32l} and \ref{tab:jp52l} show the results
for the $\Lambda$ resonances generated by the model.
\begin{itemize}
\item[i)] In the PDG there are three observed $J^P=\oh^-$ $\Lambda$
states. The lightest of them is the $\Lambda(1405)$ which has been
throughly studied \cite{lamb1,lamb2,lamb3,GarciaRecio:2003ks} and is
believed to have a two pole structure. In our model the two poles that
describe this state are at the positions $\sqrt{s}=1374-85{\rm i}\,{\rm
MeV}$ and $\sqrt{s}=1430-3{\rm i}\,{\rm MeV}$, and both of them stem from the
strongly attractive 70 irrep.  It is noteworthy that vector mesons turn
out to have important components in both poles, though channels
involving these mesons are well above the position of the poles.

We identify the four star $\Lambda(1670)$ resonance with the 56 irrep
pole at $\sqrt{s}=1691-26{\rm i}\,{\rm MeV}$. This state has decays to
$\pi\sig$, $\bar{K}N$ and $\eta\Lambda$ but the couplings are not so
strong and the resonance, despite the large phase
space available for the decays, is fairly narrow. Indeed, this state shows a
large $K\Xi$ component, but this channel is kinematically
closed. Results for this resonance here compare rather well with those 
obtained in Refs.~\cite{lamb3,GarciaRecio:2003ks,Oset:2001cn}.

We also find two poles in the region of the $\Lambda(1800)$ at
positions $\sqrt{s}=1824-115{\rm i}\,{\rm MeV}$ and
$\sqrt{s}=1870-27{\rm i}\,{\rm
MeV}$.  Both poles have small $\pi\Sigma$ couplings, and specially the
singlet one shows large ${\bar K} N $ and ${\bar K}^* N $
components. The $\Lambda(1800)$ fits nicely with these features, and
it has a broad width which covers the whole region of these two
poles. Experimentally it should be very difficult to distinguish the
contribution of each one of these two structures separately.

\item[ii)] The most prominent resonance with $J^P=\trh^-$ is the
$\Lambda(1520)$ which we associate with the pole located at
$\sqrt{s}=1542-37{\rm i}\,{\rm MeV}$. This state has important
$d$-wave components which are not taken into account by our model.  In
\cite{sourav} the authors develop a phenomenological coupling of the
$s$-wave $\pi\sigs$ and $K\Xi^*$ with the $d$-wave $\pi\sig$ and
$\bar{K}N$ channels. In that model, a subtraction constant is fitted
in order to obtain the pole with a lower mass than in our
approach. This would be equivalent to change the subtraction point in
the calculation of the $T$-matrix in the present approach which would
drastically decrease also the width of the resonance\footnote{This is
explicitly shown in Fig. 2 of Ref.~\cite{Toki:2007ab}. There, it is
also explained how the couplings to the different channels decrease
when the dynamically generated pole approaches to the $\pi \Sigma^*$
threshold. }. Including then the $d$-wave channels to the model would
create the appropriate width bringing the resonance properties closer
to the experimental values.

The decay modes of the four star $\Lambda(1690)$ resonance are more or
less equally distributed in $d$-wave pseudoscalar baryon channels and
three body channels ($\Lambda\pi\pi$ and $\sig\pi\pi$). The three body
channels can come from $s$-wave channels considered by our model like
$\sigs\pi$ or $\rho\sig$ and therefore we could associate this state
with the pole at $\sqrt{s}=1879-32{\rm i}\,{\rm MeV}$. Note that it would
not be difficult to readjust the subtraction constant to achieve a
better agreement of the position of this pole with the mass and the 
width of the physical $\Lambda(1690)$ state. We cannot
discard here a possible mixing with the pole, close in energy, at
$\sqrt{s}=1894-120{\rm i}\,{\rm MeV}$, and also generated within the
1134 irrep. However, this last pole is quite broad, significantly
wider than the $\Lambda(1690)$ state, because its large coupling to
the open channel ${\bar K}^* N$. However, this latter decay mode does
not appear in \cite{pdg} for the physical $\Lambda(1690)$, which
disfavors any relation of this pole  with the physical state.

The association of the $\Lambda(2325)$ with the pole at
$\sqrt{s}=2338-54{\rm i}$ MeV is based on the mass and width of the
state, since there is not enough data on this resonance in order to
analyze its decay channels. This is just a tentative identification,
subject to all shortcomings that we have noted above for 1134-plet
states.

\item[iii)] The poles obtained with $J^P=\fvh^-$ are too heavy to be
associated with the firmly established $\Lambda(1830)$ resonance. The
model fails to generate this state. This is not surprising because
from the data on this resonance compiled in the PDG, one expects a
fundamental role of $d-$wave interactions involving the $N\bar K$,
$\Sigma \pi$ and $\Sigma^* \pi$ pairs.

\end{itemize}

\bc
\btab
\caption{Same as Table \ref{tab:jp12n} for $\oh^-$ $\Xi$ resonances.}
\label{tab:jp12x}
\vspace{0.1cm}
\bt{c|c|c|c|c}
\hline
SU(3) &  &  &  \\
(SU(6))& Pole     & $|g_i|$    & possible ID & status PDG \\
irrep & position [MeV] & &  \\
\hline

\hline 35 & 2476-28i & $g_{\pi\Xi}=0.1$, $g_{\bar{K}\Lambda}=0.1$,
$g_{\bar{K}\sig}=0.2$, $g_{\eta\Xi}=0.2$, & \\ & & $g_{\bar{K}^*\Lambda}=0.1$,
$g_{\bar{K}^*\sig}=0.1$, $g_{\rho\Xi}=0.1$, $g_{\omega\Xi}<0.1$, & \\ (1134) &
& $g_{\bar{K}^*\sigs}=0.9$, $g_{\rho\Xi^*}=0.4$, $g_{\omega\Xi^*}=0.3$,
$g_{\phi\Xi}=0.1$
$\big\uparrow$
& \\ & & $\mathbf{g_{\phi\Xi^*}=2.9}$, $\mathbf{g_{K^*\Omega}=1.1}$ & \\

\hline 10 & 2338-67i & $g_{\pi\Xi}=0.2$, $g_{\bar{K}\Lambda}=0.1$,
$g_{\bar{K}\sig}=0.3$, $g_{\eta\Xi}=0.7$, & \\ & & $g_{\bar{K}^*\Lambda}=0.4$,
$g_{\bar{K}^*\sig}=0.8$, $g_{\rho\Xi}=0.3$, $g_{\omega\Xi}=0.3$, & \\ (1134) &
& $g_{\bar{K}^*\sigs}=0.2$, $g_{\rho\Xi^*}=0.8$, $g_{\omega\Xi^*}=0.5$,
$g_{\phi\Xi}=0.8$
$\big\uparrow$
& \\ & & $\mathbf{g_{\phi\Xi^*}=1.5}$, $\mathbf{g_{K^*\Omega}=3.5}$ &
\\ 

\hline 27 & 2244-38i & $g_{\pi\Xi}=0.3$, $g_{\bar{K}\Lambda}=0.5$,
$g_{\bar{K}\sig}=0.3$, $g_{\eta\Xi}=1.2$, & \\ & & $g_{\bar{K}^*\Lambda}=0.4$,
$g_{\bar{K}^*\sig}=0.2$, $g_{\rho\Xi}=0.3$, $g_{\omega\Xi}<0.1$
$\big\uparrow$
& \\ (1134) & & $g_{\bar{K}^*\sigs}=0.8$, $g_{\rho\Xi^*}=0.8$,
$\mathbf{g_{\omega\Xi^*}=2.9}$, $\mathbf{g_{\phi\Xi}=2.5}$, & \\ & &
$\mathbf{g_{\phi\Xi^*}=1.3}$, $\mathbf{g_{K^*\Omega}=1.3}$ & \\

\hline 10 & 2238-53i & $g_{\pi\Xi}=0.3$, $g_{\bar{K}\Lambda}=0.6$,
$g_{\bar{K}\sig}=0.3$, $g_{\eta\Xi}=1.2$, & \\ & [2229-71i] &
$g_{\bar{K}^*\Lambda}=0.7$, $g_{\bar{K}^*\sig}=0.5$, $g_{\rho\Xi}=0.6$,
$g_{\omega\Xi}=0.1$
$\big\uparrow$
& \\ (1134) & & $g_{\bar{K}^*\sigs}=0.4$, $\mathbf{g_{\rho\Xi^*}=2.0}$,
$\mathbf{g_{\omega\Xi^*}=2.2}$, $\mathbf{g_{\phi\Xi}=2.7}$, & \\ & &
$g_{\phi\Xi^*}=1.0$, $g_{K^*\Omega}=1.5$ & \\

\hline 27 & 2094-59i & $g_{\pi\Xi}=0.4$, $g_{\bar{K}\Lambda}=0.3$,
$g_{\bar{K}\sig}=0.7$, $g_{\eta\Xi}=0.6$, & \\ & [2111-111i] &
$g_{\bar{K}^*\Lambda}=0.9$, $g_{\bar{K}^*\sig}=1.0$, $g_{\rho\Xi}=1.0$
$\big\uparrow$
$g_{\omega\Xi}=0.8$, & \\ (1134) & & $\mathbf{g_{\bar{K}^*\sigs}=1.5}$,
$\mathbf{g_{\rho\Xi^*}=2.1}$, $g_{\omega\Xi^*}=1.2$,
$\mathbf{g_{\phi\Xi}=1.4}$, & \\ & & $g_{\phi\Xi^*}=0.4$,
$\mathbf{g_{K^*\Omega}=2.5}$ & \\

\hline 8 & 2037-24i & $g_{\pi\Xi}=0.6$, $g_{\bar{K}\Lambda}=0.6$,
$g_{\bar{K}\sig}=0.3$, $g_{\eta\Xi}=0.2$, & \\ & & $g_{\bar{K}^*\Lambda}=0.3$
$\big\uparrow$
$g_{\bar{K}^*\sig}=0.5$, $g_{\rho\Xi}=1.5$, $g_{\omega\Xi}=0.6$, & \\ (1134) &
& $\mathbf{g_{\bar{K}^*\sigs}=1.8}$, $\mathbf{g_{\rho\Xi^*}=2.4}$,
$g_{\omega\Xi^*}=1.1$, $g_{\phi\Xi}=0.2$, & \\ & & $g_{\phi\Xi^*}=1.0$,
$\mathbf{g_{K^*\Omega}=2.1}$ & \\

\hline {\bf 10} & 1729-46i & $g_{\pi\Xi}=0.6$, $g_{\bar{K}\Lambda}=1.4$,
$g_{\bar{K}\sig}=0.4$
$\big\uparrow$
$g_{\eta\Xi}=1.6$, & \\ & & $g_{\bar{K}^*\Lambda}=1.4$,
$\mathbf{g_{\bar{K}^*\sig}=2.1}$, $g_{\rho\Xi}=1.0$,
$g_{\omega\Xi}=0.4$, & \\ {\bf (70)} & &
$\mathbf{g_{\bar{K}^*\sigs}=3.3}$, $g_{\rho\Xi^*}=1.5$, $g_{\omega\Xi^*}=0.4$,
$g_{\phi\Xi}=0.2$, & ${\bf \Xi(1950)}$ &$\star\star\star$  \\ & & $g_{\phi\Xi^*}=1.6$, $g_{K^*\Omega}=1.0$ & \\

\hline {\bf 8} & 1651-2i & $g_{\pi\Xi}=0.2$, $g_{\bar{K}\Lambda}=0.3$
$\big\uparrow$
$\mathbf{g_{\bar{K}\sig}=2.2}$, $g_{\eta\Xi}=1.3$, & \\ & &
$g_{\bar{K}^*\Lambda}=1.0$, $\mathbf{g_{\bar{K}^*\sig}=2.6}$,
$g_{\rho\Xi}=0.2$, $g_{\omega\Xi}=0.6$, & \\ {\bf (70)} & &
$g_{\bar{K}^*\sigs}=0.9$, $g_{\rho\Xi^*}=0.4$, $g_{\omega\Xi^*}=0.2$,
$\mathbf{g_{\phi\Xi}=1.7}$, & ${\bf \Xi(1690)}$ &$\star\star\star$ \\ & & $g_{\phi\Xi^*}=0.4$,
$g_{K^*\Omega}=0.2$ & \\

\hline
{\bf 8} & 1577-139i & $\mathbf{g_{\pi\Xi}=2.6}$
$\big\uparrow$
$\mathbf{g_{\bar{K}\Lambda}=1.7}$, $g_{\bar{K}\sig}=0.5$, $g_{\eta\Xi}=0.1$, &
\\ & & $g_{\bar{K}^*\Lambda}=0.8$, $g_{\bar{K}^*\sig}=1.0$, $g_{\rho\Xi}=0.7$,
$g_{\omega\Xi}=0.1$, & \\ {\bf (56)} & & $g_{\bar{K}^*\sigs}=0.6$,
$g_{\rho\Xi^*}=1.3$, $g_{\omega\Xi^*}=0.3$, $g_{\phi\Xi}=0.1$, &
${\bf \Xi(1620)}$& $\star$  \\ & &
$g_{\phi\Xi^*}=0.2$, $g_{K^*\Omega}=1.2$ & \\

\hline
\et
\etab
\ec

\bc
\btab
\caption{Same as Table \ref{tab:jp12n} for $\trh^-$ $\Xi$ resonances.}
\label{tab:jp32x}
\vspace{0.1cm}
\bt{c|c|c|c|c}
\hline
SU(3) &  &  &  \\
(SU(6))& Pole     & $|g_i|$    & possible ID & status PDG  \\
irrep & position [MeV] & &  \\
\hline

\hline 10 & 2440-54i & $g_{\pi\Xi^*}=0.2$, $g_{\bar{K}\sigs}=0.2$,
$g_{\bar{K}^*\Lambda}=0.3$, $g_{\eta\Xi^*}=0.2$, & \\ & &
$g_{\bar{K}^*\sig}=0.1$, $g_{\rho\Xi}=0.4$, $g_{\omega\Xi}=0.3$,
$g_{K\Omega}=0.1$, & \\ (1134) & & $g_{\bar{K}^*\sigs}=0.5$,
$g_{\rho\Xi^*}=0.5$, $g_{\omega\Xi^*}=0.3$, $\mathbf{g_{\phi\Xi}=1.2}$
$\big\uparrow$
& \\ & & $\mathbf{g_{\phi\Xi^*}=2.0}$, $\mathbf{g_{K^*\Omega}=2.6}$ & \\

\hline 35 & 2414-45i & $g_{\pi\Xi^*}<0.1$, $g_{\bar{K}\sigs}<0.1$,
$g_{\bar{K}^*\Lambda}<0.1$, $g_{\eta\Xi^*}=0.9$, & \\ & &
$g_{\bar{K}^*\sig}<0.1$, $g_{\rho\Xi}=0.1$, $g_{\omega\Xi}<0.1$,
$g_{K\Omega}=0.8$, & \\ (1134) & & $g_{\bar{K}^*\sigs}=0.1$,
$g_{\rho\Xi^*}=0.1$, $g_{\omega\Xi^*}<0.1$, $g_{\phi\Xi}=0.2$
$\big\uparrow$
& \\ & & $\mathbf{g_{\phi\Xi^*}=2.9}$, $\mathbf{g_{K^*\Omega}=2.0}$ & \\

\hline 27 & 2283-27i & $g_{\pi\Xi^*}=0.1$, $g_{\bar{K}\sigs}=0.1$,
$g_{\bar{K}^*\Lambda}=0.8$, $g_{\eta\Xi^*}=0.2$, & \\ & [2265-62i] &
$g_{\bar{K}^*\sig}=0.1$, $g_{\rho\Xi}=0.7$, $g_{\omega\Xi}=0.1$,
$g_{K\Omega}=0.2$, & \\ (1134) & & $g_{\bar{K}^*\sigs}=0.2$
$\big\uparrow$
$\mathbf{g_{\rho\Xi^*}=2.0}$, $\mathbf{g_{\omega\Xi^*}=1.5}$,
$g_{\phi\Xi}=0.3$, & \\ & & $g_{\phi\Xi^*}=0.1$, $g_{K^*\Omega}=0.2$ & \\

\hline 8 & 2224-51i & $g_{\pi\Xi^*}=0.2$, $g_{\bar{K}\sigs}=0.4$,
$g_{\bar{K}^*\Lambda}=0.1$, $g_{\eta\Xi^*}=0.6$, & \\ & [2225-66i] &
$g_{\bar{K}^*\sig}=1.1$, $g_{\rho\Xi}=0.2$, $g_{\omega\Xi}=0.4$,
$g_{K\Omega}=0.8$
$\big\uparrow$
& \\ (1134) & & $g_{\bar{K}^*\sigs}=0.9$, $g_{\rho\Xi^*}=1.6$,
$\mathbf{g_{\omega\Xi^*}=2.2}$, $g_{\phi\Xi}=1.1$, & \\ & & $g_{\phi\Xi^*}=0.4$,
$g_{K^*\Omega}=0.5$ & \\

\hline 10 & 2193-46i & $g_{\pi\Xi^*}=0.3$, $\mathbf{g_{\bar{K}\sigs}=1.0}$,
$g_{\bar{K}^*\Lambda}=0.1$, $g_{\eta\Xi^*}=0.2$, & \\ & [2189-48i] &
$g_{\bar{K}^*\sig}=0.7$, $g_{\rho\Xi}=0.2$, $g_{\omega\Xi}=0.3$,
$g_{K\Omega}=0.3$
$\big\uparrow$
& \\ (1134) & & $\mathbf{g_{\bar{K}^*\sigs}=3.0}$, $g_{\rho\Xi^*}=0.3$,
$\mathbf{g_{\omega\Xi^*}=1.3}$, $g_{\phi\Xi}=0.5$, & \\ & &
$g_{\phi\Xi^*}=0.5$, $g_{K^*\Omega}=0.1$ & \\

\hline 8 & 2104-58i & $g_{\pi\Xi^*}=0.4$, $g_{\bar{K}\sigs}=0.2$,
$g_{\bar{K}^*\Lambda}=0.4$, $g_{\eta\Xi^*}=0.7$, & \\ & [2123-93i] &
$\mathbf{g_{\bar{K}^*\sig}=1.4}$, $\mathbf{g_{\rho\Xi}=1.4}$, $g_{\omega\Xi}=0.2$
$\big\uparrow$
$\mathbf{g_{K\Omega}=2.0}$, & \\ (1134) & & $g_{\bar{K}^*\sigs}=0.7$,
$\mathbf{g_{\rho\Xi^*}=1.0}$, $g_{\omega\Xi^*}=1.1$, $g_{\phi\Xi}=1.0$, & \\ & &
$g_{\phi\Xi^*}=0.2$, $\mathbf{g_{K^*\Omega}=1.4}$ & \\

\hline 27 & 1972-47i & $g_{\pi\Xi^*}=1.3$, $g_{\bar{K}\sigs}=0.6$
$\big\uparrow$
$g_{\bar{K}^*\Lambda}=1.3$, $g_{\eta\Xi^*}=0.7$, & \\ & &
$g_{\bar{K}^*\sig}=0.1$, $\mathbf{g_{\rho\Xi}=1.6}$,
$\mathbf{g_{\omega\Xi}=1.6}$, $\mathbf{g_{K\Omega}=1.8}$, & 
\\ (1134) & & $g_{\bar{K}^*\sigs}=0.4$, $g_{\rho\Xi^*}=1.0$,
$g_{\omega\Xi^*}=0.9$, $g_{\phi\Xi}=0.3$, & \\ & & $g_{\phi\Xi^*}=0.2$,
$\mathbf{g_{K^*\Omega}=1.5}$ & \\

\hline 
{\bf 10} & 1772-15i & $g_{\pi\Xi^*}=1.4$
$\big\uparrow$
$\mathbf{g_{\bar{K}\sigs}=2.7}$, $\mathbf{g_{\bar{K}^*\Lambda}=2.0}$,
$\mathbf{g_{\eta\Xi^*}=2.1}$, & \\ & & $g_{\bar{K}^*\sig}=1.8$,
$g_{\rho\Xi}=1.1$, $g_{\omega\Xi}=0.4$, $g_{K\Omega}=0.8$, & \\ {\bf (56)} & &
$\mathbf{g_{\bar{K}^*\sigs}=2.3}$, $g_{\rho\Xi^*}=0.9$, $g_{\omega\Xi^*}=0.6$,
$g_{\phi\Xi}=0.9$, & ${\bf \Xi(2250) ?}$ & {\bf $\star\star$}\\ & & $g_{\phi\Xi^*}=1.6$, $g_{K^*\Omega}=0.8$ & \\

\hline {\bf 8} & 1748-48i & $\mathbf{g_{\pi\Xi^*}=2.6}$
$\big\uparrow$
$\mathbf{g_{\bar{K}\sigs}=1.6}$, $\mathbf{g_{\bar{K}^*\Lambda}=1.5}$,
$g_{\eta\Xi^*}=1.1$, & \\ & & $g_{\bar{K}^*\sig}=1.2$,
$\mathbf{g_{\rho\Xi}=2.1}$, $g_{\omega\Xi}=0.5$, $g_{K\Omega}=1.4$, &
\\ {\bf (70)}
& & $g_{\bar{K}^*\sigs}=1.4$, $g_{\rho\Xi^*}=1.5$, $g_{\omega\Xi^*}=0.7$,
$g_{\phi\Xi}=0.4$, & ${\bf \Xi(1820)}$ & {\bf $\star\star\star$}\\ & & $g_{\phi\Xi^*}=0.8$, $g_{K^*\Omega}=1.3$ & \\

\hline
\et
\etab
\ec

\bc
\btab
\caption{Same as Table \ref{tab:jp12n} for $\fvh^-$ $\Xi$ resonances.}
\label{tab:jp52x}
\vspace{0.1cm}
\bt{c|c|c|c|c}
\hline
SU(3) &  &  &  \\
(SU(6))& Pole      & $|g_i|$    & possible ID & status PDG  \\
irrep & position [MeV] & &  \\
\hline

\hline 27 & 2529 & $g_{\bar{K}^*\sigs}=0$, $g_{\rho\Xi^*}=0$,
$g_{\omega\Xi^*}=0$
$\big\uparrow$
$\mathbf{g_{\phi\Xi^*}=1.2}$, &\\ (1134) & [2525-4i] &
$\mathbf{g_{K^*\Omega}=2.1}$ &\\

\hline 10 & 2342-38i & $\mathbf{g_{\bar{K}^*\sigs}=3.0}$,
$\mathbf{g_{\rho\Xi^*}=1.2}$, $\mathbf{g_{\omega\Xi^*}=1.4}$
$\big\uparrow$
$g_{\phi\Xi^*}=0$, &\\ (1134) & [2346-58i] & $g_{K^*\Omega}=0$ &\\

\hline 8 & 2304 & $g_{\bar{K}^*\sigs}=0.2$
$\big\uparrow$
$\mathbf{g_{\rho\Xi^*}=1.3}$, $\mathbf{g_{\omega\Xi^*}=0.9}$,
$g_{\phi\Xi^*}=0$,& \\ (1134) & [2316-118i] & $g_{K^*\Omega}=0$ &\\

\hline
\et
\etab
\ec

\subsection{Cascades ($\Xi$)}

The results for the $\Xi$ resonances generated by the model are shown in
Tables \ref{tab:jp12x}, \ref{tab:jp32x} and \ref{tab:jp52x}.

Besides the lowest-lying even parity $\Xi$ and $\Xi^*$ baryons, only
the three star $\Xi(1820)$ resonance has its $J^P$ quantum numbers
($\trh^-$) assigned in the PDG. This fact makes hard any
identification of the poles predicted by our model with any physical
state. Nevertheless, the information we provide about possible
poles and telling to which states each of them couples most strongly
could be a guiding line for the search of new resonance and/or for the
correct assignment of spin and parity to those already
compiled in \cite{pdg}. 
\begin{itemize}
\item[i)] We find two $J^P=\oh^-$ poles below 1.7 GeV related to 
  the strongly attractive 56 and 70 irreps. Here, we confirm the
  findings of Ref.~\cite{GarciaRecio:2003ks} and these two states can
  clearly be identified to the $\Xi(1690)$ and $\Xi(1620)$ resonances,
  which clarifies the spin, parity and nature of these two
  resonances. The spin-parity assignment of the $\Xi(1690)$ found here
  corroborates the evidence presented in \cite{Aubert:2008ty}.
  Of particular interest is the signal for the three star
  $\Xi(1690)$ resonance, where we find a quite small (large) coupling
  to the $\pi \Xi$ ($\bar K \Sigma$) channel, which explains the
  smallness of the experimental ratio, $\Gamma( \pi\Xi)/ \Gamma(\bar
  K\Sigma ) < $ 0.09~\cite{pdg} despite of the significant energy
  difference between the thresholds for the $\pi \Xi$ and $\bar K
  \Sigma$ channels. On the other hand the 56 irrep pole associated
  here with the $\Xi(1620)$ strongly couples to the $\pi \Xi$
  channel. This work, and that of Ref.~\cite{GarciaRecio:2003ks},
  widely improves the conclusions of Ref.~\cite{Ramos:2002xh}, since
  we also address here the $\Xi(1690)$ resonance, and determine its
  spin-parity quantum numbers ($J^P = \oh^-$).  Yet for spin-parity
  $\oh^-$, we predict another pole at $\sqrt{s}=1729-46{\rm i}\,{\rm
  MeV}$ of which we are reasonable convinced of its existence since it
  is related to the 70 irrep. It is placed in a decuplet and it would
  be partner of the $\Delta (1620)$ and $\Sigma(1750)$
  resonances. Assuming an equal spacing rule, we would expect a
  strangeness $-2$ state of around 1900 MeV that could naturally be
  the three star $\Xi(1950)$ resonance. Note that the predicted pole
  positions in the $\Delta (1620)$ and $\Sigma(1750)$ cases were also
  low by around 150 MeV. This identification would allow to set up to
  $\oh^-$ the spin-parity, still undetermined, of this state. It
  couples strongly to the $\bar K^* \Sigma$ and $\bar K^* \Sigma^*$
  (vector-baryon octet and vector-baryon decuplet), in analogous
  manner to its partners in this decuplet that had big $\rho N$ and
  $\rho \Delta$, and $\bar K^* N$ and $\bar K^* \Delta$ components,
  respectively. Moreover, the $\bar K^* \Sigma$ and $\bar K^*
  \Sigma^*$ components of the pole will lead to the only seen $\bar K
  \Lambda$ decay mode, through mechanisms like the one in the left
  panel of Fig.~\ref{fig:box} and thanks to the large $\bar K^*\bar K
  \pi$, and $\pi \Sigma \Lambda$ and $\pi\Sigma^* \Lambda$ strong
  vertices. In Ref.~\cite{angels}, the $\Xi(1950)$ resonance is
  identified to one of the states generated there, but its dynamics is
  different from that deduced within our approach since, in that work,
  coupled channel effects with vector meson-baryon decuplet are not
  considered.

  The remaining $\oh^-$ poles predicted by the model stem from the 
  weakly attractive 1134 irrep. Some of them might have some
  correspondence with some of the states compiled in the PDG, like the
  $\Xi(2120)$, $\Xi(2250)$, $\Xi(2370)$, $\Xi(2500)$, \dots, or to
  states not discovered yet. However, we cannot make any meaningful
  statement, at this stage.

\item[ii)] 
In the $J^P=\trh^-$ sector, we associate the 70-plet pole at
$\sqrt{s}=1748-48{\rm i}\,{\rm MeV}$ with the three star $\Xi(1820)$
state. As happens for other $J^P=\trh^-$ resonances, we expect that
inclusion of the $d$-wave channels and fine tuning of the subtraction
point could bring its position closer to the experimental value. The
$\Xi(1820)$ dominant modes seem to be $\bar K \Lambda$ and perhaps
$\pi\Xi^*$, but the branching fractions are very poorly determined
($30\pm 15 \%$ for both decay modes)~\cite{pdg}. Though we can
easily explain the latter decay mode, we have serious problems to
understand within our model the $\bar K \Lambda$ one. This, together
with the fact that our pole is much wider\footnote{Note, however that
because of the Flatt\`e effect~\cite{Flatte:1976xu}, with the opening
of the $\bar K\Sigma^* $ channel to which the resonance couples
strongly, the apparent width might be smaller than that deduced from
the imaginary part of the pole position.} than the actual $\Xi(1820)$
state, reveals the important role played by $d-$wave components, not
taken into account here, in the dynamics of this state. This situation
is similar to those previously discussed for the other partners
[$N(1520)$, $\Lambda(1520)$ and $\Sigma (1670)$] of the $\Xi(1820)$ in
this $8_4$ octet of the 70-plet. Neither in Ref.~\cite{bao} nor in
Ref.~\cite{angels}, this $\Xi(1820)$ state is discussed. However, it
is studied in \cite{Kolomeitsev:2003kt} and in \cite{sarkar}. In the
latter work, both the decuplet and octet poles belonging to the 56 and
70 irreps respectively, are found with couplings similar to those
compiled in the Table~\ref{tab:jp32x}. However, there the $\Xi(1820)$
is identified with the decuplet pole, because it is narrower, while the
octet pole, that in \cite{sarkar} is four times wider than here, is
ignored. This is because in the approach of Ref.~\cite{sarkar} it did not show
up in the $|T|^2$ plot of the amplitudes in the real plane, and hence
the chances of observation were thought to be not too bright.  The
approach of Ref.~\cite{Kolomeitsev:2003kt} is based on speed plots,
where two close poles cannot be disentangled, and hence the combined
signature of the octet and decuplet poles was assigned there to the
$\Xi(1820)$ resonance.

Next, we might try to identify in our scheme the 56 irrep pole with
some other cascade resonance. It would be partner of the
$\Delta(1700)$ and the $\Sigma(1940)$ resonances in an SU(3) decuplet,
and we might have the same difficulties that in these two cases to do
a proper assignment. Assuming an equal spacing rule, we would expect a
cascade state of around 2.1 GeV. It would be quite far from
the mass predicted by the model, however we have already faced up
this problem for the other two members of this decuplet. In the PDG
there exists one state in this region of energies. This is the one
star $\Xi(2120)$ resonance, however the existence of this state is
highly uncertain. Next in energy in the PDG, we find the two star $\Xi
(2250)$ state. Its spin-parity is unknown, and it decays into the
$\Xi\pi\pi$, $\Lambda \bar K \pi$ and $\Sigma \bar K \pi$ three body
states. These decay modes are in agreement with the large $\pi \Xi^*$,
$\bar K \Sigma^*$ and $\bar K^* \Lambda$ couplings of the 56-plet
pole, and we tentatively assign this $\Xi (2250)$ resonance to this
pole. This fixes the spin-parity of this resonance, which is not known
yet. Nevertheless, we must acknowledge that this identification is
not theoretically robust, and it might well be instead that this pole
should be associated with the $\Xi(1820)-$resonance or to a new
$\Xi-$state not discovered yet. The fact that it is related to the 56
irrep make us confident that it might have some counterpart in nature.

\item[iii)] In the $J^P=\fvh^-$ sector, we find poles only from the
  1134 SU(6) irrep. Experimentally, there exists one state, the three
  star $\Xi(2030)$, with $J\ge \fvh$, and parity undetermined. It has
  large $d-$wave decays into $\bar K \Lambda$ and $\bar K \Sigma$,
  around $20\%$ and $80\%$, respectively. As it was the case of the
  four star $N(1675)$, $\Sigma(1775)$ and $\Lambda(1830)$ $\fvh^-$
  resonances, this well established state can not be described in our
  scheme and it might belong, together with the latter resonances,
  to an octet of genuine $d-$wave resonances of spin $\fvh$. However,
  in \cite{Samios:1974tw} is argued that this $\Xi(2030)$ state could
  be better accommodated in a $J^P=\fvh^+$ octet that would include also
  the four star $N(1680)$, $\Sigma(1915)$ and $\Lambda(1820)$
  resonances. It is also interesting to reproduce here a warning that
  is made in the PDG related to the $\Xi(1950)$: `{\it ... the accumulated
  evidence for a $\Xi$ near 1950 MeV seems strong enough to include a 
  $\Xi(1950)$ in the main Baryon Table, but not much can be said about its
  properties. In fact, there may be more than one $\Xi$ near this
  mass'.} We have identified the $\Xi(1950)$ with a spin--parity $\oh^-$
  state related to a decuplet of the 70 irrep. However, if it would
  exist a second state at this energy, that could be the octet partner
  of  the $\fvh^-$ resonances mentioned above.

  Some of the poles listed in Table~\ref{tab:jp52x} might have some
  correspondence with some of the states compiled in the PDG, like the
  $\Xi(2120)$, $\Xi(2370)$, $\Xi(2500)$, \dots, or to
  states not yet discovered. Some of them present similarities with states
  listed  in Refs.~\cite{bao, angels}. However, as in the previous
  isospin--strangeness sectors, we cannot make any definitive 
  statement for spin 5/2 states.

\end{itemize} 

\bc \btab
\caption{Same as Table \ref{tab:jp12n} for $\oh^-$ $\Omega$ resonances.}
\label{tab:jp12o}
\vspace{0.1cm}
\bt{c|c|c|c|c}
\hline
SU(3) &  &  &  \\
(SU(6))& Pole      & $|g_i|$    & possible ID & status PDG\\
irrep & position [MeV] & &  \\
\hline

\hline 35 & 2557-40i & $g_{\bar{K}\Xi}=0.4$, $g_{\bar{K}^*\Xi}=0.3$,
$\mathbf{g_{\bar{K}^*\Xi^*}=1.2}$, $g_{\omega\Omega}=0.2$
$\big\uparrow$
& \\ (1134) & & $\mathbf{g_{\phi\Omega}=3.4}$ & \\

\hline 10 & 2364-26i & $g_{\bar{K}\Xi}=0.8$, $g_{\bar{K}^*\Xi}=0.5$
$\big\uparrow$
$g_{\bar{K}^*\Xi^*}=0.7$, $\mathbf{g_{\omega\Omega}=3.0}$, & \\ (1134) & &
$\mathbf{g_{\phi\Omega}=1.0}$ & \\

\hline 10 & 2230-62i & $g_{\bar{K}\Xi}=0.4$, $\mathbf{g_{\bar{K}^*\Xi}=2.1}$
$\big\uparrow$
$\mathbf{g_{\bar{K}^*\Xi^*}=3.0}$, $g_{\omega\Omega}=0.8$, & \\ (1134) &
[2245-73i] & $\mathbf{g_{\phi\Omega}=2.0}$ & \\

\hline 
{\bf 10} & 1798 (*) & $\big\uparrow$ $g_{\bar{K}\Xi}=3.6$, $\mathbf{g_{\bar{K}^*\Xi}=5.5}$
$\mathbf{g_{\bar{K}^*\Xi^*}=6.8}$, $g_{\omega\Omega}=2.3$, &
${\bf \Omega(2250)}$ & $\star\star\star$\\  {\bf (70)} &
 & $\mathbf{g_{\phi\Omega}=4.7}$ & \\

\hline
\et
\etab
\ec

\bc
\btab
\caption{Same as Table \ref{tab:jp12n} for $\trh^-$ $\Omega$ resonances.}
\label{tab:jp32o}
\vspace{0.1cm}
\bt{c|c|c|c|c}
\hline
SU(3) &  &  &  \\
(SU(6))& Pole      & $|g_i|$    & possible ID & status PDG \\
irrep & position [MeV] & &  \\
\hline

\hline 35 & 2519-41i & $g_{\bar{K}\Xi^*}<0.1$, $g_{\bar{K}^*\Xi}=0.1$,
$g_{\eta\Omega}=1.1$, $g_{\bar{K}^*\Xi^*}=0.1$, & \\ (1134) & &
$g_{\omega\Omega}<0.1$
$\big\uparrow$
$\mathbf{g_{\phi\Omega}=3.7}$ & \\

\hline 10 & 2391-25i & $g_{\bar{K}\Xi^*}=0.1$,
$\mathbf{g_{\bar{K}^*\Xi}=1.1}$, $g_{\eta\Omega}=0.1$
$\big\uparrow$
$\mathbf{g_{\bar{K}^*\Xi^*}=1.3}$, & \\ (1134) & &
$\mathbf{g_{\omega\Omega}=2.5}$, $g_{\phi\Omega}=0.2$ & \\

\hline 10 & 2322-45i & $\mathbf{g_{\bar{K}\Xi^*}=1.2}$, $g_{\bar{K}^*\Xi}=0.2$,
$g_{\eta\Omega}=0.2$
$\big\uparrow$
$\mathbf{g_{\bar{K}^*\Xi^*}=2.9}$, &  \\ (1134) & & $g_{\omega\Omega}=0.6$,
$g_{\phi\Omega}=0.4$ & \\

\hline 
{\bf 10} & 1928 &
$\big\uparrow$
$\mathbf{g_{\bar{K}\Xi^*}=1.9}$, $\mathbf{g_{\bar{K}^*\Xi}=2.4}$,
$\mathbf{g_{\eta\Omega}=2.1}$, $g_{\bar{K}^*\Xi^*}=1.4$, & 
${\bf  \Omega(2380)}$ & $\star\star$  \\ {\bf (56)} & &
$g_{\omega\Omega}=0.2$, $g_{\phi\Omega}=1.6$ & \\

\hline
\et
\etab
\ec

\bc
\btab
\caption{Same as Table \ref{tab:jp12n} for $\fvh^-$ $\Omega$ resonances.}
\label{tab:jp52o}
\vspace{0.1cm}
\bt{c|c|c|c|c}
\hline
SU(3) &  &  &  \\
(SU(6))& Pole      & $|g_i|$    & possible ID & status PDG \\
irrep & position [MeV] & &  \\
\hline

\hline 10 & 2416 & 
$\big\uparrow$
$\mathbf{g_{\bar{K}^*\Xi^*}=1.1}$, $\mathbf{g_{\omega\Omega}=2.1}$,
$g_{\phi\Omega}=0$ & \\ (1134) & [2415-4i] & & \\

\hline
\et
\etab
\ec

\subsection{Omegas ($\Omega$)}

Finally, Tables \ref{tab:jp12o}, \ref{tab:jp32o} and \ref{tab:jp52o} show the
results for the $\Omega$ resonances generated by the model.

The strangeness $-3$ sector of the SU(6) model has been investigated in
\cite{juanomega}. In this previous work the meson decay constant for vector
and pseudoscalar mesons has been taken as equal. In the present work,
however, the vector meson decay constants used are higher and therefore the
interaction is weakened. As a result the binding of the resonances is reduced
and their mass increased with respect to \cite{juanomega}.

In this sector also the experimental data is very poor and more
information is needed in order to do a proper identification of the
poles obtained in our model. We will concentrate only in two states
that are placed in two SU(3) decuplets associated with the strongly
attractive 70 (in this case, we find a virtual state) and 56 irreps,
respectively, and that we will identify to the $\Omega(2250)$ and
$\Omega(2380)$ states. These resonances are not generated in the works
of Refs.~\cite{bao,angels}.
\begin{itemize}
\item[i)] For the 70 irrep pole, the spin is $\oh$ and from the various
discussions above, this state would be partner of the $\Delta(1620)$,
$\Sigma(1750)$ and $\Xi(1950)$ resonances. Following the pattern of
flavor breaking, we expect its real mass to be around 2.2 GeV. Thus,
we find a clear candidate in the PDG: the $\Omega(2250)$. Moreover, we
can see that this resonance shares many features in common with the other
resonances mentioned above. Among its main decay modes, we pay first
attention to the $\bar K \Xi^*$ one. It is of the 
pseudoscalar--baryon decuplet type, and it would be similar to the $\pi
\Delta$, $\pi\Sigma^*$ and $\pi\Xi^*$ modes for the $\Delta(1620)$,
$\Sigma(1750)$ and $\Xi(1950)$ resonances, respectively. If all these
resonances have spin $\oh$, these components are genuinely $d-$wave
and produce some distortion between the predicted masses and widths
for these states in our scheme and the actual ones of the physical
resonances.  The other decay mode for the $\Omega(2250)$ resonance is
the three body one $\Xi \pi \bar K$, which is analogous to the $\Delta
\pi \pi $ for the $\Delta(1620)$ and that can be naturally explained
from the large vector-baryon octet ${\bar K}^* \Xi$ and vector-baryon
decuplet ${\bar K}^* \Xi^*$ couplings of the 70-plet pole.

Thus we conclude that it is fair to identify this 70-plet pole with
the $\Omega(2250)$ state, which in turn also determines the spin-parity of
this resonance.

\item[ii)] For the 56 irrep pole, the spin is $\trh$ and from the various
discussions above, this state would be partner, in a decuplet, of the
$\Delta(1700)$, $\Sigma(1940)$ and $\Xi(2250)$ resonances. From the
pattern of flavor breaking, we expect its real mass to be around 2.5 
GeV. In the PDG are listed two other omega resonances: the two star 
$\Omega(2380)$ and  $\Omega(2470)$. The latter one is seen to decay
into $\Omega \pi\pi$, while the main decay modes of the former one
are $\Xi \pi \bar K$, $\bar K \Xi^*$ and $\bar K^* \Xi$ in perfect
agreement with the couplings of our predicted 56-plet pole, and
exhibiting some similarities with the features of the other members of
this decuplet. Hence, it seems natural to identify this pole with the
$\Omega(2380)$ resonance, which  allows again to determine its
spin-parity.

Similar poles were found in Refs.~\cite{Kolomeitsev:2003kt} and
\cite{sarkar}. The dynamics of this state in these two references is
different to that found here, since in both schemes the interplay with
the vector--baryon decuplet $\bar K^* \Xi^*$ channel was not
considered. While in the former work the state was not identified to
any resonance, in the latter work it was tentatively assigned to the
$\Omega(2250)$ baryon. For the reasons given above, we disagree with
such identification. 

\end{itemize}

\subsection{Exotics}

If we look at Table~\ref{tab:states}, there are many states that do
not have $N,\Delta,\Sigma,\Lambda, \Xi$ or $\Omega$ quantum
numbers. These are, what we will call here exotic states. All of them
stem from the evolution of the weakly attractive 1134 irrep. We find
several poles, but they might be subject to larger relative
corrections or even disappear by the consideration of higher order
terms and higher order even waves, as we have been discussing for all
non-exotic poles belonging to the 1134 in the previous subsections.
Given this uncertain scenario, we feel that we cannot draw any robust
conclusion on exotic states at this point. It is, however, a valuable
piece of information that exotic states are not related with the
strongly attractive 70 and 56 plets. As we have seen, the bulk of
$J=\oh,\trh$ odd parity three and four star baryon resonances listed
in the PDG can comfortably be associated with these two SU(6)
multiplets.  Moreover, we would like to draw the attention here to
some of the findings of Ref.~\cite{GarciaRecio:2006wb} when the number
of colors $N_c$ departs from 3.  There, it is shown that the in the 70
SU(6) irreducible space, the SU(6) extension of the WT $s-$wave
meson-baryon interaction scales as ${\cal O}(1)$, instead of the
well-known ${\cal O}(N_c^{-1})$ behavior for its SU(3)
counterpart. However, the WT interaction behaves as ${\cal
O}(N_c^{-1})$ within the 56 and 1134 meson-baryon spaces.  This
presumably implies that 1134 states do not appear in the large $N_c$
QCD spectrum, since both excitation energies and widths grow with an
approximate $\sqrt{N_c}$ rate.

Finally, just mention that in previous works \cite{su6model,juangs},
we advocated for the existence of some exotic states. In particular,
we paid an special attention to the existence of an pentaquark of spin
$\trh$, isospin zero and strangeness $+1$. Indeed, it naturally showed
up as $K^* N$ bound state with a mass around 1.7--1.8 GeV and it was
part of an SU(3) anti-decuplet of the 1134 irrep. In these previous
works, as mentioned above, the meson decay constant for vector and
pseudoscalar mesons were taken as equal. In the present work,
however, the vector meson decay constants used are higher and
therefore the interaction is weakened. As a result, this pole
disappears within the RS employed here. However, it is true that by
using instead a cutoff to renormalize the ultraviolet loops, and
keeping it greater than 1.3 GeV, one still finds such state 
within the pattern of spin symmetry breaking assumed here. 

To conclude, we cannot discard the existence of exotic states, because
to large extent this is a RS dependent issue. However we can say that the
SU(6) extension of the WT presented here does not provide robust
theoretical hints of their actual existence.

\subsection{Assignation of SU(6) and SU(3) labels}

As explained at the beginning of this Section, we have attached definite SU(6)
and SU(3) labels to each pole found by paying attention to how this pole is
generated in SU(6) or SU(3) symmetric scenarios. This procedure allows to
uncover the pattern of SU(6) or SU(3) multiplets in the final physical
results, where these symmetries are broken.\footnote{An alternative procedure
  to reveal the genesis of each pole under SU(6) would be to study its
  response under changes of the eigenvalues $\lambda_r$ in \Eq{eq:xi}. The
  analysis can be extended to SU(3) in the obvious way.}

The previous procedure reveals the nature of the pole from the {\em genetic}
point of view. Alternatively, one can study the {\em structure} of the
resonance in the final scenario with broken symmetry. This can be done by
analyzing the wave function of the resonance in coupled channels
space. Following \cite{Gamermann:2009uq}, we note that the pole condition on
the $T$-matrix\footnote{We work in a given sector of coupled channel space
  throughout, so we drop the sector label $SIJ$.}  is equivalent to the
Schr\"odinger equation-like condition
\be
(G^{-1}- V)\psi = 0
\ee
where $G$ and $V$ are matrices in coupled channel space and $\psi$ is a column
vector. The condition is on $\sqrt{s}$ for the matrix $GV$ to have an
eigenvector with eigenvalue equal to unity, namely, $GV\psi=\psi$. $\psi$ is
related to the wave function of the resonance in coupled channel space (more
specifically to the wave function for small baryon-meson separation
\cite{Gamermann:2009uq}).

Up to a factor, the quantities $V\psi$ (a column vector) are the coupling
constants $g_i$ (modulus and phase) appearing in the residue of the resonance
pole. So these couplings give us information on the structure of the
resonance, however, this is not directly the wave function, rather,
$g_i=\langle i|V|\psi\rangle$ are the transition matrix elements related to
the probability of formation and decay of the resonance. Working instead with
the wave function\footnote{There is a subtlety here since the quantity $\psi$
  obtained as above depends on conventions on how precisely $G$ and $V$ are
  normalized (all the conventions having the same poles in the plane
  $\sqrt{s}$). The proper definition of the wave function is such that the
  propagator is normalized as $G_s=(E-H_0)^{-1}$. In our case
  $G_i=c_i^2G_{s,i}$ with $c_i=\sqrt{(M_i+m_i)/(m_i 16\pi^3\sqrt{s})}$ from
  Eqs.~(66,67) of \cite{Gamermann:2009uq} (note that our $G_i= 2M_i G^{\rm
    FT}$ for $G^{\rm FT}$ of \cite{Gamermann:2009uq}). The potential that
  combined with $G_s= c^{-1}G c^{-1}$ (matrix notation) gives the same poles
  is $V_s=cVc$, with corresponding $T$-matrix $T_s=cTc$. The couplings from
  the residues of $T_s$ at the poles are $g_s=cg$. For the (unnormalized) wave
  function, $g_s=V_s\psi_s$, so $\psi_s=G_sg_s=c^{-1}Gg=c^{-1}\psi$.
  Therefore, up to normalization, the wave function is
  $\sqrt{\frac{m_i}{m_i+M_i}} G_ig_i$, with $G_i$ evaluated at the pole.} we
can analyze the resonance from the point of view of its SU(6) and SU(3)
composition. The coupled channel space baryon-meson basis is the basis
attached to $56\otimes 35$. In terms of product of representations, this is is
the ``uncoupled basis''. Using the appropriate scalar factors of ${\rm
  SU}(6)\supset {\rm SU}(3)_f\otimes {\rm SU}(2)_J$ \cite{GarciaRecio:2010vf},
one can express the same state in the ``coupled basis'', with well defined
${\rm SU}(6)\supset {\rm SU}(3)_f\otimes {\rm SU}(2)_J$ labels. This gives us
for instance how much of the resonance belongs to each of the SU(6) irreps 56,
70, 700 and 1134.

As it turns out, it is found that the irrep 1134 has an important weight in
almost all resonances. This reflects that masses and meson decay constants
break SU(6) (as well as SU(3)). In a purely random state the 1134 multiplet
would be expected to dominate from statistical considerations. Valuable
information follows from deviations from statistics. The analysis shows that
the SU(6) irrep 700 has a small (in fact almost always negligible) role in the
wave function of the resonances; presumably a consequence of the repulsive
character of the interaction in that sector. It follows that all
$\frac{5}{2}^-$ poles are nearly 100\% 1134.

After the evolution from the SU(6) and SU(3) symmetric points, we find that
considerable mixing of irreps is achieved for resonances originally in the 56
or 70. The mixing takes place with the 1134 and also between 56 and 70. This
is true in many cases and particularly in the $\Lambda$ sector. Notably, the
$(\Omega,\frac{3}{2}^-)$ with pole at $1928\,\MeV$ is a very pure
$(56,10)$ state with very small mixing.  On the other hand, as a rule, the
heavier resonances originally in the 1134 stay in that multiplet with small or
no mixing.

From the point of view of classification of resonances, we stick to the
prescription given above, based on how the resonance is originated.  Due to
the breaking of symmetry it would not make sense to expect the resonances to
form clear and distinct multiplets at the end of the evolution different from
the initial ones. Rather one expects to find the same multiplets plus
breaking. As a rule, we find that the final structure of the resonances
reflect the SU(6) and SU(3) multiplets assigned to them. The are some
exceptions. Thus, the pole $(1867-36\,{\rm i})\,\MeV$ in the
$(\Sigma,\frac{1}{2}^-)$ sector has dominant structure of $(56,8_2)$ rather than
$(1134,8_2)$. And the following poles in the $(\Xi,\frac{1}{2}^-)$ sector:
$(1577-139\,{\rm i})$  and $(2037-24\,{\rm i})\,\MeV$ would have
structure of $(1134,8_2)$ and $(56,8_2)$, respectively. It is
interesting that the large coupling of $(1577-139\,{\rm i})\,\MeV$ to the
channel $\pi\Xi$, attending to the corresponding Clebsh-Gordan coefficients,
cannot be easily achieved through the 56 or 70 irreps and requires an important
weight of the 1134. This pole is identified to $\Xi(1620)$, and the same large
coupling is obtained in other models \cite{Ramos:2002xh,GarciaRecio:2003ks}.

%%%%%%%%%%%%%%%%%%%%%%%%%%%%%%%%%

\section{Summary}

We have studied the light baryon resonances based on the SU(6) model
introduced in \cite{su6model,GarciaRecio:2006wb}. The model assumes
that the light-quark interactions are approximately spin and SU(3)
flavor independent. With this assumption the usual SU(3)
Weinberg-Tomozawa interaction is extended to SU(6). This allows to
construct the elementary amplitudes for the $s$-wave scattering of
mesons with baryons including in a systematic way low-lying $0^-$ and
$1^-$ mesons and $1/2^+$ and $3/2^+$ baryons. With these amplitudes, 
the $T$-matrix is calculated and poles are identified. Each pole is
associated with a resonance. The information obtained for each pole
includes its mass, width and couplings of each resonance and
comparison of the information obtained allows us to associate some of
the theoretical states from the model with observed experimental
states.

We have studied the possible N, $\Delta$, $\Sigma$, $\Lambda$, $\Xi$ and
$\Omega$ states generated by the model.  Most of the experimental $J^P=\oh^-$
states fit within our approach fairly well. For the $J^P=\trh^-$ one should
bear in mind that $d$-wave channels, which are not considered in the model,
might play an important role, but even with this handicap, the model describes
many of the observed states. Most of the low-lying three and four star odd
parity baryon resonances with spin $\oh$ and $\trh$ are generated in our
scheme, and they can be related to the 70 and 56 multiplets of
the spin-flavor symmetry group SU(6), as sketched in
Fig.~\ref{fig:summ}. Indeed, the spin-flavor WT interaction turns out to be
quite attractive in these two irreps, specially in the first one, and thus we
believe these results are robust, except perhaps for those concerning the spin
$\oh^-$ octet of the 56-plet which are subject to larger uncertainties, as
argued in the previous subsections.  The spin-parity of the $\Xi(1620)$,
$\Xi(1690)$, $\Xi(1950)$, $\Xi(2250)$, $\Omega(2250)$ and $\Omega(2380)$
resonances, not experimentally determined yet, can be read off the figure and
thus are predictions of our scheme. More precise experiments on this issue
would be very welcome in order to test these assignments.

It should be stressed that we have chosen not adjust any parameter;
the RS used here completely fixes the subtraction constant to some specific
quantity determined by the masses of the hadrons in each $IS$ sector (see
Eq.~(\ref{eq:musi})). This is in contrast with the
RS advocated in other works~\cite{lamb1, lamb3, inoue, juan3,
Oset:2001cn, sarkar,bao,angels}, which allows for some free variation
in the subtraction constants of each of the coupled channels that
enter in any $JIS-$sector.  Such freedom makes possible to achieve a
better phenomenological description of data. However in some sense,
this flexibility dilutes the predictive power of the scheme, and also
it might happen that some more freedom than that allowed by the
underlying symmetry is being used.

We predict also many states associated with the weakly attractive 1134
irrep, some of them with exotic quantum numbers. In order to do a
proper identification in these cases, it is essential to have accurate
data because slight changes in the RS might change drastically the
position and main features of the predicted states, and some of them
might even disappear. That is the reason why we have tried to
associate these poles with resonances in the PDG only in few cases,
mostly for those which could be related to firmly established resonances 
(three or four stars).
In this context, we mention here that
the four star $\Lambda(1690)$ and the three star $N(1700)$, $\Delta
(1930)$ and $\Lambda(1800)$ resonances could be also accommodated
within the model. Thus, considering the states included in
Fig.~\ref{fig:summ} and these four resonances above, all three and
four star odd parity baryons listed in the PDG, except for the
$N(1675)$, $\Sigma(1775)$, $\Lambda(1830)$ and $\Xi(2030)$ resonances,
are dynamically generated within this scheme. These latter five states
have spin $J=\fvh$, their existence is firmly established, and in all
cases their dominant decay modes always involve $d-$wave interactions,
which are beyond the scope of this work. We observe that they can be
cast into an SU(3) octet, though some other possibilities cannot be
discarded.

The hidden gauge scheme of Refs.~\cite{bao,angels} leads to a
distinctive pattern, where the higher energy states are degenerate in
spin~\cite{eulogio,bao,angels}.  The authors of these works found in
some cases candidates in the PDG that seem to follow this pattern. For
instance, the triplet of resonances $\Delta (1900)$, $\Delta (1940)$,
$\Delta (1930)$ that have spin-parity $J^P=\oh^-, \trh^-$, and
$\fvh^-$, respectively.\footnote{Note the scarce experimental evidence
on the actual existence of the first two $\Delta-$states that are
classified in the PDG as one and two stars, respectively.} Many of the states
 predicted in~\cite{eulogio,bao,angels} are missing, however this does not mean
that the pattern deduced there is necessarily incorrect,
since the predicted states are at the frontier of the experimental
research. It would not be difficult to fine tune the subtraction
constants of some of the 1134 states, which have always energies
higher than those associated with the more attractive 56 and 70 irreps,
obtained in our model to meet the results of \cite{bao,angels}. Indeed,
in Tables~\ref{tab:jp12d}, \ref{tab:jp32d} and \ref{tab:jp52d}, we have
identified the triplet of $\Delta$'s mentioned above without introducing
to modifications in the RS.

As noted, the 1134-plet contains exotic states which require pentaquark
configurations. On the contrary, spin-flavor wavefunctions in the 56- and
70-plets can be obtained with $qqq$ states, i.e., from the product $6\otimes
6\otimes 6$. The $70^-$, which has the strongest attraction in our model and
becomes dominant in the large $N_c$ limit \cite{GarciaRecio:2006wb}, is also
natural in quark model approaches with $qqq$ configurations
\cite{Hey:1982aj}. The $70^-$ corresponds to the symmetric combination with
two quarks in $s$-wave (the lowest state) and the other quark in an excited
$p$-wave state of the bag. (The color wavefunction is antisymmetric and so the
spin-flavor-orbital wavefunction must be symmetric.) In this view, one would
expect some mixing between our dynamically generated states and the
$qqq$-states. The situation is different for the $56^-$ states. For a 56 the
$qqq$ spin-flavor wavefunction is completely symmetric and this requires a
completely symmetric wavefunction in the orbital space. The states obtained
from putting one of the quarks in $p$-wave are spurious and disappear after
center of mass projection. Hence $56^-$ is not natural as a $qqq$ state as it
requires more complicated, and so heavier, quark configurations.

\begin{acknowledgments}

We thank M. Pav\'on Valderrama 
for useful discussions. Research supported by DGI under contract
FIS2008-01143, Junta de Andaluc\'{\i}a grant FQM-225, Generalitat
Valenciana contract PROMETEO/2009/0090, the Spanish Consolider-Ingenio
2010 Programme CPAN contract CSD2007-00042, and the European
Community-Research Infrastructure Integrating Activity {\em Study of
Strongly Interacting Matter} (HadronPhysics2, Grant Agreement
no. 227431) under the 7th Framework Programme of EU.
\end{acknowledgments}

\appendix
\section{Matrix elements between SU(3)$\otimes$SU(2) multiplets}
\label{app:tables}

This appendix presents Tables \ref{rep12}-\ref{rep356} for the matrix
elements $\xi$ of \Eq{eq:pot}. Because there are in all 38 sectors
$S,I,J$ (counting only those with negative eigenvalues; see
Table~\ref{tab:states}) and some of them with many channels, we
provide here tables for the sixteen $(R,J)$ sectors where $R$ denotes
an SU(3) irreducible representation. The tables display the matrix
elements between SU(3) multiplets. To obtain the matrix element for a
concrete channel one should use the well known ${\rm SU}(3)\supset{\rm
SU}(2)_I\otimes{\rm U}(1)_Y$ isoscalar factors. We note that the
tables have been constructed using the phase convention in
\cite{Baird:1964zm} which, unlike that of \cite{deSwart:1963gc}, is
suitable for more than three flavors. The corresponding isoscalar
factors using this convention can be found in
\cite{Rabl:1975zy,GarciaRecio:2010vf}. The tables of this Appendix can
be obtained using the ${\rm SU}(6)\supset{\rm SU}(3)\otimes{\rm
SU}(2)$ scalar factors in \cite{GarciaRecio:2010vf}. The same scalar
factors but with the convention of \cite{deSwart:1963gc} can be found
in \cite{Carter:1965,Cook65}.

For instance, from \cite{Rabl:1975zy,GarciaRecio:2010vf} and including
the appropriate spin SU(2) Clebsch-Gordan coefficients, we learn that
\begin{eqnarray}
|N\bar{K}^*\rangle^{S=-1,I=1,J=3/2}
&=&
-\sqrt{\frac{3}{10}}|(\mathbf{8_2},\mathbf{8_3}),\mathbf{8_{s,4}}\rangle
-\frac{1}{\sqrt{6}}|(\mathbf{8_2},\mathbf{8_3}),\mathbf{8_{a,4}}\rangle
-\frac{1}{\sqrt{6}}|(\mathbf{8_2},\mathbf{8_3}),\mathbf{10_4}\rangle
\nonumber \\
&&
+\frac{1}{\sqrt{6}}|(\mathbf{8_2},\mathbf{8_3}),\mathbf{10^*_4}\rangle
+\frac{1}{\sqrt{5}}|(\mathbf{8_2},\mathbf{8_3}),\mathbf{27_4}\rangle
,
\\
|\Sigma^*\eta\rangle^{S=-1,I=1,J=3/2}
&=&
\frac{1}{\sqrt{5}} |(\mathbf{10_4},\mathbf{8_1}),\mathbf{8_4}\rangle
+\sqrt{\frac{3}{10}}|(\mathbf{10_4},\mathbf{8_1}),\mathbf{27_4}\rangle
+\frac{1}{\sqrt{2}}|(\mathbf{10_4},\mathbf{8_1}),\mathbf{35_4}\rangle
.
\nonumber 
\end{eqnarray}

Henceforth, using the tables  provided for
$\mathbf{8_4}$ and $\mathbf{27_4}$, one easily obtains

\begin{eqnarray}
 \xi^{-1,1,3/2}_{\Sigma^*\eta,N\bar{K}^*} &=&
\langle \Sigma^*\eta|\mathbf{8_4}\rangle
\langle (\mathbf{10_4},\mathbf{8_1}) |\xi^{\mathbf{8_4}}
|(\mathbf{8_2},\mathbf{8_3})_s\rangle
\langle \mathbf{8_{s,4}}| N\bar{K}^*\rangle
\nonumber \\ &&+
\langle \Sigma^*\eta|\mathbf{8_4}\rangle
\langle (\mathbf{10_4},\mathbf{8_1}) |\xi^{\mathbf{8_4}}
|(\mathbf{8_2},\mathbf{8_3})_a\rangle
\langle \mathbf{8_{a,4}}| N\bar{K}^*\rangle
\nonumber \\ &&+
\langle \Sigma^*\eta|\mathbf{27_4}\rangle
\langle (\mathbf{10_4},\mathbf{8_1}) |\xi^{\mathbf{27_4}}
|(\mathbf{8_2},\mathbf{8_3})\rangle
\langle \mathbf{27_4}| N\bar{K}^*\rangle
\nonumber \\
&=&
\frac{1}{\sqrt{5}}
\underbrace{(2\sqrt{3})}_{\rm Table~\protect\ref{rep84}}
\left(-\sqrt{\frac{3}{10}}\right)+ \frac{1}{\sqrt{5}}
\underbrace{(0)}_{\rm Table~\protect\ref{rep84}}
\left(-\sqrt{\frac{1}{6}}\right)
+
\sqrt{\frac{3}{10}}
\underbrace{\left(-\frac{4}{\sqrt{3}}\right)}_{\rm Table~\protect\ref{rep274}}
\frac{1}{\sqrt{5}}
\nonumber \\
&=&
-\sqrt{2} .  
\end{eqnarray}

\begin{table}[ht]
\caption{Matrix elements for $\mathbf{1}$ and $J=1/2$. 
Eigenvalues: $-18,-2$.}
\label{rep12}
\begin{center}
\begin{tabular}{c|cc}
$ \mathbf{1_2}$ & $(\mathbf{8_2} ,\mathbf{8_1} ) $&$ (\mathbf{8_2} ,\mathbf{8_3} )$ \\\hline
 $(\mathbf{8_2} ,\mathbf{8_1} )$ &$ -6$ &$ 4 \sqrt{3} $\\
 $(\mathbf{8_2} ,\mathbf{8_3} ) $&$ 4 \sqrt{3} $&$ -14 $
\end{tabular}
\end{center}
\end{table}

\begin{table}[ht]
\caption{Matrix elements for $\mathbf{1}$ and $J=3/2$. 
Eigenvalue: $-2$.}
\label{rep14}
\begin{center}
\begin{tabular}{c|c}$
 \mathbf{1_4} $&$ (\mathbf{8_2} ,\mathbf{8_3} ) $\\\hline$
 (\mathbf{8_2} ,\mathbf{8_3} ) $&$ -2
$\end{tabular}
\end{center}
\end{table}

\begin{table}[ht]
\caption{Matrix elements for $\mathbf{8}$ and $J=1/2$. 
Eigenvalues: $-18,-12,6,-2,-2,-2$.}
\label{rep82}
\begin{center}
\begin{tabular}{c|cccccc}$
 \mathbf{8_2} $&$ (\mathbf{8_2} ,\mathbf{8_1} )_s $&$ (\mathbf{8_2} ,\mathbf{8_1} )_a $&$ (\mathbf{8_2} ,\mathbf{8_3} )_s $&$ (\mathbf{8_2} ,\mathbf{8_3} )_a $&$
   (\mathbf{8_2} ,\mathbf{1_3} ) $&$ (\mathbf{10_4} ,\mathbf{8_3} ) $\\\hline$
 (\mathbf{8_2} ,\mathbf{8_1} )_s $&$ -3 $&$ 0 $&$ 2 \sqrt{3} $&$ -\sqrt{15} $&$ 0 $&$ -2 \sqrt{6} $\\$
 (\mathbf{8_2} ,\mathbf{8_1} )_a $&$ 0 $&$ -3 $&$ -\sqrt{15} $&$ 2 \sqrt{3} $&$ 0 $&$ 0 $\\$
 (\mathbf{8_2} ,\mathbf{8_3} )_s $&$ 2 \sqrt{3} $&$ -\sqrt{15} $&$ -\frac{7}{3} $&$ \frac{4 \sqrt{5}}{3} $&$ -\frac{4 \sqrt{10}}{3} $&$
   -\frac{4 \sqrt{2}}{3} $\\$
 (\mathbf{8_2} ,\mathbf{8_3} )_a $&$ -\sqrt{15} $&$ 2 \sqrt{3} $&$ \frac{4 \sqrt{5}}{3} $&$ -\frac{23}{3} $&$ \frac{8 \sqrt{2}}{3} $&$ -\frac{2
   \sqrt{10}}{3} $\\$
 (\mathbf{8_2} ,\mathbf{1_3} ) $&$ 0 $&$ 0 $&$ -\frac{4 \sqrt{10}}{3} $&$ \frac{8 \sqrt{2}}{3} $&$ -\frac{4}{3} $&$ \frac{4 \sqrt{5}}{3} $\\$
 (\mathbf{10_4} ,\mathbf{8_3} ) $&$ -2 \sqrt{6} $&$ 0 $&$ -\frac{4 \sqrt{2}}{3} $&$ -\frac{2 \sqrt{10}}{3} $&$ \frac{4 \sqrt{5}}{3} $&$
   -\frac{38}{3}
$\end{tabular}
\end{center}
\end{table}

\begin{table}[ht]
\caption{Matrix elements for $\mathbf{8}$ and $J=3/2$. 
Eigenvalues: $-18,6,-2,-2,-2$.}
\label{rep84}
\begin{center}
\begin{tabular}{c|ccccc}$
 \mathbf{8_4} $&$ (\mathbf{8_2} ,\mathbf{8_3} )_s $&$ (\mathbf{8_2} ,\mathbf{8_3} )_a $&$ (\mathbf{8_2} ,\mathbf{1_3} ) $&$ (\mathbf{10_4} ,\mathbf{8_1} ) $&$
   (\mathbf{10_4} ,\mathbf{8_3} ) $\\\hline$
 (\mathbf{8_2} ,\mathbf{8_3} )_s $&$ -\frac{10}{3} $&$ -\frac{2 \sqrt{5}}{3} $&$ \frac{2 \sqrt{10}}{3} $&$ 2 \sqrt{3} $&$ -\frac{4
   \sqrt{5}}{3} $\\$
 (\mathbf{8_2} ,\mathbf{8_3} )_a $&$ -\frac{2 \sqrt{5}}{3} $&$ -\frac{2}{3} $&$ -\frac{4 \sqrt{2}}{3} $&$ 0 $&$ -\frac{10}{3} $\\$
 (\mathbf{8_2} ,\mathbf{1_3} ) $&$ \frac{2 \sqrt{10}}{3} $&$ -\frac{4 \sqrt{2}}{3} $&$ \frac{2}{3} $&$ 0 $&$ \frac{10 \sqrt{2}}{3} $\\$
 (\mathbf{10_4} ,\mathbf{8_1} ) $&$ 2 \sqrt{3} $&$ 0 $&$ 0 $&$ -6 $&$ 2 \sqrt{15} $\\$
 (\mathbf{10_4} ,\mathbf{8_3} ) $&$ -\frac{4 \sqrt{5}}{3} $&$ -\frac{10}{3} $&$ \frac{10 \sqrt{2}}{3} $&$ 2 \sqrt{15} $&$ -\frac{26}{3}
$\end{tabular}
\end{center}
\end{table}

\begin{table}[ht]
\caption{Matrix elements for $\mathbf{8}$ and $J=5/2$. 
Eigenvalue: $-2$.}
\label{rep86}
\begin{center}
\begin{tabular}{c|c}$
 \mathbf{8_6} $&$ (\mathbf{10_4} ,\mathbf{8_3} ) $\\\hline$
 (\mathbf{10_4} ,\mathbf{8_3} ) $&$ -2
$\end{tabular}
\end{center}
\end{table}

\begin{table}[ht]
\caption{Matrix elements for $\mathbf{10}$ and $J=1/2$. 
Eigenvalues: $-18,6,-2,-2$.}
\label{rep102}
\begin{center}
\begin{tabular}{c|cccc}$
 \mathbf{10_2} $&$ (\mathbf{8_2} ,\mathbf{8_1} ) $&$ (\mathbf{8_2} ,\mathbf{8_3} ) $&$ (\mathbf{10_4} ,\mathbf{8_3} ) $&$ (\mathbf{10_4} ,\mathbf{1_3} ) $\\\hline$
 (\mathbf{8_2} ,\mathbf{8_1} ) $&$ 0 $&$ -2 \sqrt{3} $&$ -4 \sqrt{3} $&$ 0 $\\$
 (\mathbf{8_2} ,\mathbf{8_3} ) $&$ -2 \sqrt{3} $&$ -\frac{4}{3} $&$ -\frac{4}{3} $&$ \frac{8}{3} $\\$
 (\mathbf{10_4} ,\mathbf{8_3} ) $&$ -4 \sqrt{3} $&$ -\frac{4}{3} $&$ -\frac{34}{3} $&$ \frac{20}{3} $\\$
 (\mathbf{10_4} ,\mathbf{1_3} ) $&$ 0 $&$ \frac{8}{3} $&$ \frac{20}{3} $&$ -\frac{10}{3}
$\end{tabular}
\end{center}
\end{table}

\begin{table}[ht]
\caption{Matrix elements for $\mathbf{10}$ and $J=3/2$. 
Eigenvalues: $-12,6,-2,-2$.}
\label{rep104}
\begin{center}
\begin{tabular}{c|cccc}$
 \mathbf{10_4} $&$ (\mathbf{8_2} ,\mathbf{8_3} ) $&$ (\mathbf{10_4} ,\mathbf{8_1} ) $&$ (\mathbf{10_4} ,\mathbf{8_3} ) $&$ (\mathbf{10_4} ,\mathbf{1_3} ) $\\\hline$
 (\mathbf{8_2} ,\mathbf{8_3} ) $&$ \frac{2}{3} $&$ 2 \sqrt{6} $&$ -\frac{2 \sqrt{10}}{3} $&$ \frac{4 \sqrt{10}}{3} $\\$
 (\mathbf{10_4} ,\mathbf{8_1} ) $&$ 2 \sqrt{6} $&$ -3 $&$ \sqrt{15} $&$ 0 $\\$
 (\mathbf{10_4} ,\mathbf{8_3} ) $&$ -\frac{2 \sqrt{10}}{3} $&$ \sqrt{15} $&$ -\frac{19}{3} $&$ \frac{8}{3} $\\$
 (\mathbf{10_4} ,\mathbf{1_3} ) $&$ \frac{4 \sqrt{10}}{3} $&$ 0 $&$ \frac{8}{3} $&$ -\frac{4}{3}
$\end{tabular}
\end{center}
\end{table}

\begin{table}[ht]
\caption{Matrix elements for $\mathbf{10}$ and $J=5/2$. 
Eigenvalues: $6,-2$.}
\label{rep106}
\begin{center}
\begin{tabular}{c|cc}$
 \mathbf{10_6} $&$ (\mathbf{10_4} ,\mathbf{8_3} ) $&$ (\mathbf{10_4} ,\mathbf{1_3} ) $\\\hline$
 (\mathbf{10_4} ,\mathbf{8_3} ) $&$ 2 $&$ -4 $\\$
 (\mathbf{10_4} ,\mathbf{1_3} ) $&$ -4 $&$ 2
$\end{tabular}
\end{center}
\end{table}

\begin{table}[ht]
\caption{Matrix elements for $\mathbf{10^*}$ and $J=1/2$. 
Eigenvalues: $6,-2$.}
\label{rep10c2}
\begin{center}
\begin{tabular}{c|cc}$
 \mathbf{10^*_2} $&$ (\mathbf{8_2} ,\mathbf{8_1} ) $&$ (\mathbf{8_2} ,\mathbf{8_3} ) $\\\hline$
 (\mathbf{8_2} ,\mathbf{8_1} ) $&$ 0 $&$ 2 \sqrt{3} $\\$
 (\mathbf{8_2} ,\mathbf{8_3} ) $&$ 2 \sqrt{3} $&$ 4
$\end{tabular}
\end{center}
\end{table}

\begin{table}[ht]
\caption{Matrix elements for $\mathbf{10^*}$ and $J=3/2$. 
Eigenvalues: $-2$.}
\label{rep10c4}
\begin{center}
\begin{tabular}{c|c}$
 \mathbf{10^*_4} $&$ (\mathbf{8_2} ,\mathbf{8_3} ) $\\\hline$
 (\mathbf{8_2} ,\mathbf{8_3} ) $&$ -2
$\end{tabular}
\end{center}
\end{table}

\begin{table}[ht]
\caption{Matrix elements for $\mathbf{27}$ and $J=1/2$. 
Eigenvalues: $6,-2,-2$.}
\label{rep272}
\begin{center}
\begin{tabular}{c|ccc}$
 \mathbf{27_2} $&$ (\mathbf{8_2} ,\mathbf{8_1} ) $&$ (\mathbf{8_2} ,\mathbf{8_3} ) $&$ (\mathbf{10_4} ,\mathbf{8_3} ) $\\\hline$
 (\mathbf{8_2} ,\mathbf{8_1} ) $&$ 2 $&$ -\frac{4}{\sqrt{3}} $&$ 4 \sqrt{\frac{2}{3}} $\\$
 (\mathbf{8_2} ,\mathbf{8_3} ) $&$ -\frac{4}{\sqrt{3}} $&$ -\frac{2}{3} $&$ -\frac{4 \sqrt{2}}{3} $\\$
 (\mathbf{10_4} ,\mathbf{8_3} ) $&$ 4 \sqrt{\frac{2}{3}} $&$ -\frac{4 \sqrt{2}}{3} $&$ \frac{2}{3}
$\end{tabular}
\end{center}
\end{table}

\begin{table}[ht]
\caption{Matrix elements for $\mathbf{27}$ and $J=3/2$. 
Eigenvalues: $6,-2,-2$.}
\label{rep274}
\begin{center}
\begin{tabular}{c|ccc}$
 \mathbf{27_4} $&$ (\mathbf{8_2} ,\mathbf{8_3} ) $&$ (\mathbf{10_4} ,\mathbf{8_1} ) $&$ (\mathbf{10_4} ,\mathbf{8_3} ) $\\\hline$
 (\mathbf{8_2} ,\mathbf{8_3} ) $&$ \frac{10}{3} $&$ -\frac{4}{\sqrt{3}} $&$ -\frac{4 \sqrt{5}}{3} $\\$
 (\mathbf{10_4} ,\mathbf{8_1} ) $&$ -\frac{4}{\sqrt{3}} $&$ -1 $&$ \sqrt{\frac{5}{3}} $\\$
 (\mathbf{10_4} ,\mathbf{8_3} ) $&$ -\frac{4 \sqrt{5}}{3} $&$ \sqrt{\frac{5}{3}} $&$ -\frac{1}{3}
$\end{tabular}
\end{center}
\end{table}

\begin{table}[ht]
\caption{Matrix elements for $\mathbf{27}$ and $J=5/2$. 
Eigenvalues: $-2$.}
\label{rep276}
\begin{center}
\begin{tabular}{c|c}$
 \mathbf{27_6} $&$ (\mathbf{10_4} ,\mathbf{8_3} ) $\\\hline$
 (\mathbf{10_4} ,\mathbf{8_3} ) $&$ -2
$\end{tabular}
\end{center}
\end{table}

\begin{table}[ht]
\caption{Matrix elements for $\mathbf{35}$ and $J=1/2$. 
Eigenvalues: $-2$.}
\label{rep352}
\begin{center}
\begin{tabular}{c|c}$
 \mathbf{35_2} $&$ (\mathbf{10_4} ,\mathbf{8_3} ) $\\\hline$
 (\mathbf{10_4} ,\mathbf{8_3} ) $&$ -2
$\end{tabular}
\end{center}
\end{table}

\begin{table}[ht]
\caption{Matrix elements for $\mathbf{35}$ and $J=3/2$. 
Eigenvalues: $6,-2$.}
\label{rep354}
\begin{center}
\begin{tabular}{c|cc}$
 \mathbf{35_4} $&$ (\mathbf{10_4} ,\mathbf{8_1} ) $&$ (\mathbf{10_4} ,\mathbf{8_3} ) $\\\hline$
 (\mathbf{10_4} ,\mathbf{8_1} ) $&$ 3 $&$ -\sqrt{15} $\\$
 (\mathbf{10_4} ,\mathbf{8_3} ) $&$ -\sqrt{15} $&$ 1
$\end{tabular}
\end{center}
\end{table}

\begin{table}[ht]
\caption{Matrix elements for $\mathbf{35}$ and $J=5/2$. 
Eigenvalues: $6$.}
\label{rep356}
\begin{center}
\begin{tabular}{c|c}$
 \mathbf{35_6} $&$ (\mathbf{10_4} ,\mathbf{8_3} ) $\\\hline$
 (\mathbf{10_4} ,\mathbf{8_3} ) $&$ 6
$\end{tabular}
\end{center}
\end{table}

\clearpage
\section{$\pi N$ $S_{11}$ phase shift and inelasticities}
\label{app:s11}
\begin{figure}[tbh]
\begin{center}
\makebox[0pt]{\hspace{-0.35cm}\includegraphics[height=7.5cm]{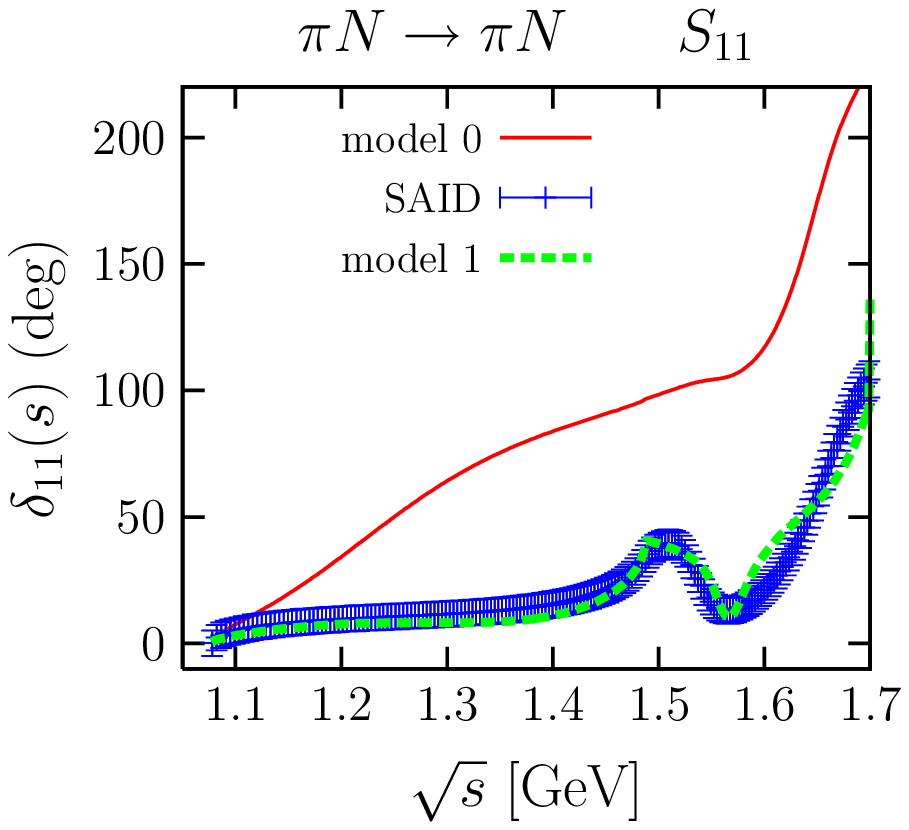}\hspace{0.15cm}\includegraphics[height=7.5cm]{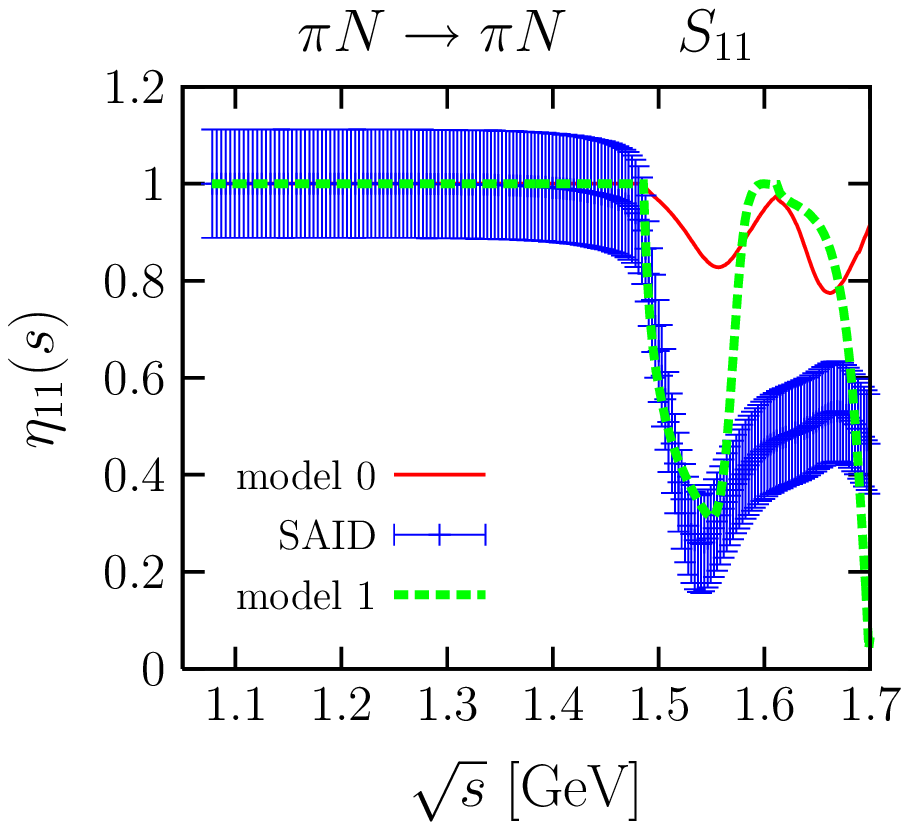}}\\ \vspace{0.5cm}
\makebox[0pt]{\hspace{-0.9cm}\includegraphics[height=7.5cm]{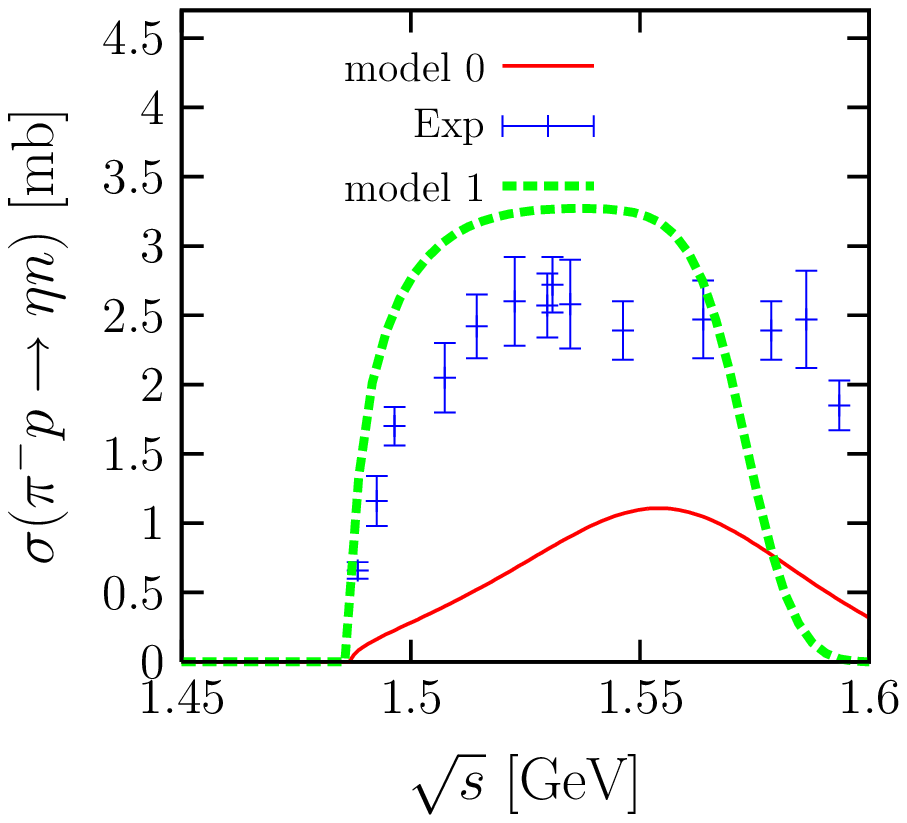}\hspace{-0.9cm}\includegraphics[height=7.5cm]{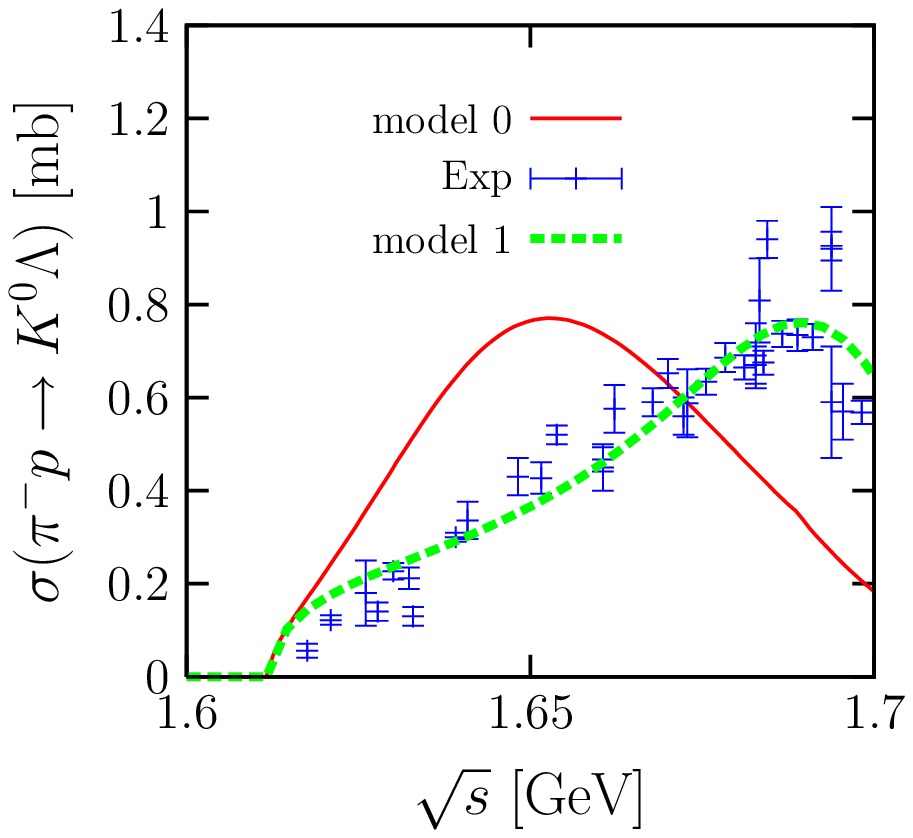}}
\end{center}
\caption{Top panels: $S_{11}$-elastic $\pi N$ phase shifts (left) and
inelasticities (right) as a function of the C.M. energy
$\sqrt{s}$. Data are from Ref.~\cite{Arndt:1995bj}, and to better
appreciate the discrepancies, we have
assumed in both cases a 5\% systematic error and a statistical
uncertainty of $5^{\rm o}$ and 0.1 for $\delta's$ and $\eta's$,
respectively (errors have been added in quadrature). Bottom panels:
$\pi^- p \to \eta n$ and $\pi^- p \to K^0 \Lambda$ total cross sections as a
function of $\sqrt{s}$. Data are from
Ref.~\cite{Ba88}.  Solid lines (model 0) stand for the predictions
obtained within the scheme presented here, where no parameters have
been adjusted to data. Dashed lines (model 1) show results from a modified
model, where in Eq.~(\ref{eq:subs}), the subtraction constants
$\bar{J}_0(\mu=\sqrt{m_\pi^2+M_N^2}; M_i,m_i))$ are multiplied 
by the factors 
$-0.039, 0.007, 2.228, -0.539, 0.443,0.757,2.855, 0.710,1.460, 2.677,
0.333$, for the $\pi N$, $\eta N$, $K \Lambda$, $K \Sigma$, $\rho N$,
$\omega N$, $\phi N$, $\rho \Delta$, $K^* \Lambda$, $K^* \Sigma$,
$K^*\Sigma^*$ channels, respectively.}
\label{fig:s11}
\end{figure}

In this appendix, we pay an special attention to $\pi N$ elastic and
inelastic scattering, and in particular to phase shifts,
inelasticities and some total inelastic cross sections in the $S_{11}$
wave (notation $L_{2I 2J}$, with $L$ the $\pi N$ orbital angular
momentum).  We will restrict our discussion to
relatively low energies ($\sqrt{s}< 1.7$ GeV), where the $N(1535)$ and 
$N(1650)$ four star resonances generated by the interaction in the 
56 and 70 irreps  should play a central role. Phase shifts,
inelasticities and inelastic cross sections are evaluated using
Eqs.(18), (19) and (35) of Ref.~\cite{juan3}, but replacing this
latter equation   by
\begin{equation}
[f_0^{\frac12}(s)]_{BA} = - \frac{1}{8\pi\sqrt{s}}
\sqrt{\frac{|\vec{k}_B|}{|\vec{k}_A|}}\frac{T_{BA}}{\sqrt{2M_A}\sqrt{2M_B}}
\end{equation}
for the transition $ B \leftarrow A$. This is necessary to account for
minor differences between the normalizations used here and those
employed in \cite{juan3}.

At first sight, the model presented up to here (solid lines in
Fig.~\ref{fig:s11}, labeled model 0 there) leads to a poor description
of data, though it already explains their gross features\footnote{Note
that we do not show results for the $\pi^- p \to K^0 \Sigma^0$ total
cross section because of the likely sizable isospin 3/2 contribution,
and that as the C.M. energy increases, higher $\pi N$ wave
contributions (neglected here) to the two total inelastic cross
sections showed in Fig.~\ref{fig:s11} become much more relevant. Yet
the three body final state $\pi N \to N \pi \pi$ process, not
considered here, will affect to the inelasticities, as
well~\cite{juan3, inoue}.}. Indeed, we could appreciate the changes of
curvarture in the phase-shifts and inelasticities, which hints the
existence of both the $N(1535)$ and $N(1650)$ resonances. Those states
show clearly up in the $\pi^- p \to \eta n$ and $\pi^- p \to K^0
\Lambda$ total cross sections, as well.

This raw description of the data should not be surprising, since we
have not fitted any parameter and we have just retained here the SU(3)
WT lowest order contribution to fix the SU(6) interaction. Accurate
descriptions of data have been achieved in previous
works~\cite{juan3,Bruns:2010sv,Lutz:2001mi}, but always within more
general schemes, where a large number of parameters are fitted to
data. Thus for instance in \cite{juan3}, though vector meson and
decuplet baryon degrees of freedom are not incorporated and the SU(3)
WT is taken as the kernel to solve a Bethe-Salpeter Equation (BSE),
the consideration of off-shell effects in \cite{juan3} led to a total
of 12 counter-terms which are fitted to data. Four of them (one for
each of the four channels, $\pi N$, $\eta N$, $K \Lambda$ and $K
\Sigma$, included in \cite{juan3}) are the subtraction constants needed
in Eq.~(\ref{eq:subs}) to renormalize the ultraviolet divergences in
the loop function. Here, we have not only neglected the counter-terms
that arise from off shell effects in the solution of the BSE, but
moreover, those counter-terms that appear in the on-shell scheme
adopted here have been also totally fixed, by means of the
prescription of Eq.~(\ref{eq:musi}), instead of fitting
them to data. In Ref.~\cite{Bruns:2010sv}, there is a total of 17 free
parameters, given by the 14 low energy constants that appear when one
goes beyond the SU(3) WT term, and includes all dimension two contact
terms, as well as three subtraction constants for the regularized loop
integrals\footnote{In \cite{Bruns:2010sv}, it is assumed the same
subtraction constant for both the $K\Lambda$ and $K\Sigma$
channels.}. Finally, in \cite{Lutz:2001mi} an even larger number of
parameters is fitted to data.

We could adopt here also a more flexible RS, and relax the
prescription of Eq.~(\ref{eq:musi}) to achieve a better agreement to
data. Fitting the subtraction constants to data is a difficult task,
since there are likely many local minima, and it requires a careful
analysis. Besides, it would require also to work in parallel possible
off-shell effects~\cite{juan3} and next-to-leading
contributions~\cite{Bruns:2010sv} to the kernel of the BSE. These
latter ones should account also for additional SU(6) and SU(3)
breaking terms to be considered on top of those already incorporated
in our simplified scheme. This is an ambitious and formidable task,
which is out of the scope of the present work. Here, we just aim to
show how the underlying chiral symmetry of the WT term induces a
qualitative SU(6) classification pattern, where most of the lowest
lying odd parity three and four star resonances of the PDG fill into
70 and 56 irrep SU(6) multiplets.  We would like however to point out
that there are regions in the parameter space which lead to better
descriptions of the scattering data. As a matter of example, we show
also in Fig.~\ref{fig:s11} results (dashed lines, labeled model 1
there) which look like more phenomenological acceptable, and that have
been obtained by modifying the prescription of Eq.~(\ref{eq:musi})
(see figure caption, for some more details\footnote{In some cases,
there are appear large deviations when compared to the prescription of
Eq.~(\ref{eq:musi}), which however do not attribute much physical
relevance, because of the complexity of the parameter space, as we
already mentioned.}).  For this particular set of parameters the state
identified with the $N(1535)$ becomes wider whereas that identified
with the $N(1650)$ becomes lighter than the corresponding states in
the model~0.

The conclusions of the above discussion are similar for other sectors
of strangeness, spin and isospin. 

In summary the simple scheme advocated in this work, where no
parameters are being fitted, provides the main features of the
lowest-lying odd parity baryon resonances. However, one should expect
a not too good, actually it could be  poor, description of data
but that however hints their major features. This situation is similar for 
other simple models, like those of Refs.~\cite{sarkar, eulogio,
bao,angels}, where the $\Delta$ baryon decuplet and the vector meson
nonet degrees of freedom are taken into account. Accurate
descriptions of data, beyond masses, widths and the main couplings of
the relevant low lying resonances in each sector,   can been achieved,
but it requires much more physics to enter in the form of undetermined
counter-terms.

%%%%%%%%%%%%%%%%%%%%%%%%%%%%%%%%%%%%%%%%%%%%

\end{document}